\documentclass[12pt]{article}

\pdfoutput=1
\usepackage[usenames,dvipsnames]{color}
\usepackage{amsmath, amssymb}
\usepackage{epsfig, palatino}
\usepackage{pstricks,pst-node,pst-tree}
\usepackage{epic}
\usepackage{ulem}

\usepackage[font={small}]{caption}

\usepackage{lscape}

\usepackage{mathrsfs}

\usepackage{ae} 
\usepackage[T1]{fontenc}
\usepackage[ansinew]{inputenc}
\usepackage{amsmath}
\usepackage{amssymb}
\usepackage{graphicx}
\usepackage{color}
\definecolor{darkblue}{cmyk}{0.9,0.9,0,0}
\definecolor{applegreen}{rgb}{0.55, 0.71, 0.0}
\usepackage[colorlinks=true,linkcolor=darkblue,citecolor=darkblue,urlcolor=darkblue]{hyperref}
\usepackage{cite}
\usepackage{hyperref}
\usepackage{wasysym}

\def\w{{\omega}}

\newcommand{\comment}[1]{}

\newcommand{\beq}{\begin{equation}}
\newcommand{\eeq}{\end{equation}}
\newcommand{\beqq}{\begin{equation*}}
\newcommand{\eeqq}{\end{equation*}}
\newcommand\beqa{\begin{eqnarray}}
\newcommand\eeqa{\end{eqnarray}}
\newcommand\beqaa{\begin{eqnarray*}}
\newcommand\eeqaa{\end{eqnarray*}}
\newcommand\bea{\begin{array}}
\newcommand\eea{\end{array}}

\def\XXint#1#2#3{{\setbox0=\hbox{$#1{#2#3}{\int}$}
\vcenter{\hbox{$#2#3$}}\kern-.5\wd0}}

\newcommand{\nn}{\nonumber}

\newcommand{\neqa}{\nonumber\end{eqnarray}} 
\newcommand{\la}[1]{\label{#1}}

\renewcommand{\d}{\partial}

\newcommand{\<}{{\langle}}
\renewcommand{\>}{{\rangle}}

\newcommand{\cC}{{\cal C}}

\newcommand{\re}{\relax{\rm I\kern-.18em R}}

\renewcommand{\sp}{p\hspace{-.40em}/}

\newcommand{\ft}[2]{{\textstyle\frac{#1}{#2}}}

\def\su2{{SU(2)}}

\def\[{\left[}
\def\]{\right]}

\def\({\left(}
\def\){\right)}
\def\[{\left[}
\def\]{\right]}

\def\<{\langle}
\def\>{\rangle}

\def\i2{\frac{i}{2}}

\def\spi{\relax{\rm \pi\kern-0.5em /}}
\def\sA{\relax{\rm A\kern-0.5em /}}
\def\sp{\relax{\rm p\kern-0.5em /}}
\def\sd{\relax{\rm \d\kern-0.5em /}}
\def\sk{\relax{\rm k\kern-0.5em /}}
\def\sn{\relax{\rm n\kern-0.5em /}}
\def\sl{\relax{\rm l\kern-0.5em /}}
\def\sP{\relax{\rm P\kern-0.7em /}}
\def\sBethe{\relax{\rm \Bethe\kern-0.5em /}}

\def\cF{{\cal F}}

\def\cF{{\cal F}}
\def\cC{{\cal C}}
\def\cR{{\cal R}}
\def\cO{{\cal O}}

\def\cW{{\cal W}}

\def\cD{{\cal D}}
\def\2F1{\,_2{\rm F}_1}

 \newcommand{\Blue}[1]{{\color{blue}#1\color{black}}}
\newcommand{\Red}[1]{{\color{red}#1\color{black}}}

\usepackage{varioref}
\usepackage{makeidx}
\makeindex

\usepackage[english]{babel}
\begin{document}

\thispagestyle{empty}

\renewcommand{\thefootnote}{\fnsymbol{footnote}}
\setcounter{page}{1}
\setcounter{footnote}{0}
\setcounter{figure}{0}
\begin{center}
$$$$
{\Large\textbf{\mathversion{bold}
Space-time S-matrix and Flux-tube S-matrix III.\\ 
The two-particle contributions
}\par}

\vspace{1.0cm}

\textrm{Benjamin Basso$^{{\displaystyle\hexagon}}$, Amit Sever$^{{\displaystyle\Box}}$ and Pedro Vieira$^{\displaystyle\pentagon}$}
\\ \vspace{1.2cm}
\footnotesize{\textit{
$^{\displaystyle\pentagon}$Perimeter Institute for Theoretical Physics,
Waterloo, Ontario N2L 2Y5, Canada\\
$^{\displaystyle\hexagon}$Laboratoire de Physique Th\'eorique, \'Ecole Normale Sup\'erieure, Paris 75005, France\\
$^{\displaystyle\Box}$School of Natural Sciences, Institute for Advanced Study, Princeton, NJ 08540, USA
}  
\vspace{4mm}
}

\par\vspace{1.5cm}

\textbf{Abstract}\vspace{2mm}
\end{center}
We consider light-like Wilson loops with hexagonal geometry in the planar limit of $\mathcal{N}=4$ Super-Yang-Mills theory. Within the Operator-Product-Expansion framework these loops receive contributions from all states that can propagate on top of the colour flux tube sourced by any two opposite edges of the loops. Of particular interest are the two-particle contributions. They comprise virtual effects like the propagation of a pair of scalars, fermions, and gluons, on top of the flux tube. Each one of them is thoroughly discussed in this paper. Our main result is the prediction of all the twist-2 corrections to the expansion of the dual 6-gluons MHV amplitude in the near-collinear limit at finite coupling. At weak coupling, our result was recently used by Dixon, Drummond, Duhr and Pennington to predict the full amplitude at four loops. At strong coupling, {it allows us to make} contact with the classical string description and {to recover the (previously elusive) $\rm AdS_3$ mode from the continuum of two-fermion states}. {More generally,} the two-particle contributions serve as an exemplar for all the multi-particle corrections.

\noindent

\setcounter{page}{1}
\renewcommand{\thefootnote}{\arabic{footnote}}
\setcounter{footnote}{0}

 \def\nref#1{{(\ref{#1})}}

\newpage

\tableofcontents

\parskip 5pt plus 1pt   \jot = 1.5ex
\newpage

\section{Introduction}

In the planar $\mathcal{N}=4$ Super-Yang-Mills theory, null polygonal Wilson loops can be computed at any value of the coupling using the Operator Product Expansion~\cite{OPEpaper}. In this approach, which is analogous to the usual OPE for local operators, the Wilson loop is decomposed into sums over the color flux-tube eigenstates $\psi$ propagating in the consecutive OPE channels~\cite{heptagonPaper}. The simplest example, which will also be the focus of this paper, is the hexagonal Wilson loop or, more precisely, the conformally invariant finite ratio $\mathcal{W}\equiv \mathcal{W}_\text{hexagon}$ introduced in \cite{short,long}. It is given as a sum over a single OPE channel~\cite{short}
\beq\la{hexagon2}
\begin{array}{l}
\def\svgwidth{9cm}
\begingroup%
  \makeatletter%
  \providecommand\color[2][]{%
    \errmessage{(Inkscape) Color is used for the text in Inkscape, but the package 'color.sty' is not loaded}%
    \renewcommand\color[2][]{}%
  }%
  \providecommand\transparent[1]{%
    \errmessage{(Inkscape) Transparency is used (non-zero) for the text in Inkscape, but the package 'transparent.sty' is not loaded}%
    \renewcommand\transparent[1]{}%
  }%
  \providecommand\rotatebox[2]{#2}%
  \ifx\svgwidth\undefined%
    \setlength{\unitlength}{997.45273438bp}%
    \ifx\svgscale\undefined%
      \relax%
    \else%
      \setlength{\unitlength}{\unitlength * \real{\svgscale}}%
    \fi%
  \else%
    \setlength{\unitlength}{\svgwidth}%
  \fi%
  \global\let\svgwidth\undefined%
  \global\let\svgscale\undefined%
  \makeatother%
  \begin{picture}(1,0.35415388)%
    \put(0,0){\includegraphics[width=\unitlength]{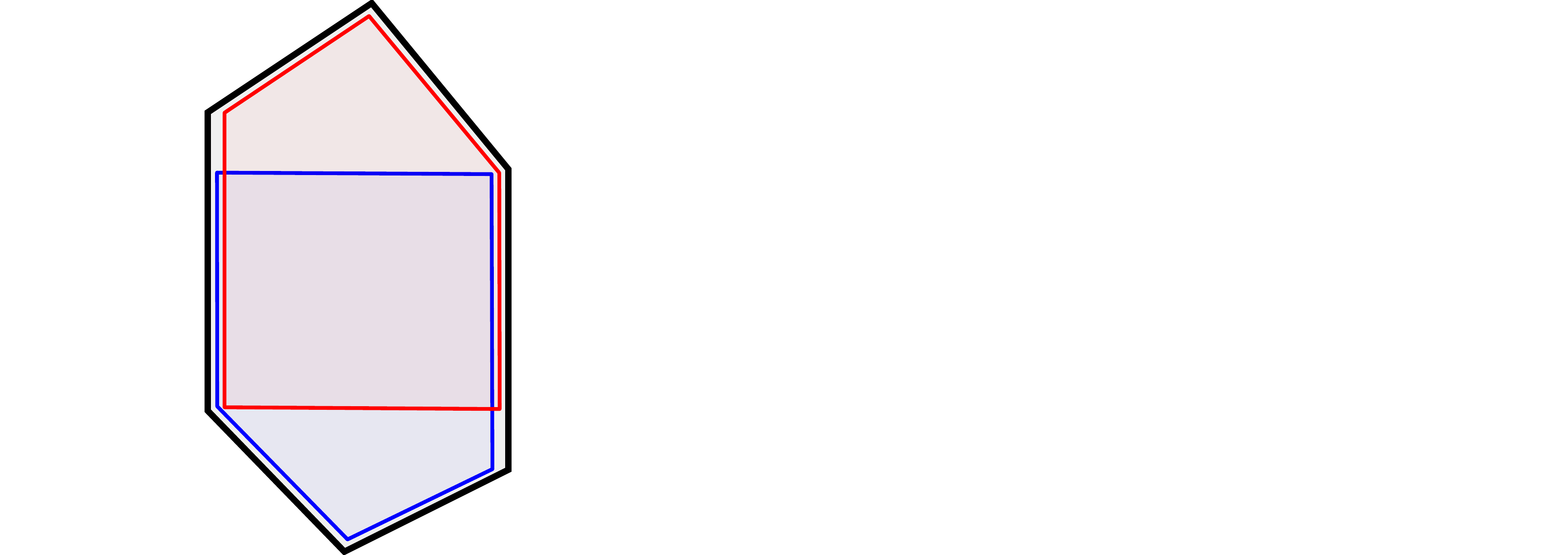}}%
    \put(-0.00161,0.16202654){\color[rgb]{0,0,0}\makebox(0,0)[lb]{\smash{$\cW=$}}}%
    \put(0.35128893,0.16202654){\color[rgb]{0,0,0}\makebox(0,0)[lb]{\smash{$=\sum\limits_\psi   P(0|\psi)e^{-E\tau+ip\sigma+im\phi}P({\psi}|0)$}}}%
  \end{picture}%
\endgroup%
 
\end{array}
\eeq
where $\tau,\sigma$ and $\phi$ parameterize the three independent (conformal) cross ratios of the hexagon, while $E,p$ and $m$ are the conjugate energy, momentum and angular momentum of the state~$\psi$ flowing in the middle square. The most non-trivial ingredients in this expression are the form factors $P(0|\psi)$ and $P(\psi|0)$. They stand for the creation and annihilation amplitudes of the state $\psi$ at the bottom and top of the hexagon, respectively, and are the Wilson loop analog of the familiar structure constants for correlation functions. They describe the transition between two consecutive null squares and were dubbed \textit{pentagon transitions} in~\cite{short,long}.

In the near collinear limit, the flux-tube time coordinate $\tau$ is large and the least energetic states $\psi$ dominate. These are the \textit{single-particle} states that were studied in detail in \cite{short,long}. Beyond it stands the realm of multi-particle corrections which for the most part is uncharted territory.
Among all these contributions, those associated to the \textit{two-particle} states are dominating at large $\tau$, see figure~\ref{hexagonFigure}, and are the simplest ones. 
The analysis performed in this paper is chiefly dedicated to them.

This paper is further intended to unveil the structure of multi-particle corrections and the two-particle contributions serve as an exemplar in this regard. {We shall see that} the latter already exhibit most of the new physics associated to multi-particle states.
In the end, the fact that such a reduction from the many- to two-body problem operates should be no surprise. It is indeed the hallmark of {\textit{all}} integrable models, of which the flux-tube theory is merely one instance among many others. 
The sharpest illustration is provided by the factorization of the many-body S-matrix into a sequence of two-to-two scattering events, see \cite{Dorey} for a review. Closer to our discussion is the example of the multi-particle form factors which are often built out of the two-particle ones, see e.g. \cite{Watson}.  The application of this program to the pentagon transitions was initiated in~\cite{short} and this paper stands as its continuation. 

\begin{figure}
\def\svgwidth{15cm}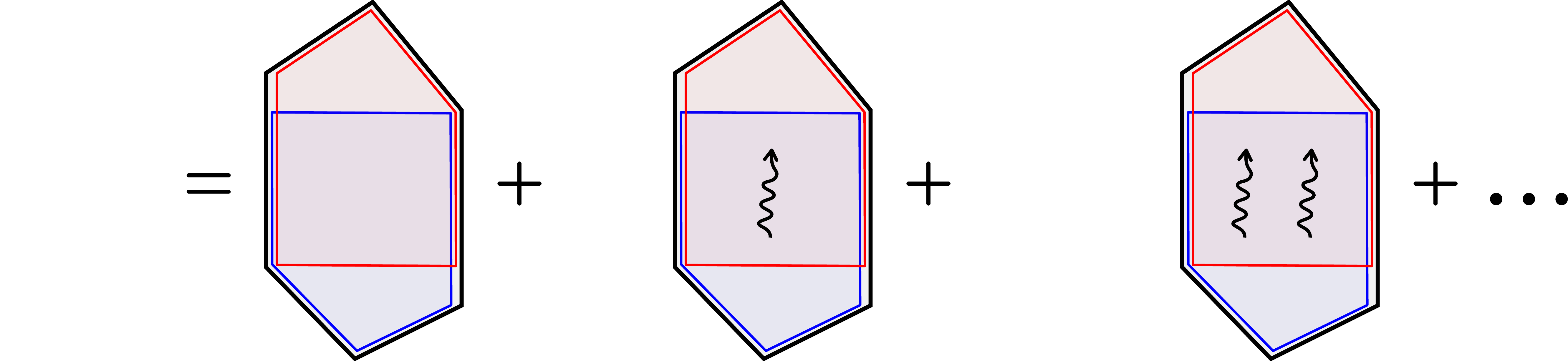
\caption{In the near collinear limit $\tau \to \infty$, the Wilson loop $\mathcal{W}$ has an expansion in the number of particles flowing in the color flux tube. Here we depict the three leading contributions corresponding to the vacuum, single-particle and two-particle states, respectively.}
\la{hexagonFigure}
\end{figure}

Our discussion shall be mostly non-perturbative and our main results for the two-particle contributions will hold at any value of the 't Hooft coupling $g\equiv \sqrt{\lambda}/(4\pi)$, with $\lambda = g^2_{YM}N$.

At weak coupling, $g^2\to0$, our conjecture predicts the next-to-next-to-leading collinear behaviour,
\beq
\mathcal{W} =1+e^{-\tau} f_1(\tau,\sigma,\phi)+e^{-2\tau} f_2(\tau,\sigma,\phi)+\mathcal{O}(e^{-3\tau}) \la{fgexpression}\, ,
\eeq
to all loops. At any given order, the twist-one and twist-two corrections, $f_1$ and $f_2$, are polynomials in both $\tau$ and $e^{i\phi}$, but non-trivial functions of $\sigma$. The leading one, $f_1$, is governed by the contributions of the lightest single-particle states and was studied in \cite{short}. Here we will predict $f_2$ which receives most contributions from the two-particle states. As discussed in greater length in section \ref{WeakSec}, the hexagon functions program pushed forward by Dixon et al.~in~\cite{Lance,4Loops} provides a plethora of valuable checks of our predictions and vice-versa.

At strong coupling, $\sqrt{\lambda}\gg 1$, we shall pin down the contributions corresponding to the three modes of the dual string in $\rm AdS_5$. Two of them were previously uncovered in~\cite{short} and attributed to the single-particle states. The missing one, which describes the transverse fluctuation in the $\rm AdS_3$ subspace, will be found here to emerge from the continuum of two-fermion states. Perhaps less intuitive is the important role played at strong coupling by the {stringy fluctuations along the sphere $\rm S^5$}, that we shall also highlight.

The paper is articulated as follows. In section~\ref{reviewSec}, we review the flux-tube spectrum of particles. In sections~\ref{Gluonssec}, \ref{scalar-section}, and~\ref{fermionssection}, we consider in turn the two gluons, two scalars, and two fermions contributions at finite coupling. In section~\ref{fullSec}, we collect all the contributions together and elaborate on the proper contours of integrations. In sections~\ref{WeakSec} and~\ref{strongcouplingsec}, we expand our result at weak and strong coupling, respectively. At weak coupling, we unveil an effective description in terms of emergent flux-tube excitations and confront our findings with the perturbative results in the literature. At strong coupling,
we make contact with the string theory description in terms of minimal surface in $\rm AdS_5$. We end in section \ref{discussion-sect} with a discussion of our results and an overview of the multi-particle OPE program. Three appendices complement the main text. Appendix~\ref{paradoxSec} can be read on its own. It contains a thorough discussion of the fermionic excitations and their associated S-matrices. Appendix~\ref{appenpixB} complements  sections \ref{Gluonssec}, \ref{scalar-section}, and \ref{fermionssection}, with more details on the transitions. Finally, appendix~\ref{fundamentaltransitions} contains a summary of all the finite-coupling transitions (as well as their analytic continuations).

\section{Review of the flux-tube spectrum} \la{reviewSec}

\begin{figure}
\centering
\def\svgwidth{13cm}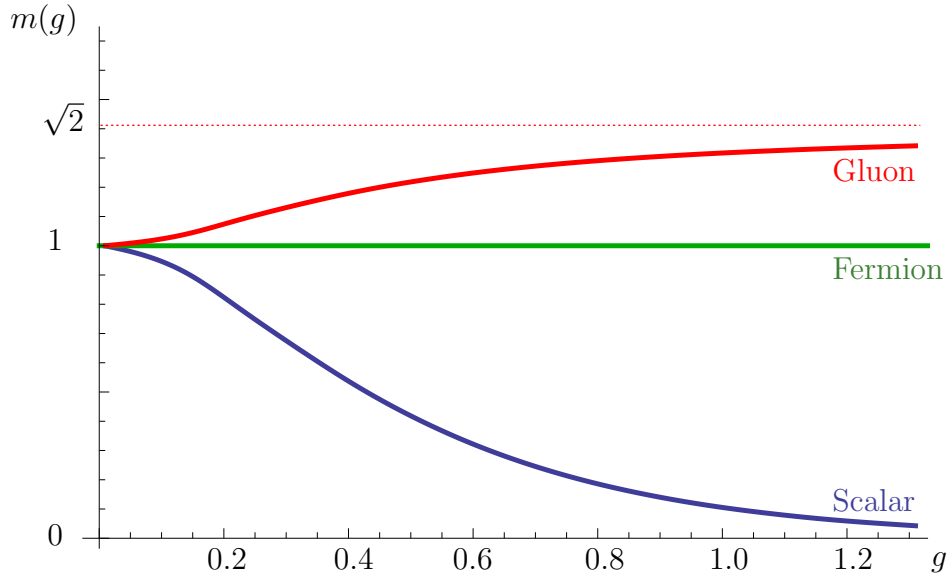
\caption{Masses $m(g)$ of the twist-one excitations as functions of the coupling $g = \sqrt{\lambda}/(4\pi)$. The lightest excitations are the scalar ones. Their mass defines the mass gap of the theory and it becomes exponentially small at strong coupling~\cite{AldayMaldacena}. Its plot above agrees with the one in~\cite{Fioravanti:2008rv} which studied the related problem of solving the Freyhult-Rej-Staudacher equation~\cite{Freyhult:2007pz}. The mass of the fermions is protected by supersymmetry~\cite{AldayMaldacena} while the one of the gluons interpolates between $1$ and~$\sqrt{2}$.}\label{masses}
\end{figure}

As alluded to before, in the OPE approach, one has to sum over all the excitations $\psi$ of the colour flux tube. This set of states forms the Fock space of a two dimensional theory whose structure can be unravelled at any coupling, thanks to the integrability of the flux-tube dynamics in the  planar limit. In this section we review its main features.

As recently discussed in~\cite{long}, in relation to our problem, there are many equivalent ways of thinking about the flux tube and its excitations. One of them operates in terms of large-spin {single trace} operators which are dual to excitations of the Gubser-Klebanov-Polyakov (GKP) string~\cite{GKP}. This picture gives us with both a classification of the flux-tube eigenstates and a diagonalization of the flux-tube Hamiltonian, by means of the asymptotic Bethe ansatz equations~\cite{BS}. The relevant analysis was carried out in~\cite{BenDispPaper} (see also~\cite{Straps} and~\cite{Dorey:2010iy}) building on previous studies~\cite{Belitsky:2006en, BES}. The outcome is that the spectrum of large-spin operators, {or} equivalently the flux-tube Hilbert space, is {\textit{entirely}} built out of the lightest {(= twist-one)} excitations and their bound states, as we will now explain.

All states in the spectrum can be classified according to {their transformations under the} symmetries of the flux tube, see~\cite{OPEpaper} for a discussion. The latter consist of the three kinematical (and commuting abelian) transformations generated by $\partial_{\tau}, \partial_{\sigma}, \partial_{\phi}$, on the one hand, and of the internal $SU(4)$ R-symmetry of the gauge theory, on the other hand. (Note in particular that none of the supersymmetries remains on the flux tube background.) Flux-tube eigenstates can then be organized, at any value of the coupling, {according to their
energy $E$, momentum $p$, angular momentum (a.k.a.~helicity or $U(1)$ charge), and $SU(4)$ weights.}

It is also convenient to keep track of the bare twist of the excitation, {which gives} the mass $E(p=0)$ in the free theory, i.e., at $g=0$. At weak coupling, the lightest excitations are then in one-to-one correspondence with the twist-one fields of the gauge theory. (This field-excitation dictionary becomes {clearer} after recalling that excitations can be created on the flux tube by field insertions along the edges of a null Wilson loops, see~\cite{long,Wang,Belitsky:2011nn}). {The lightest excitations thus comprises $6$ scalars, $4+4$ fermions, and $1+1$ gluons. Adopting the notations of~\cite{long}, they shall be written as $(\phi_{1, \ldots, 6})$, $(\psi_{1, \ldots, 4}, \bar{\psi}^{1, \ldots , 4}),$ and $(F, \bar{F})$, respectively,
with  $F = F_{-z}$ the twist-one component of the Faraday tensor, $\bar{F} = F_{-\bar{z}}$ its complex conjugate, and $\psi = \psi_{-}, \bar{\psi} = \bar{\psi}^{-}$ the twist-one components of the Weyl fields. The $U(1)\times SU(4)$ quantum numbers of all these excitations are given in figure~\ref{fundamentalExcitations}.

Importantly enough, all the twist-one excitations are stable and each one of them has a definite dispersion relation $E(p)$, at any coupling~\cite{BenDispPaper}. For illustration, one easily derives at weak coupling that
\beq\label{OLEn}
E(p) = 1+2g^2(\psi(s+\frac{ip}{2})+\psi(s-\frac{ip}{2})-2\psi(1)) + O(g^4)\, ,
\eeq
with $\psi(z) = \Gamma'(z)/\Gamma(z)$ the digamma function and $s$ the conformal spin of the excitation (with $s = \frac{1}{2}, 1,$ and $\frac{3}{2}$, for scalars, fermions, and gluons, respectively). We verify here that all these states are degenerate in the free theory, as they carry the same twist. The degeneracy is lifted at one loop, as seen in~(\ref{OLEn}), and to a larger extent at higher values of the coupling, as shown by the running of the masses depicted in figure~\ref{masses}.} (We also check this way that none of these excitations are superpartners of one another.) 

The twist-one particles play a very important role in our analysis, but they are not the only stable flux-tube excitations. {We also have bound states}. Whether {and which} bound states can form is always a dynamical question. In our case, it can be addressed at finite coupling using integrability. The analysis of \cite{BenDispPaper} provided the following simple picture.

\begin{figure}[t]
\centering
\def\svgwidth{16cm}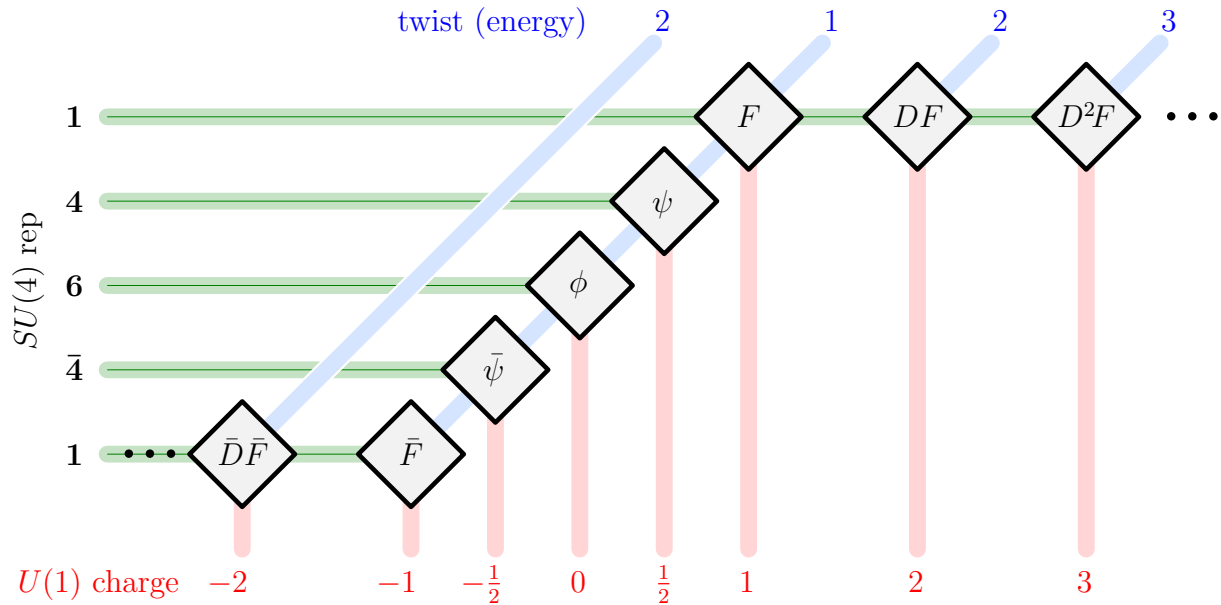
\caption{Table of fundamental excitations, represented by the squares in this figure. The twist-one excitations sit on the diagonal and their multiplicities follow from the dimensions of their $SU(4)$ representations. The two towers of gluon bound states, with twist $2, 3, \ldots$, and $U(1)$ charge $\pm 2, \pm 3, \ldots$, are depicted on the two semi-infinite lines at the top and bottom of the table.  Multi-particle states made out of these excitations span the flux-tube Hilbert space at any coupling. 
} \la{fundamentalExcitations}
\end{figure}

The only bound states are made out of gluons of the same helicity. {For example, we can have a bound state of two positive helicity gluons ${F}$ which we denote as $D {F}$, a bound state of three such gluons which we denote as $D^2F$, and so on.} The total number of gluons involved is arbitrarily large, such that we end up with two semi-infinite towers of gluonic bound states, {each} parametrized by an integer $n=1, 2, 3, \ldots$, as depicted in figure~\ref{fundamentalExcitations}. The energy and momentum of these bound-state were computed in \cite{BenDispPaper}. They are given by a simple sum over the energies and momenta of their constituents. For example, for the {two-gluon bound state} $DF$, one has
\beq
E_{DF}(u)= E_F(u+i/2)+E_F(u-i/2)\,, \qquad p_{DF}(u)= p_F(u+i/2)+p_F(u-i/2)\, , \la{EDF}
\eeq
where we introduced the so-called Bethe rapidity $u$. The latter {variable provides} a very convenient parametrization for all dispersion relations and will be {largely} used throughout this paper {in place of the momentum: $(p, E(p))\rightarrow (p(u), E(u))$}. {To some extent, it plays the same role for the flux-tube theory as does the hyperbolic rapidity for 2D relativistic models.} In particular, one nice feature of the Bethe parametrization is that the fusion into bound-states (which requires knowledge of the positions of the poles of the S-matrices) is trivially described in $u$ space at any value of the coupling. It simply amounts to certain (coupling independent) shifts of the rapidity, as in~(\ref{EDF}) for instance.

There is a subtlety, discussed in detail in section 4.4 of \cite{BenDispPaper}, that is worth recalling. Namely, to properly obtain the energy of the bound state $DF$, we \textit{must} compute the sum in~(\ref{EDF}) for rapidity $u$ with real part lying between $-2g$ and $2g$. Only afterwards, can we analytically continue it toward the full $u$ plane and take the small $g$ limit to obtain $E_{DF}(u)$ in perturbation theory. We \textit{cannot} construct $E_{DF}(u)$ out of $E_F(u)$ by applying~(\ref{EDF}) at a given order in perturbation theory, i.e., at small $g$ for fixed $u=O(1)$. The two operations simply do not commute. Since, for the latter reason, computing $E_{DF}$ and similarly $p_{DF}$ is not totally straightforward, we recall their final expressions in appendix \ref{summaryBS}. 

The twist-one states and the two gluonic towers of bound states {support the entire spectrum of flux-tube excitations} -- see figure \ref{fundamentalExcitations} for a graphical summary. We dub them as the \textit{fundamental} excitations. The claim is that it is possible to write the complete OPE in terms of multi-particle states made out of these excitations {and these excitations only}. Said otherwise, they provide us with a complete basis of asymptotic states for the flux tube. {Thereby}, if one knew the pentagon transitions for all these states, one would have the full amplitude in its OPE series representation.

That the set of fundamental excitations sketched before is enough {for our purposes} is not obvious at all.
{Naively, any adjoint (local) field of the theory, modded out by the field's equations of motion, is candidate for being a flux-tube excitation. A description in these terms would clearly involve many more excitations than we have in our set.} Whether the additional excitations are stable and visible asymptotically {(i.e. can propagate over large distance and time on the flux tube)} relies however on the flux-tube dynamics. We should stress, therefore, that our above claim is not that {evident} and already incorporates non-trivial aspects of the flux-tube dynamics.

Just as a final illustration, one could consider various components of the Faraday tensor, besides those with twist equals one, like the $F_{+-}, F_{z\bar{z}}$, and $F_{+z}$ fields (with twist 2 and 3, respectively) that arose in the weak coupling analysis of~\cite{Straps}. There is nothing wrong about that and, as a matter of fact, these excitations (or field insertions) are necessary to fully decompose the WL in the weak coupling OPE. One of the goals of this paper is actually to reveal the precise mechanism that leads to the appearance of these additional excitations in the weak coupling description. As we shall see, these extra excitations and the fundamental ones spelled out before are not on a same footing. E.g., the former cease to have a real dispersion relation at a high enough loop order and are better thought of as some sort of `long-lived' resonances. 

\section{Gluons}\la{Gluonssec}

With this section we initiate the study of the various contributions to the hexagon OPE with twist smaller or equal to two. We begin here with the gluonic contributions, which are the easiest ones. They comprise the contributions associated to single gluons,  two-gluon states and bound states of two gluons.

We start by reviewing the single-gluon case, that is, the twist-one contributions entering the decomposition (\ref{hexagon2}).  
They can be written as~\cite{short}
\beqa
\mathcal{W}^\text{\,1-gluon}=  e^{ i \phi} \,\mathcal{W}_{F} +e^{-i \phi}\, \mathcal{W}_{\bar{F}}\, ,  \la{hexagonSingle}
\eeqa
where 
\beq
\mathcal{W}_{F}=\mathcal{W}_{\bar F} = \int \frac{du}{2\pi} \, \hat\mu_F(u)\, .  \la{WF}
\eeq
The two terms in~(\ref{hexagonSingle}) correspond to the two possible polarizations, $F$ and~$\bar F$, of our gluonic excitation. These two states carry opposite $U(1)$ charges, as made manifest by their couplings to the angular ``cross-ratio" $\phi$ in~(\ref{hexagonSingle}). The associated two contributions both take the form of a single integral~(\ref{WF}) over the rapidity $u$. This is nothing but the sum over the (flux-tube) momentum $p$ of the gluon {that is propagating} through the middle square of the hexagon Wilson loop~(\ref{hexagon2}). This process, whenever it happens, comes with a specific weight~\cite{short,long} given by the product of the so-called square measure $\mu_{F}(u)$ with the propagating factor $\exp{(-\tau E_{F}(u)+i\sigma p_{F}(u))}$. The former is a function of the rapidity and coupling only, while the latter obviously incorporates the dependence on the space and time ``cross-ratios" $\sigma$ and~$\tau$. The measure and propagating factors always come along. This is why, throughout this paper, we shall use the compact and convenient notation
\beq\la{mueffective}
\hat{\mu}_{a}(u) \equiv \mu_{a}(u)e^{-\tau E_{a}(u)+i p_{a}(u) \sigma} \, , \qquad a = F, \phi, \psi, \ldots\, ,
\eeq
as in~(\ref{WF}). The finite-coupling expression for the measure $\mu_F$ was proposed in~\cite{short} and is summarized in appendix~\ref{fundamentaltransitions}. 

We move next to the contribution of the gluonic states with (total) twist equal to two. The simplest are those made out of two (separate) gluonic excitations. They are given by
\beqa
\mathcal{W}^\text{\,2-gluon}=  e^{2 i \phi} \,\mathcal{W}_{FF} +\mathcal{W}_{F\bar{F}}+  e^{-2i \phi}\, \mathcal{W}_{\bar F \bar F} \, ,
\eeqa
where  
\beqa
\mathcal{W}_{FF}&=& \frac{1}{2} \int \frac{du\, dv}{(2\pi)^2}P_{FF}(0|u,v) \hat\mu_{F}(u)\hat\mu_{F}(v)P_{FF}(-v,-u|0) \la{WFF}\, ,\\
\mathcal{W}_{F\bar F}&=&\ \ \ \int \frac{du\, dv}{(2\pi)^2}P_{F\bar F}(0|u,v) \hat\mu_{F}(u)\hat\mu_{F}(v)P_{F\bar F}(-v,-u|0)\, ,\la{WFbarF}
\eeqa
and $\cW_{\bar F\bar F}=\cW_{FF}$. The factor $1/2$ in $\cW_{FF}$ is the symmetry factor for two identical particles. 
The new ingredients here are the creation and annihilation amplitudes for two gluonic excitations on the top and bottom of a pentagon, e.g., $P_{FF}(0|u,v)$ and $P_{FF}(-v,-u|0)$. Determining them is what this section is mostly about.

We shall notice before that the vexing minus signs in~(\ref{WFF}),~(\ref{WFbarF}) are mainly conventional, see, e.g.,~equation (3) in \cite{short}. They are relevant for polygons with seven or more edges, where we have intermediate pentagon transitions with excitations on both the top and bottom. Here, for the hexagon, we can always {trade them for a re-ordering of the rapidities, using the reflection symmetries of a pentagon:
\beq
P_{F\bar F}(-v,-u|0)= P_{F\bar F}(u,v|0)= P_{F\bar F}(0|v,u)= P_{F\bar F}(0|-u,-v) \,, \la{chain}
\eeq
and similarly for $P_{FF}$. This is telling us along the way that the creation and annihilation form factors are not independent from one another.

The creation amplitudes of interest are actually not new objects.
Instead, in disguise, they are the same as the pentagon transitions studied in~\cite{short,long}. The latter, we recall, describe the two transitions of a single gluon from the bottom to top of a pentagon,
\beq \la{singleGluonTransition}
\begin{array}{l}
\def\svgwidth{14cm}
\begingroup%
  \makeatletter%
  \providecommand\color[2][]{%
    \errmessage{(Inkscape) Color is used for the text in Inkscape, but the package 'color.sty' is not loaded}%
    \renewcommand\color[2][]{}%
  }%
  \providecommand\transparent[1]{%
    \errmessage{(Inkscape) Transparency is used (non-zero) for the text in Inkscape, but the package 'transparent.sty' is not loaded}%
    \renewcommand\transparent[1]{}%
  }%
  \providecommand\rotatebox[2]{#2}%
  \ifx\svgwidth\undefined%
    \setlength{\unitlength}{1016.10820313bp}%
    \ifx\svgscale\undefined%
      \relax%
    \else%
      \setlength{\unitlength}{\unitlength * \real{\svgscale}}%
    \fi%
  \else%
    \setlength{\unitlength}{\svgwidth}%
  \fi%
  \global\let\svgwidth\undefined%
  \global\let\svgscale\undefined%
  \makeatother%
  \begin{picture}(1,0.1854929)%
    \put(0,0){\includegraphics[width=\unitlength]{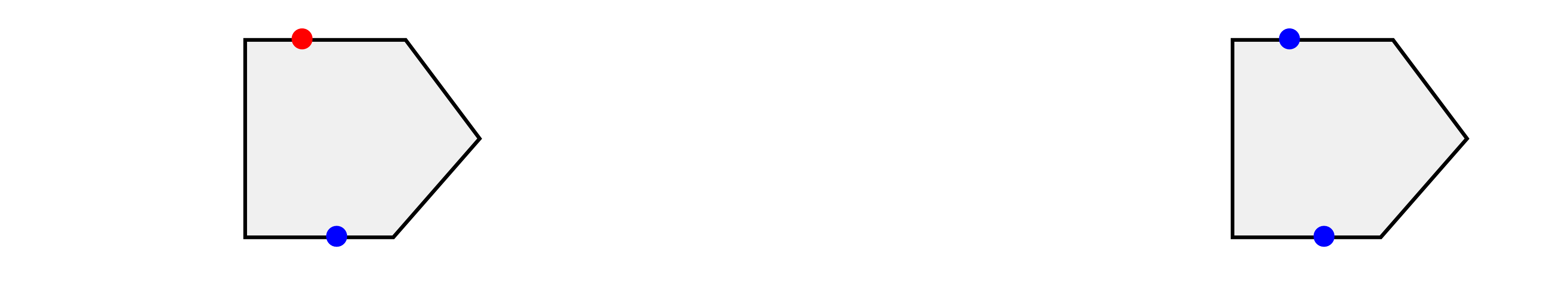}}%
    \put(-0.03247301,0.09204868){\color[rgb]{0,0,0}\makebox(0,0)[lb]{\smash{$\displaystyle P_{FF}(u|v)\ =$}}}%
    \put(0.20512279,0.04934827){\color[rgb]{0,0,0}\makebox(0,0)[lb]{\smash{$u$}}}%
    \put(0.18381896,0.13442358){\color[rgb]{0,0,0}\makebox(0,0)[lb]{\smash{$v$}}}%
    \put(0.20555091,0.00173571){\color[rgb]{0,0,0}\makebox(0,0)[lb]{\smash{$F$}}}%
    \put(0.18193137,0.17652024){\color[rgb]{0,0,0}\makebox(0,0)[lb]{\smash{$\bar F$}}}%
    \put(0.59738119,0.09204868){\color[rgb]{0,0,0}\makebox(0,0)[lb]{\smash{$\displaystyle P_{F\bar F}(u|v)\ =$}}}%
    \put(0.834977,0.04934827){\color[rgb]{0,0,0}\makebox(0,0)[lb]{\smash{$u$}}}%
    \put(0.8136731,0.13442358){\color[rgb]{0,0,0}\makebox(0,0)[lb]{\smash{$v$}}}%
    \put(0.83540506,0.00173571){\color[rgb]{0,0,0}\makebox(0,0)[lb]{\smash{$F$}}}%
    \put(0.81178548,0.17652024){\color[rgb]{0,0,0}\makebox(0,0)[lb]{\smash{$F$}}}%
    \put(0.32969314,0.09204868){\color[rgb]{0,0,0}\makebox(0,0)[lb]{\smash{$,$}}}%
    \put(0.95954735,0.09204868){\color[rgb]{0,0,0}\makebox(0,0)[lb]{\smash{$,$}}}%
  \end{picture}%
\endgroup%
 
\end{array}
\eeq
where blue/red dots represent insertions of a positive/negative helicity gluon field. Note that, in our conventions, the transition $P_{FF}$ corresponds to inserting the field $F$ on the bottom and its conjugate $\bar F$ on the top. This is thus the one preserving the $U(1)$ charge of the excitation that is flowing inbetween, whereas $P_{F\bar F}$ violates it. In \cite{short,long} we presented conjectures for these single-particle transitions at any coupling. Their explicit expressions are recalled in Appendix~\ref{fundamentaltransitions} for completeness.  

To relate the pentagon transitions~(\ref{singleGluonTransition}) to the amplitudes entering~(\ref{WFF}),~(\ref{WFbarF}) we should use the mirror transformation $u\to u^{\gamma}$. It permits to move excitations around, from one edge of the pentagon to the neighbouring one on its left, precisely. Whenever a gluon is rotated in this manner it also flips its helicity~\cite{short,long}. To get the transition with two $F$'s on the top, i.e., $P_{FF}(0|u,v)$, we can thus start from $P_{F\bar F}(u|v)$ and move the bottom gluon to the top through the left edge. If, instead, we bring it up through the right, by means of three opposite moves, we would end up with an $F$ and an $\bar F$ excitations on the top. In summary, \\
\beq
\begin{array}{l}
\def\svgwidth{15cm}
\begingroup%
  \makeatletter%
  \providecommand\color[2][]{%
    \errmessage{(Inkscape) Color is used for the text in Inkscape, but the package 'color.sty' is not loaded}%
    \renewcommand\color[2][]{}%
  }%
  \providecommand\transparent[1]{%
    \errmessage{(Inkscape) Transparency is used (non-zero) for the text in Inkscape, but the package 'transparent.sty' is not loaded}%
    \renewcommand\transparent[1]{}%
  }%
  \providecommand\rotatebox[2]{#2}%
  \ifx\svgwidth\undefined%
    \setlength{\unitlength}{1476.55478516bp}%
    \ifx\svgscale\undefined%
      \relax%
    \else%
      \setlength{\unitlength}{\unitlength * \real{\svgscale}}%
    \fi%
  \else%
    \setlength{\unitlength}{\svgwidth}%
  \fi%
  \global\let\svgwidth\undefined%
  \global\let\svgscale\undefined%
  \makeatother%
  \begin{picture}(1,0.17550579)%
    \put(0,0){\includegraphics[width=\unitlength]{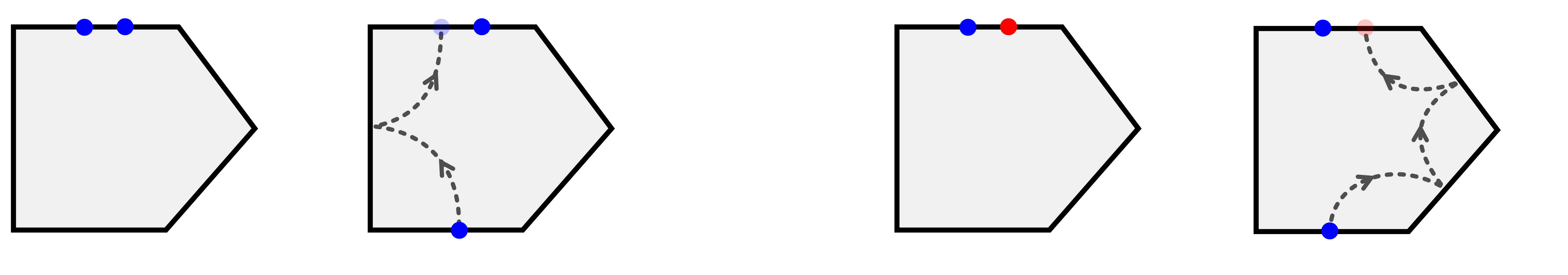}}%
    \put(0.75050157,0.08728014){\color[rgb]{0,0,0}\makebox(0,0)[lb]{\smash{$\displaystyle =$}}}%
    \put(0.63550742,0.12792387){\color[rgb]{0,0,0}\makebox(0,0)[lb]{\smash{$v$}}}%
    \put(0.60733373,0.12792387){\color[rgb]{0,0,0}\makebox(0,0)[lb]{\smash{$u$}}}%
    \put(0.80647068,0.0537324){\color[rgb]{0,0,0}\makebox(0,0)[lb]{\smash{$v^{-3\gamma}$}}}%
    \put(0.83597408,0.12792387){\color[rgb]{0,0,0}\makebox(0,0)[lb]{\smash{$u$}}}%
    \put(0.18702773,0.08728014){\color[rgb]{0,0,0}\makebox(0,0)[lb]{\smash{$\displaystyle =$}}}%
    \put(0.29959032,0.03798478){\color[rgb]{0,0,0}\makebox(0,0)[lb]{\smash{$u^{2\gamma}$}}}%
    \put(0.29742312,0.12792387){\color[rgb]{0,0,0}\makebox(0,0)[lb]{\smash{$v$}}}%
    \put(0.07203358,0.12792387){\color[rgb]{0,0,0}\makebox(0,0)[lb]{\smash{$v$}}}%
    \put(0.04385989,0.12792387){\color[rgb]{0,0,0}\makebox(0,0)[lb]{\smash{$u$}}}%
    \put(0.0481943,0.17018441){\color[rgb]{0,0,0}\makebox(0,0)[lb]{\smash{$\Blue{F}$}}}%
    \put(0.06986637,0.17018441){\color[rgb]{0,0,0}\makebox(0,0)[lb]{\smash{$\Blue{F}$}}}%
    \put(0.28441987,0.00114226){\color[rgb]{0,0,0}\makebox(0,0)[lb]{\smash{$\Blue{F}$}}}%
    \put(0.30067393,0.17018441){\color[rgb]{0,0,0}\makebox(0,0)[lb]{\smash{$\Blue{F}$}}}%
    \put(0.60950094,0.17018441){\color[rgb]{0,0,0}\makebox(0,0)[lb]{\smash{$\Blue{F}$}}}%
    \put(0.83922489,0.00114226){\color[rgb]{0,0,0}\makebox(0,0)[lb]{\smash{$\Blue{F}$}}}%
    \put(0.83814129,0.17018441){\color[rgb]{0,0,0}\makebox(0,0)[lb]{\smash{$\Blue{F}$}}}%
    \put(0.63550742,0.17018441){\color[rgb]{0,0,0}\makebox(0,0)[lb]{\smash{$\Red{\bar F}$}}}%
    \put(0.40374844,0.08728014){\color[rgb]{0,0,0}\makebox(0,0)[lb]{\smash{$\displaystyle ,$}}}%
    \put(0.96722228,0.08728014){\color[rgb]{0,0,0}\makebox(0,0)[lb]{\smash{$\displaystyle .$}}}%
  \end{picture}%
\endgroup%
 \la{crossingtotop}
\end{array}
\eeq
That is, we obtain 
\beq 
\la{tosimp} P_{FF}(0|u,v)=P_{F \bar F}(u^{2\gamma}|v) \, , \qquad P_{F\bar F}(0|u,v)= P_{F \bar F}(v^{-3\gamma}|u) \, ,
\eeq
with, we stress again, the single-particle transitions in the right-hand sides being the same as the ones studied in~\cite{short,long}. 

We can simplify the expressions~(\ref{tosimp}) even further by using the explicit solutions for the pentagon transitions presented in \cite{short}. The procedure is explained in detail in appendix \ref{manipG}. At the end, we arrive at the remarkably simple relations
\beq\la{creationFromTransition}
P_{FF}(0|u,v)={1\over P_{FF}(u|v)}\ ,\qquad P_{F\bar F}(0|u,v)={1\over  P_{F\bar F}(u|v)} \,. 
\eeq 
They, of course, match with the more general multi-particle conjecture for the pentagon transition involving any number of gluons, as given by equation~(9) in \cite{short}, see also \cite{andrei} for a nice recent discussion at Born level. {The fundamental relations between the pentagon transitions $P_{FF}(u|v)$, $P_{F\bar{F}}(u|v)$ and the two gluonic scattering phases $S_{FF}$, $S_{F\bar{F}}$~\cite{short} guarantee that the Watson equations}
\beq
P_{FF}(0|u, v) = S_{FF}(v, u)P_{FF}(0|v, u)\, , \qquad P_{F\bar{F}}(0|u, v) = S_{F\bar{F}}(v, u)P_{F\bar{F}}(0|v, u)\, ,
\eeq
{are fulfilled by the form factors~(\ref{creationFromTransition}), in agreement with}~\cite{short,long}. 

Putting all the pieces together, we can now write down concise formulae for the two-gluon contributions, 
\beq
\mathcal{W}_{FF} = \frac{1}{2} \int \frac{du\, dv}{(2\pi)^2} \frac{\hat\mu_{F}(u)\hat\mu_{F}(v)}{P_{FF}(u|v)P_{FF}(v|u)} \,, \qquad 
\mathcal{W}_{F\bar {F}}=  \int \frac{du\, dv}{(2\pi)^2} \frac{\hat\mu_F(u)\hat\mu_F(v)}{ {P}_{F\bar F}(u|v)  {P}_{F\bar F}(v|u)}\, ,  \la{WFF2}
\eeq
which elegantly combine all the dynamical ingredients bootstrapped in~\cite{short}. 

Finally, we must include the bound-state contributions to the hexagonal Wilson loop. At twist smaller or equal to two, there are only two bound states -- of two $F$'s or two $\bar F$'s. They were denoted $DF$ and $\bar D\bar F$ in figure~\ref{fundamentalExcitations} and they carry $U(1)$ charge $+2$ and~$-2$, respectively. Their contributions to the OPE series read therefore
\beqa
\mathcal{W}^\text{BS}=  e^{2 i \phi} \,\cW_{DF}+e^{-2 i \phi} \,\cW_{\bar D\bar F}\, ,
\eeqa
where 
\beq\la{WDF}
\cW_{DF} = \cW_{\bar D\bar F} = \int \frac{du}{2\pi} \,\hat{\mu}_{DF}(u) \,,
\eeq
in strict analogy with~(\ref{hexagonSingle}). As already mentioned in the previous section, the energy and momentum of the bound states are known. 
We thus turn our attention to the bound-state measure $\mu_{DF}(u) = \mu_{\bar{D}\bar{F}}(u)$.
There are two natural routes to find it. 

One option is to bootstrap the transition $P_{D\!FD\!F}(u|v)$ directly, as we did by for the twist-one gluon, by postulating a set of axioms that this one should obey. The measure could then be obtained from the square limit of this transition~\cite{short}, that is, from the residue at coinciding rapidities $u=v$.  The other option is to exploit the close relationship between the bound states and their constituents. This means constructing the measure directly by fusing two twist-one gluons into a bound state, similarly to the way the energy and momentum of the bound state were found by adding up the energies and momenta of the gluons in~(\ref{EDF}). When applied to the measure, this philosophy leads us to the expectation that the integrand (\ref{WFF2}) for producing two gluons in the middle square, i.e.,
\beq
\frac{\hat\mu_F(u)\hat\mu_F(v)}{P_{FF}(u|v)P_{FF}(v|u)} \,,
\eeq
should have a pole at $u-v=i$ with the residue related to the measure $\hat\mu_{DF}(\ft{1}{2}(u+v))$. 

\begin{figure}[t]
\centering
\def\svgwidth{16cm}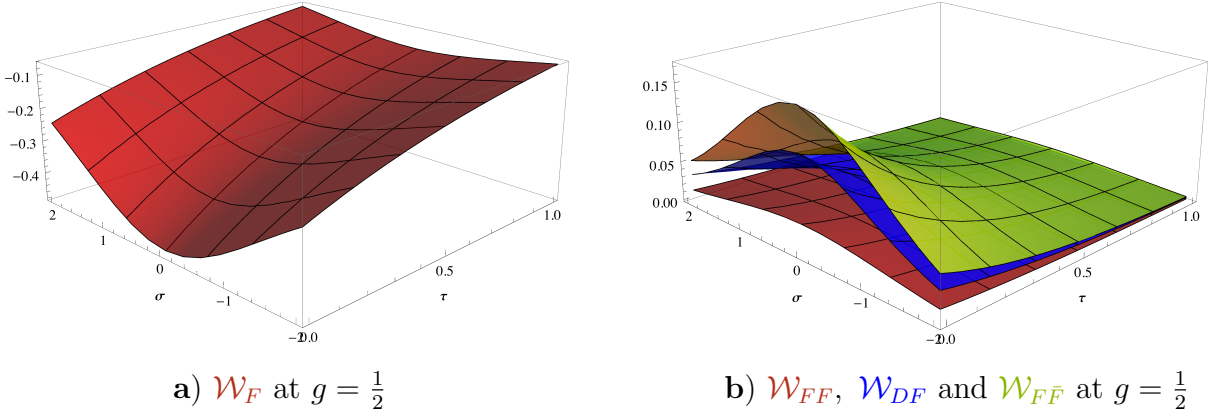
\caption{ {\bf a}) {The single-gluon contribution $\cW_F$ at $g=1/2$. {\bf b}) The two-gluon contributions $\cW_{FF}$  (bottom/red), $\cW_{DF}$ (middle/blue) and $\cW_{F\bar F}$ (top/yellow) at $g=1/2$. Contributions ({\bf b}) comprising heavier excitations are seen to be smaller than the single-particle one ({\bf a}), as expected (for positive $\tau$). As we increase $\tau$ or $\sigma$, the excitations have to propagate over larger distance and their contributions get more suppressed, as shown in both {\bf a)} and {\bf b)}.}}\label{gluonsPlot}
\end{figure}

These two routes ought to give the same result and shall be discussed in more detail in~\cite{toappear}. Here we merely quote the outcome which reads
\beq
\mu_{DF}(u)= i  \,\underset{v=u}{\operatorname{{\rm residue}}} \,\frac{\mu_F(u^+)\mu_F(v^-)}{P_{FF}(u^+|v^-)P_{FF}(v^-|u^+)}\, ,  \, \la{wdf} 
\eeq
with $u^{\pm}=u\pm i/2 $ and similarly for $v$. 
For evaluating this measure in perturbation theory, one must be careful with the order-of-limit issue mentioned earlier -- see discussion below~(\ref{EDF}). In particular, it is not enough to plug the weak coupling expression for each of the ingredients in~(\ref{wdf}). Doing this, one would not even find a pole in the right-hand side of (\ref{wdf}). The right procedure is to perform the fusion~(\ref{wdf}) at finite coupling and only after take the weak coupling limit.
The measure $\mu_{DF}(u)$ -- written in a form convenient for the perturbative expansion -- is presented in appendix~\ref{summaryBS}

The above construction clearly shows that the bound state $DF$ and the two-gluon state $FF$ come hand in hand. They both have the same bare twist $two$, the same $U(1)$ charge, and therefore contribute at the same order $\sim e^{-2\tau+2i \phi}$ in the weak coupling OPE sum~(\ref{fgexpression}).
Similarly, when considering more than two gluons, it would be natural to take into account the heavier bound states, $D^n F$ with $n\ge 2$, as well. The set of all possible multi-gluon states together with their bound-states form a nice subsector of the full OPE series which will be analyzed in detail in~\cite{toappear}. 

{We conclude this discussion with figure \ref{gluonsPlot} where the various gluonic contributions are plotted as function of $\sigma$ and $\tau$ for $g=1/2$.}

\section{Scalars}\label{scalar-section}

In this section we study the scalar excitations. Scalars are charged under the $SU(4)$ R-charge symmetry of the $\mathcal{N}=4$ SYM theory. As such, a single scalar {can not appear} in the OPE series~(\ref{hexagon2}) for the bosonic hexagonal Wilson loop. However, a pair of scalars can form an R-charge singlet {and hence contributes to~(\ref{hexagon2})}. The purpose of this section is to unveil this important contribution. 

The pentagon transitions for scalars{, in contrast to the gluonic ones,} must involve SU(4) matrix structures, see conclusions in \cite{long}. The direct transition of a scalar $\phi_i(u)$ at the bottom of the pentagon to a scalar $\phi_j(v)$ located at the top provides us with the simplest possible example. We denote it as $P_{ij}(u|v)$ with $i,j=1,2,\dots,6$ the two $SO(6)$ vector indices.
This case is however so elementary that disentangling the indices from the rest is straightforward: for symmetry reasons, this transition ought to be proportional to the only available $SU(4)$ invariant tensor, that is, to the Kronecker delta $\delta_{ij}$. We can thus immediately write down
\beq
P_{ij}(u|v)=\delta_{ij} \,P_{\phi\phi}(u|v) \,, \la{scalarTransition}
\eeq 
which, again, simply encodes charge conservation. The single-particle transition $P_{\phi\phi}(u|v)$ is the most interesting part at the end. It was presented in \cite{long} and is recalled in Appendix~\ref{fundamentaltransitions} for convenience.

By the same token, the creation amplitude for a pair of scalars $\phi_i(u)\phi_j(v)$ must take the form
\beq
P_{ij}(0|u,v)=\delta_{ij} \,P_{\phi\phi}(0|u,v) \,, \la{scalarCreation}
\eeq 
meaning that only an R-charge singlet can be produced, as anticipated.
Here, as for gluons, to obtain the creation form factor  $P_{\phi\phi}(0|u,v)$, we can apply the crossing transformation to the direct pentagon transition $P_{\phi\phi}(u|v)$. We find
\beq\la{scalarCase}
P_{\phi\phi}(0|u,v)= P_{\phi\phi}(u^{2\gamma}|v) = \frac{1}{g^2 (u-v+2i)(u-v+i)P_{\phi\phi}(u|v)}  \, ,
\eeq 
as explained in appendix~\ref{scalar-manip}, and
observe the same phenomenon as for the gluons. Namely,  the form factor appears as being the inverse of the direct transition. However, in contradistinction to the gluon case~(\ref{creationFromTransition}) we also obtained the rational prefactor $(u-v+2i)(u-v+i)$, sitting in the denominator. This is not accidental as this factor is actually needed to convert the fundamental relation for the direct pentagon transition $P_{\phi\phi}(u|v)$ into the Watson relation for the form factor $P_{\phi\phi}(0|u,v)$,
\beq
\frac{P_{\phi\phi}(0|u,v)}{P_{\phi\phi}(0|v,u)} = \texttt{rational} \times \frac{P_{\phi\phi}(v|u)}{P_{\phi\phi}(u|v)} = \texttt{rational} \times S_{\phi\phi}(v, u) = S_{\phi\phi}(v, u)_{\textrm{singlet}}\, .
\eeq
It accommodates precisely for the difference among scattering phases in the symmetric and singlet channels (see discussion in \cite{long} for more details).
For the annihilation amplitudes, we can use
\beq 
P_{ij}(u,v|0)=P_{ji}(0|v,u)=P_{ij}(0|-v,-u)\,. 
\eeq

Putting all form factors together, we get the contribution of the two-scalar states to the hexagonal Wilson loop (\ref{hexagon2}) in the form 
\beq\la{Twoscalars}
\begin{array}{l}
\centering
\def\svgwidth{15cm}
\begingroup%
  \makeatletter%
  \providecommand\color[2][]{%
    \errmessage{(Inkscape) Color is used for the text in Inkscape, but the package 'color.sty' is not loaded}%
    \renewcommand\color[2][]{}%
  }%
  \providecommand\transparent[1]{%
    \errmessage{(Inkscape) Transparency is used (non-zero) for the text in Inkscape, but the package 'transparent.sty' is not loaded}%
    \renewcommand\transparent[1]{}%
  }%
  \providecommand\rotatebox[2]{#2}%
  \ifx\svgwidth\undefined%
    \setlength{\unitlength}{1321.70410156bp}%
    \ifx\svgscale\undefined%
      \relax%
    \else%
      \setlength{\unitlength}{\unitlength * \real{\svgscale}}%
    \fi%
  \else%
    \setlength{\unitlength}{\svgwidth}%
  \fi%
  \global\let\svgwidth\undefined%
  \global\let\svgscale\undefined%
  \makeatother%
  \begin{picture}(1,0.18475527)%
    \put(0,0){\includegraphics[width=\unitlength]{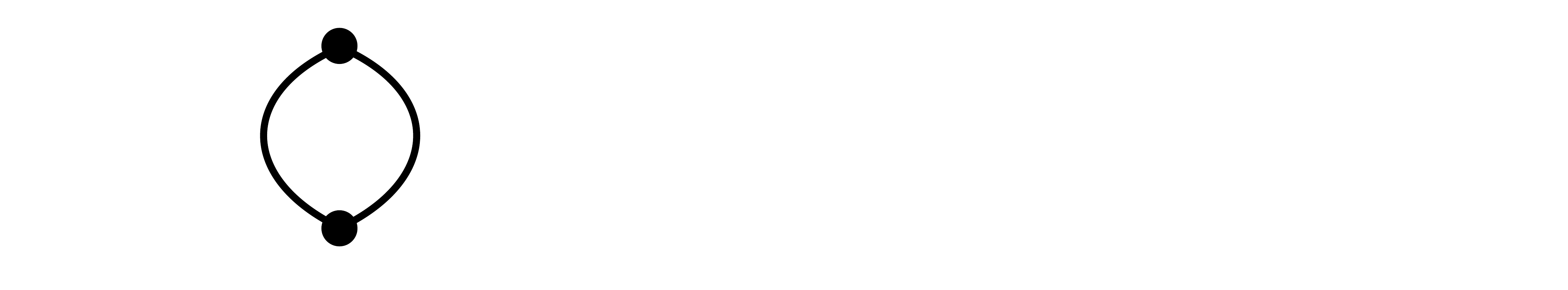}}%
    \put(0.20095221,0.00184535){\color[rgb]{0,0,0}\makebox(0,0)[lb]{\smash{$P_{ii}$}}}%
    \put(-0.00074831,0.09076197){\color[rgb]{0,0,0}\makebox(0,0)[lb]{\smash{$\displaystyle \mathcal{W}_{\phi \phi}={1\over2}\sum_{i=1}^6$}}}%
    \put(0.19890182,0.17881043){\color[rgb]{0,0,0}\makebox(0,0)[lb]{\smash{$P_{ii}$}}}%
    \put(0.27933262,0.09068605){\color[rgb]{0,0,0}\makebox(0,0)[lb]{\smash{$\displaystyle = \frac{1}{2} \times 6 \, \int  \frac{du \,dv\,\hat\mu_\phi(u)\hat\mu_\phi(v)}{g^4\((u-v)^2+4\)\((u-v)^2+1\)P_{\phi\phi}(u|v)P_{\phi\phi}(v|u)}$}}}%
    \put(0.17560341,0.09211218){\color[rgb]{0,0,0}\makebox(0,0)[lb]{\smash{$i$}}}%
    \put(0.24823692,0.09211218){\color[rgb]{0,0,0}\makebox(0,0)[lb]{\smash{$i$}}}%
  \end{picture}%
\endgroup%
 
\end{array}
\eeq
The overall factor $1/2$ above is the usual symmetry factor while the factor $6$ stands for the number of scalars and arises from the contraction of two Kr\"onecker deltas. The measures, transitions and dispersion relation, necessary for evaluating this expression, are all summarized in appendix \ref{fundamentaltransitions}.

In the strong coupling section below we present a plot of the scalar contribution at a large (but finite) value of the coupling, see figure \ref{scalar plot}.

\section{Fermions}\la{fermionssection}

In this section we consider the contribution from fermions to the hexagonal Wilson loop. At twist two, it can only be a singlet pair of fermions contributing. The discussion of the fermion $SU(4)$ R-charge indices mimics that of the scalars -- see  (\ref{scalarTransition}) and (\ref{scalarCreation}) -- and will therefore be {kept implicit in the following}.

{As for the bosons,} there are two pentagon transitions of interest here.
One is the direct transition of a fermion from the bottom to the top, i.e., $P_{\psi\psi}(u|v)$ in our notations. 
 It is the transition whose residue at $u=v$ yields the fermion measure $\mu_\psi(u)$,
\beq\label{Pff-mu}
P_{\psi\psi}(u|v) \sim \frac{i}{\mu_{\psi}(u)(v-u)}\, .
\eeq
The second fermionic transition $P_{\psi \bar \psi}(0|u,v)$ is the form factor for creation of a singlet pair of fermions. These two transitions can be depicted as
\beq \la{goalF}
\begin{array}{l}
\def\svgwidth{14cm}
\begingroup%
  \makeatletter%
  \providecommand\color[2][]{%
    \errmessage{(Inkscape) Color is used for the text in Inkscape, but the package 'color.sty' is not loaded}%
    \renewcommand\color[2][]{}%
  }%
  \providecommand\transparent[1]{%
    \errmessage{(Inkscape) Transparency is used (non-zero) for the text in Inkscape, but the package 'transparent.sty' is not loaded}%
    \renewcommand\transparent[1]{}%
  }%
  \providecommand\rotatebox[2]{#2}%
  \ifx\svgwidth\undefined%
    \setlength{\unitlength}{1016.10820313bp}%
    \ifx\svgscale\undefined%
      \relax%
    \else%
      \setlength{\unitlength}{\unitlength * \real{\svgscale}}%
    \fi%
  \else%
    \setlength{\unitlength}{\svgwidth}%
  \fi%
  \global\let\svgwidth\undefined%
  \global\let\svgscale\undefined%
  \makeatother%
  \begin{picture}(1,0.19093762)%
    \put(0,0){\includegraphics[width=\unitlength]{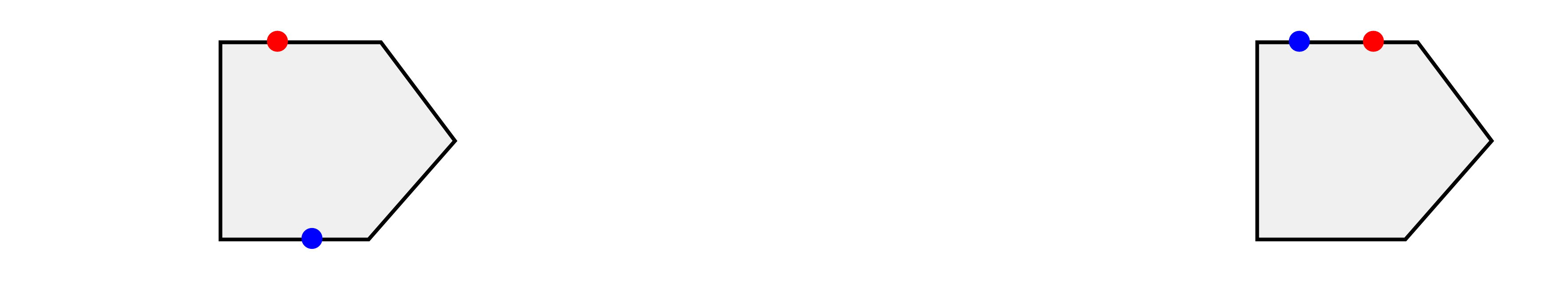}}%
    \put(-0.04821937,0.09591876){\color[rgb]{0,0,0}\makebox(0,0)[lb]{\smash{$\displaystyle P_{\psi\psi}(u|v)\ =$}}}%
    \put(0.18937644,0.05321835){\color[rgb]{0,0,0}\makebox(0,0)[lb]{\smash{$u$}}}%
    \put(0.16807261,0.13829366){\color[rgb]{0,0,0}\makebox(0,0)[lb]{\smash{$v$}}}%
    \put(0.18980455,0.00560579){\color[rgb]{0,0,0}\makebox(0,0)[lb]{\smash{$\psi$}}}%
    \put(0.16618502,0.18196496){\color[rgb]{0,0,0}\makebox(0,0)[lb]{\smash{$\bar\psi$}}}%
    \put(0.61312755,0.09591876){\color[rgb]{0,0,0}\makebox(0,0)[lb]{\smash{$\displaystyle P_{\psi\bar\psi}(0|u,v)\ =$}}}%
    \put(0.8192306,0.1398233){\color[rgb]{0,0,0}\makebox(0,0)[lb]{\smash{$u$}}}%
    \put(0.81965866,0.18196496){\color[rgb]{0,0,0}\makebox(0,0)[lb]{\smash{$\psi$}}}%
    \put(0.86646971,0.1398233){\color[rgb]{0,0,0}\makebox(0,0)[lb]{\smash{$v$}}}%
    \put(0.86689777,0.18196496){\color[rgb]{0,0,0}\makebox(0,0)[lb]{\smash{$\bar\psi$}}}%
  \end{picture}%
\endgroup%
 
\end{array}
\eeq
and they should satisfy the relations
\beq\label{fund-rel-fermions}
P_{\psi\psi}(u|v) = -S_{\psi\psi}(u, v)P_{\psi\psi}(v|u)\, , \qquad P_{\psi\bar{\psi}}(0|u,v) = -S^{\textrm{singlet}}_{\psi\bar{\psi}}(v, u)P_{\psi\bar{\psi}}(0|v,u)\, .
\eeq
The first equation in (\ref{fund-rel-fermions}) is our fundamental relation for the fermionic transition~\cite{short} while the second one has the more direct meaning of the Watson equation for the form factor~\cite{Watson}. They obviously differ from their bosonic counterparts by the minus signs on their right-hand sides. Extra minus signs, when dealing with fermions, are an expected nuisance. 
Technically, one can understand them by observing that $S_{\psi\psi}(u, u) = -S_{\psi\bar{\psi}}(u, u)_{\textrm{singlet}}= 1$ for fermions, such that, were the minus signs not there, the relations~(\ref{fund-rel-fermions}) would be in conflict with~(\ref{Pff-mu}) and $P_{\psi\bar{\psi}}(0|u,u) \sim O(1)$. 

Clearly the above two pentagon transitions are not independent from one another. Applying the same logic as for the bosonic excitations one could imagine rotating the fermions around to relate them. The problem is that there is no simple way of doing this for fermions. As opposed to bosons, fermions transform somewhat anomalously under mirror rotation. This is what is explained in greater detail in Appendix~\ref{fermion-bootstrap}. Basically, what happens as we rotate a fermion is that  we produce an excitation on the neighbouring edge {\textit{that is not}} the original fermion. This {new} excitation is actually a twist-two fermion {and is of a more exotic type}. It can be thought of as a composite object made out of an anti-fermion and a scalar, bound together in the appropriate $SU(4)$ representation. {It is not a regular bound state, however, and,} as opposed to the fundamental excitations discussed in Section~\ref{reviewSec}, the latter exotic excitation does not have a real dispersion relation{, for instance}. We shall see shortly that many excitations of this type are expected to show up in the OPE analysis at weak coupling. Decomposing the twist-two fermion into its fundamental constituents allows one to achieve a mirror transformation of sort for fermions. The price to pay is that one starts introducing more matter into the game (for example the scalar inside the composite twist-two excitation). In other words, the fermion bootstrap is not closed and takes instead the form of a hierarchy of equations tying together a much larger set of excitations.
In appendix~\ref{paradoxSec}, we illustrate how this bootstrap works at the level of the fermion S-matrix, but its generalization to the pentagon transitions stands beyond the scope of this paper. 

In this paper we shall proceed differently and conjecture the fermion transitions using as a main guide
the expressions for the scalar and gluon transitions. We shall first write down the latter bosonic transitions in a manner that is as independent as possible of their kind. This ``universal way" of writing all the bosonic transitions will then be employed to hint at a fermionic generalization and give support to our educated guesses for the fermion transitions. The reader ready to accept these conjectures on faith can jump directly to the final expressions~(\ref{WFerm}) and~(\ref{twofermions}). 

We start with the direct transitions. One striking feature of the gluon and scalar transitions is that they can both be written in the form
\beq\la{trans-formMain}
P_{XX}(u|v)^2 = \frac{f_{XX}(u,v)S_{XX}(u, v)}{(u-v)(u-v+i)S_{\star XX}(u, v)}\, , \qquad X = \textrm{either} \, \, \, \phi \, \, \, \textrm{or} \, \, \, F\, ,
\eeq
where $S_{XX}(u, v)$ is the phase for a (symmetric) real scattering event and  $S_{\star XX}(u, v)$ its mirror counterpart. 
In all cases the function $f_{XX}(u,v)$ is symmetric and much simpler than the rest. Its expression for gluon and scalar was fixed in~\cite{short,long} by demanding that the transitions have the right mirror transformations. This led to
\beq\label{f-typeMain}
\begin{aligned}
&f_{\phi\phi} = \frac{1}{g^2}\, ,\\
&f_{FF} = \frac{x^{+}x^{-}y^{+}y^{-}}{g^2}\Big(1-\frac{g^2}{x^{+}y^{-}}\Big)\Big(1-\frac{g^2}{x^{-}y^{+}}\Big)\Big(1-\frac{g^2}{x^{+}y^{+}}\Big)\Big(1-\frac{g^2}{x^{-}y^{-}}\Big)\, ,
\end{aligned}
\eeq
as the simplest possible expressions, here written in terms of the Zhukowski variable
\beq
x(u) = \frac{1}{2}\( u + \sqrt{u^2-4g^2}\) \,, \la{Zhu}
\eeq 
and with the notations $x^{\pm}=x(u\pm i/2)$ and $ y^{\pm} = x(v\pm i/2)$. Our proposal for the fermion case is somewhat intermediate to the two in~(\ref{f-typeMain}): it assumes that the fermion transition takes the form~(\ref{trans-formMain}) with the symmetric factor
\beq\label{f-fermions}
f_{\psi\psi} = \frac{xy}{g^2}\Big(1-\frac{g^2}{xy}\Big)\, ,
\eeq
where $x=x(u)$ and $y=x(v)$. This factor is of the type~(\ref{f-typeMain}) taking into account the kinematical difference between the various excitations. It is natural for instance to have unshifted Zhukowsky variables for the fermion, whose most important kinematical cut is centered around zero. It was however natural to have them centered around $\pm \ft{i}{2}$ for the gluon, as in~(\ref{f-typeMain}), since these are the emplacements of the mirror cuts for a gluon and $f_{FF}$ ought to have such a cut for consistency with mirror transformation. 

The above comment concludes the discussion of our expression for $P_{\psi\psi}(u|v)$. Next we move to the creation amplitude $P_{\psi\bar\psi}(0|u,v)$ where the guess work is slightly more acrobatic.
As a first step we parametrize all the creation amplitudes as
\beq
P_{X\bar{X}}(0|u, v) = \frac{1}{\texttt{rational} \times P_{X\bar{X}}(u|v)} \, . \la{ansatzC}
\eeq
In this expression $P_{X\bar{X}}(u|v)$ is a transition that obeys the standard fundamental relation to the S-matrix. Namely, we have that $P_{X\bar{X}}(u|v) = \pm \,S_{X\bar{X}}(u|v)P_{X\bar{X}}(v|u)$, with $+/-$ for bosons/fermions and with $S_{X\bar{X}}(u, v)$ the $X(u)\bar{X}(v)$ scattering phase in the symmetric channel. The Watson equation for the creation amplitudes~(\ref{ansatzC}) involves the S-matrix in the singlet representation rather than in the symmetric one. The \texttt{rational}  factor in  (\ref{ansatzC}) is designed to take into account the translation between the two. For a scalar this was the $(u-v+i)(u-v+2i)$ factor that appeared in the transition amplitude (\ref{scalarCase}) while it is just equal to $1$ for the gluons. For the fermions, the simple factor $\texttt{rational} =(u-v+2i)/i$ does the job, such that
\beq
P_{\psi\bar \psi}(0|u,v) = \frac{i}{(u-v+2i){P}_{\psi\bar\psi}(u|v)}\, . \la{fermionCase}
\eeq
This leaves us with the more complicated task of guessing the most dynamical part, that is the transition $P_{\psi\bar{\psi}}(u|v)$ itself.  It could be given a physical meaning on its own as a charged transition for the OPE decomposition of the super Wilson loop. For the current discussion it is enough to think of it as being the analogue of the factors $g^2{P}_{\phi\phi}(u|v)$ and ${P}_{F\bar F}(u|v)$ that parameterize the creation amplitudes~(\ref{scalarCase}) and~(\ref{creationFromTransition}), respectively. The main observation here is that the two bosonic transitions, $g^2{P}_{\phi\phi}(u|v)$ and ${P}_{F\bar F}(u|v)$, can be written succinctly as
\beq
P_{X\bar X}(u|v)^2 =( (u-v)(u-v+i) )^{\eta } \frac{S_{X\bar X}(u, v)}{f_{XX}(u,v)S_{\star X\bar X}(u, v)}\, , \la{finFermions}
\eeq
where $\eta=+1$ for the gluon $X=F$ and $\eta=-1$ for the scalar $X=\phi$. As above, we expect the fermion expression to lie somewhere in between. A simple guess is to try $\eta=0$. In Appendix~\ref{conjectureFermions} we verify that this is indeed in perfect agreement with the Watson relation in~(\ref{fund-rel-fermions}).

To summarize, our conjectures read
\beq
P_{\psi \psi}(u|v)^2 =\frac{f_{\psi\psi}(u,v)S_{\psi\psi}(u, v)}{(u-v)(u-v+i)S_{\star \psi\psi}(u, v)}\, , \qquad P_{\psi\bar \psi}(u|v)^2 =  \frac{S_{\psi\bar \psi}(u, v)}{f_{\psi\psi}(u,v)S_{\star \psi\bar \psi}(u, v)}\, , \la{WFerm}
\eeq
together with (\ref{fermionCase}) and~(\ref{f-fermions}). Their principal ingredients are the real and mirror S-matrices for two fermions, which are constructed and extensively studied in appendix~\ref{paradoxSec}.
As a particular application of these formulae, we can now write down the two-fermion contribution to the hexagon Wilson loop, which reads 
\beq\la{twofermions}
\mathcal{W}_{\psi \bar{\psi}}=4 \, \int \frac{du \,dv}{(2\pi)^2} \frac{\hat\mu_\psi(u)\hat\mu_\psi(v)}{\((u-v)^2+4\){P}_{\psi\bar{\psi}}(u|v){P}_{\psi\bar{\psi}}(v|u)} \, ,
\eeq
with the overall factor $4$ counting the number of fermions. 
This concludes the discussion of the very last two-particle contribution. See figure~\ref{fermionsPlot} below for some representative plots of the two-fermion contribution at several values of the coupling.

\section{The full two-particle contribution} \la{fullSec}

We can now combine all the pieces together and obtain the complete twist-two contribution to the hexagonal Wilson loop. It is given by 
\beq
\mathcal{W}_\text{twist-two}= \(\mathcal{W}_{\phi\phi}+\mathcal{W}_{\psi{\bar \psi}}+\mathcal{W}_{F\bar F} \) + 2\cos(2\phi) \( \mathcal{W}_{FF}+\mathcal{W}_{DF} \)\, , \la{full2pt}
\eeq
where the first brackets encompass the $U(1)$ neutral part, coming from all the particle-antiparticle pairs of elementary excitations, {while the last one is the $U(1)$ charge $2$ part}, coming from two identical gluons or bound states thereof. The explicit expressions for the several terms entering in~(\ref{full2pt}) are those derived in (\ref{WFF2}), (\ref{WDF}), (\ref{Twoscalars}) and (\ref{twofermions}), respectively. They all have in common that they are given as some (multiple) integrals over the rapidities of the excitations. To complete our result we should then explain what the appropriate contours of integration are for all the excitations involved.

The strategy to finding these contours is quite simple. First, we should recall that the Bethe rapidity $u$ was introduced as a convenient way of parametrizing the energy and the momentum of the excitations. In particular, we can always trade the integration over $u$ for an integration over the momentum $p$. In momentum space, the choice of the contour is immediate as it should cover all possible real values of $p$. From the knowledge of the $u\leftrightarrow p$ map for the fundamental excitations~\cite{BenDispPaper}, it is immediate to conclude that the integrals should be taken over the full rapidity real axis for scalars and gluons. This is the integration contour for both $u$ and $v$ in (\ref{WFF2}), (\ref{WDF}) and (\ref{Twoscalars}), for instance. For fermions, the story is more interesting and we shall spend some time now spelling out the details of their contour of integration.

\begin{figure}[t]
\def\svgwidth{15cm}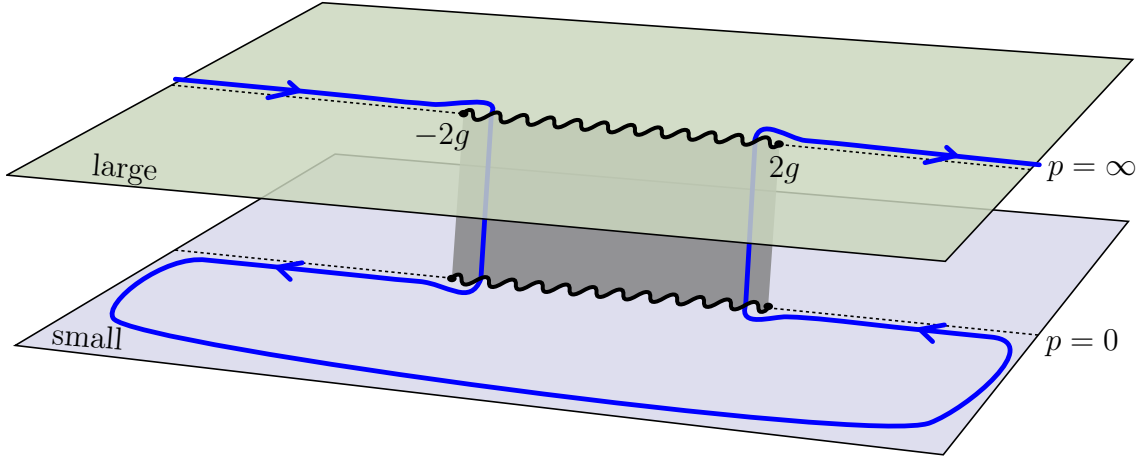
\caption{The fermion integration contour covers all real (non-zero) momenta which in terms of the Bethe rapidity $u$ amounts to a contour in a two-sheeted Riemann surface. The wavy line stretching between $u=\pm 2g$ stands for the cut connecting the two $u$-planes.}
\la{fermionsintegration}
\end{figure}

One noticeable aspect of the fermion kinematics is that it involves two copies of the $u$ plane. The reason is that the good rapidity for fermions~\cite{BenDispPaper} is not the Bethe rapidity $u$ but instead its image $x$ under the Zhukowsky map,
\beq\label{Zm}
u = x+\frac{g^2}{x}\, .
\eeq
Said differently, over the real line, energy and momentum of a fermion are smooth functions of the rapidity $x$, not $u$. (Also, energy and momentum both admit convergent Taylor series in $x$ around the point $x=0$ which corresponds to a fermion at rest.) Still, for many reasons, it is better to employ the same rapidity for both the fermions and the other excitations. (The finite-coupling expressions for all dynamical quantities are also most easily written in terms of the Bethe rapidity, for instance.) The price to pay is that we should introduce the two aforementioned copies of the $u$ plane in order to cover the full $x$ plane. They obviously correspond to the two branches $x(u)$ and $g^2/x(u)$ of the quadratic relation~(\ref{Zm}), where $x(u)$ will always stand in this paper for the branch~(\ref{Zhu}), which we recall here for convenience,
\beq\label{xofu}
x(u) = \ft{1}{2}(u+\sqrt{u^2-4g^2}) = u + O(g^2)\, .
\eeq
We call ``large (fermion) sheet'' the copy of the $u$ plane that covers, through this map, the outer domain $|x| \geqslant g$ of the $x$ plane. The origin of this terminology is that the fermion momentum can achieved arbitrarily large values on this sheet.
The points $u = \pm \infty$ map to $p =\pm \infty$ on this sheet, for instance. Similarly, we dub ``small (fermion) sheet'' the copy of the $u$ plane that covers the inner domain $|x| \leqslant g$ of the $x$ plane, through the second branch $u\to  \ft{1}{2}(u-\sqrt{u^2-4g^2})$. It is on this sheet, indeed, that the smaller values of the momentum are found. It contains notably the point $p=0$ which is at $u = \infty$. The small and large sheets are glued together by means of the cut stretching between the two branch points $u=\pm 2g$ of the map~(\ref{Zm}).

To sum over all the real positive (negative) values of the momentum we should then integrate from $u = \infty$ ($u =-\infty$) on the small sheet up to $u=\infty$ ($u =-\infty$) on the large sheet. This implies passing through the cut connecting the sheets, as depicted in figure~\ref{fermionsintegration}. For a reason that shall be given shortly, we should also avoid integrating through the point $p=0$.  This is why we closed the integration contour in the lower half of the small sheet in figure~\ref{fermionsintegration}.
In the end, we observe that our contour in figure~\ref{fermionsintegration} is equivalent to the one depicted in figure~\ref{fermionsintegration2}. Slightly abusing the terminology, we could say that we have two kinds of fermions: small and large. The latter is integrated over the half-moon contour $\mathcal{C}_{\textrm{small}}$ and the former over the almost straight line $\mathcal{C}_{\textrm{large}}$, both plotted in figure~\ref{fermionsintegration2}. All the fermion integrals can be split into components corresponding to this bi-partition of the fermion kinematics. This way of doing is well-suited for the weak coupling analysis because at weak coupling the cut closes and pinches the contour in figure~\ref{fermionsintegration}. It is then no longer possible to navigate from one sheet to the other and thus more useful to treat them separately, as in figure~\ref{fermionsintegration2}.

\begin{figure}[t]
\centering
\def\svgwidth{15cm}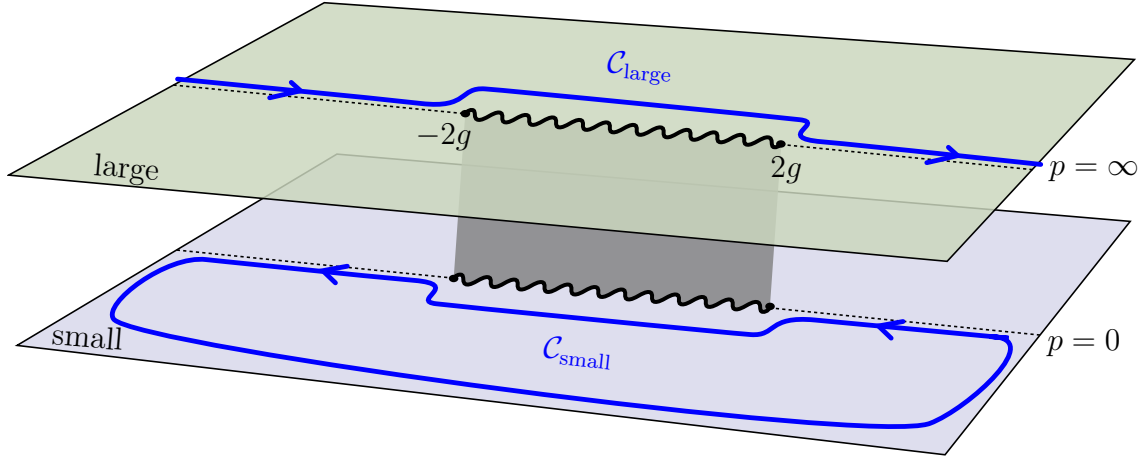
\caption{The fermions integration contour plotted in figure \ref{fermionsintegration} can be split into two contours, each living on a different sheet. Large fermion momenta are located on the large sheet and small fermion momenta on the small one}
\la{fermionsintegration2}
\end{figure}

\begin{figure}[h]
\centering
\def\svgwidth{12cm}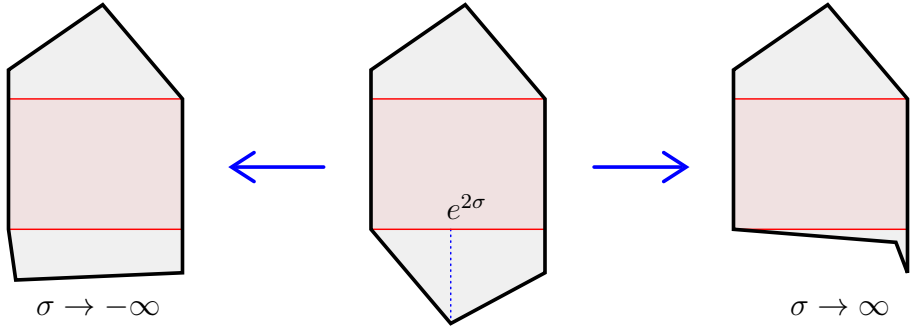
\caption{The reflection symmetry $\sigma\leftrightarrow-\sigma$ of the middle square is broken by the embedding inside the hexagon. The reason is that our finite ratio $\cW$ is based on a particular tessellation of the Wilson loop into sequences of squares and pentagon. In this framing, two cusps of the middle square are distinguished from the others by the fact that they coincide with two (opposite) cusps of the hexagon. It follows notably that the limits $\sigma\to\pm\infty$ looks very different. As shown in this figure, both $\sigma\to\pm\infty$ are collinear limits but only at $\sigma\to\infty$ does the bottom of the hexagon coincide with the bottom of the square. }
\la{largeSsmallS}
\end{figure}

\begin{figure}[h]
\centering
\def\svgwidth{16cm}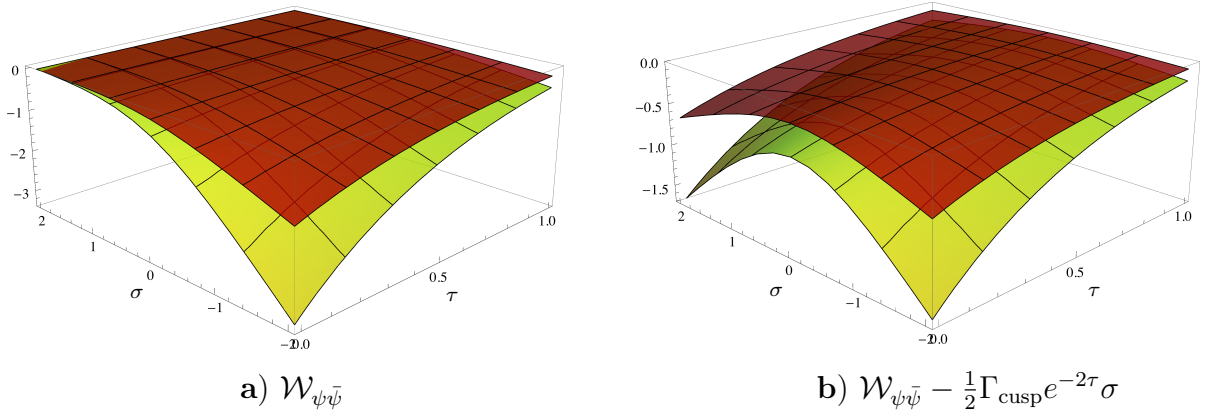
\caption{{\bf a}) The two-fermion contribution $\mathcal{W}_{\psi\bar\psi}$ for $g=1/2$ (\Red{top/red}) and $g=1$ ({\color{LimeGreen} bottom/yellow}). The double pole in the fermion measure at zero momentum leads to a contribution that grows linearly at large negative $\sigma$ and break the $\sigma \to -\sigma$ symmetry. This contribution is controlled by $\Gamma_\text{cusp}$ and is associated with our choice of reference square, see figure \ref{largeSsmallS}. {\bf b}) The subtracted contribution $\mathcal{W}_{\psi\bar\psi}-\frac{1}{2} \Gamma_{\text{cusp}} e^{-2\tau} \sigma$, obtained by removing half of the zero-momentum residue, for $g=1/2$ (\Red{top/red}) and $g=1$ ({\color{LimeGreen} bottom/yellow}). We see that the $\sigma \to -\sigma$ symmetry is restored as expected. Incidentally, this is how the two-fermion contribution shows up in the remainder function.}\label{fermionsPlot}
\end{figure}

We come back finally to our prescription for integrating around $u=\infty$ in the lower sheet. Around this point, the momentum of a fermion scales as $p \sim \Gamma_{\textrm{cusp}}(g)/(2u)$~\cite{BenDispPaper} while (our conjecture for) the measure behaves like
\beq\label{Cusp-behav}
\mu_{\psi}du \sim -\frac{1}{2}\Gamma_{\textrm{cusp}}(g) \frac{dp}{p^2}\, ,
\eeq
where $\Gamma_{\textrm{cusp}}(g) = 4g^2 + O(g^4)$ {is} the cusp anomalous dimension. Our choice of contour in figure~\ref{fermionsintegration2} corresponds therefore to integrating slightly above $p=0$ in the complex momentum plane; hence avoiding the double-pole in~(\ref{Cusp-behav}). This $i\epsilon$ prescription is not arbitrary and has its roots in our construction of the Wilson-loop ratio $\cW$. The latter requires introducing overlapping sequences of squares and pentagons which, together, define a particular tessellation of the Wilson loop~\cite{short}. Our prescription for integrating around $p=0$ implements the fact that the two collinear limits $\sigma \rightarrow \pm \infty$ are treated differently within this tessellation, as illustrated in figure~\ref{largeSsmallS}. Essentially, the limit $\sigma\rightarrow \infty$ is softer in our tessellation than the opposite one, hence our choice of prescription. The dependence on the choice of the tesselation is actually the same for the ratios $\mathcal{W}$ as for their relative $\mathcal{W}_{U(1)}$ computed in an abelian gauge theory with coupling $g^2_{U(1)}= \ft{1}{4}\Gamma_{\textrm{cusp}}(g)$. Equivalently, these $\mathcal{W}_{U(1)}$ are given by the BDS ansatz which is governed by the cusp anomalous dimension. This is in neat agreement with what we observe in (\ref{Cusp-behav}), namely, that the contribution from the $p\sim 0$ domain is controlled by $\Gamma_{\textrm{cusp}}(g)$ only. We should stress again that these subtleties are proper to the fermions. This is so because these excitations are very special at zero momentum -- they become generators of the broken supersymmetries~\cite{AldayMaldacena}.

For illustration, we plotted in figure \ref{fermionsPlot}.a the total two-fermion contribution at finite coupling. As expected, it is not $\sigma\leftrightarrow-\sigma$ symmetric. Similarly to the gauge field plot \ref{gluonsPlot}, it is suppressed at large and positive $\tau, \sigma$. However, the $\cW_{\psi\bar\psi}$ grows at large and negative $\sigma$. As shown in figure \ref{gluonsPlot}.b, the $\sigma\leftrightarrow-\sigma$ symmetry is restored once we subtract half of the zero-momentum contribution (so it is now growing in both directions). This is the way this contribution appears in the remainder function.

\section{Weak coupling} \la{WeakSec}
In this section we discuss the weak coupling expansion of our finite coupling result (\ref{full2pt}). For most of the contributions in (\ref{full2pt}) this step is straightforward. It follows closely the analysis done in \cite{long} at the one-particle level and shall be reviewed shortly. For the moment we shall focus on the case of fermions that requires a more careful study.

The starting point is
\beq\la{Wpsipsi}
\cW_{\psi \bar{\psi}}=4 \, \int_\mathcal{C} \frac{du}{2\pi}  \int_\mathcal{C} \frac{dv}{2\pi} \, \frac{\hat \mu_\psi(u)\hat \mu_\psi(v)}{\((u-v)^2+4\){P}_{\psi\bar{\psi}}(u|v){P}_{\psi\bar{\psi}}(v|u)} \, ,
\eeq
where, as explained in the previous section, the contour of integration splits into two distinct contributions, $\mathcal{C}= \mathcal{C}_\text{large} \cup \mathcal{C}_\text{small}$. We recall that here the contour $\mathcal{C}_\text{large}$ is along the real axis in the large sheet, with a small positive imaginary part, while $\mathcal{C}_\text{small}$ is a counter-clockwise half moon contour in the lower half-plane of the small sheet, see figure \ref{fermionsintegration2}. As an immediate consequence, computing the integrals~(\ref{Wpsipsi}) entails having a full control over the analytic continuation between large and small sheets of both the measure and transitions. This is the subject of appendix \ref{continuationAp}.

The splitting of the contour of integration $\mathcal{C}$ immediately yields a decomposition of~(\ref{Wpsipsi}) into a sum of various terms.  The simplest one corresponds to the configuration where both rapidities, $u$ and $v$, are on the large sheet. It is given by
\beq
\cW_{\psi \bar{\psi}}^{LL}=4 \, \int\limits_{\mathbb{R}+i0} {du\over2\pi} \int\limits_{\mathbb{R}+i0} {dv\over2\pi} \, \frac{\hat \mu_\psi({u})\hat \mu_\psi({v})}{\((u-v)^2+4\){P}_{\psi\bar{\psi}}({u}|{v}){P}_{\psi\bar{\psi}}({v}|{u})}\, ,  \la{largelarge}
\eeq
where the upper-script $LL$ reminds us that both fermions are large. The integrals are then of the same kind as the ones we shall encounter below for scalars or gluons and shall be computed in the same way in perturbation theory. As we will see later, all these contributions start at two loops, i.e., are of order $\cO(g^4)$ at weak coupling.

Next we have the contributions where at least one of the two fermions is small. They come in two types and read
\beq\la{small-int}
\cW_{\psi \bar{\psi}}^{LS}=2 \times 4 \, \int\limits_{\mathbb{R}+i0} \frac{du}{2\pi}\, \hat \mu_\psi({u}) I_\text{small}({u}) \, , \qquad \cW_{\psi \bar{\psi}}^{SS}=4 \, \int\limits_{\mathcal{C}_\text{small}} \frac{du}{2\pi}\, \hat \mu_\psi({\color{red}\check u}) I_\text{small}({\color{red}\check u})\, ,
\eeq
where 
\beq\la{int-small}
I_\text{small}(u)=  \int\limits_{\mathcal{C}_\text{small}} \frac{dv}{2\pi} \, \frac{\hat \mu_\psi({\color{red}\check v})}{\((u-v)^2+4\){P}_{\psi\bar{\psi}}(u|{\color{red}\check v}){P}_{\psi\bar{\psi}}({\color{red}\check v}|u)} \, .
\eeq
The overall factor $2$, in the right-hand side of the first equality in~(\ref{small-int}), simply comes from the fact that, for the $LS=$ ``large-small" configuration of the $\psi\bar{\psi}$ pair, we have two equivalent choices for who is small.
In these integrals, the arguments of the functions evaluated on the small sheet are indicated by a check mark on them. For illustration, in the rightmost integral in (\ref{small-int}) the notation $\hat \mu_\psi({\color{red}\check u})$ means that the fermion measure $\hat{\mu}_{\psi}$ is evaluated at rapidity $u$ on the small sheet. This convention shall be used throughout the paper. Furthermore, to help the reader, we sometimes colour in red these marked rapidities. 

An important difference between the integral in~(\ref{int-small}) and those in~(\ref{largelarge}) is that the former involves integration over the closed contour $\mathcal{C}_\text{small}$. The only reason why this integral is not zero is because its integrand has a pole at $v=u -2i$, in the lower half plane. This pole is manifest in~(\ref{int-small}) and is the only singularity encircled by $\mathcal{C}_\text{small}$. This is so because the pentagon transition $P_{\psi\bar{\psi}}$ (as well as its inverse) and the measure $\hat{\mu}_{\psi}$ are regular throughout the small sheet away from the central cut -- see discussion below (\ref{smallMeasure}) in Appendix \ref{continuationAp}.
The integration in the small sheet is then straightforward and by picking up the residue at $v = u-2i$ we find that 
\beq
I_\text{small}(u)=  -\frac{1}{4} \frac{\hat \mu_\psi({\color{red}\check{u}-2i})}{{P}_{\psi\bar{\psi}}(u|{\color{red}\check{u}-2i}){P}_{\psi\bar{\psi}}({\color{red}\check{u}-2i}|u)} \la{Ismall}
\eeq
It also immediately follows that $\mathcal{W}^{SS}_{\psi\bar{\psi}} =0$ since, after performing integration in~(\ref{int-small}), there are no longer any singularities enclosed by the integration contour of the rightmost integral in~(\ref{small-int}). This leaves us with the contribution 
\beq
\cW_{\psi \bar{\psi}}^{LS}=-2 \, \int\limits_{\mathbb{R}+i0} \frac{du}{2\pi} \frac{ \hat \mu_\psi({u}) \hat \mu_\psi({\color{red}\check{u}-2i})}{{P}_{\psi\bar{\psi}}({u}|{\color{red}\check{u}-2i}){P}_{\psi\bar{\psi}}({\color{red}{ \check{u}-2i}}|{u})} \equiv 2 \int\limits_{\mathbb{R}+i0}  du \,\hat \mu_{\mathcal{F}_{12}}({u}) \equiv 2\, \cW_{\mathcal{F}_{12}}\, . \la{largesmall}
\eeq
Its integrand is manifestly of the single-particle type, being of the form
\beq
\hat \mu_{\mathcal{F}_{12}}({u})= \mu_{\mathcal{F}_{12}}({u})\,e^{-E_{\mathcal{F}_{12}}(u)\tau+ip_{\mathcal{F}_{12}}(u)\sigma}\, ,
\eeq
with energy and momentum given by
\beq
E_{\mathcal{F}_{12}}(u) = E_{\psi}(u) + E_{\psi}({\color{red}\check{u}-2i})\, , \qquad p_{\mathcal{F}_{12}}(u) = p_{\psi}(u) + p_{\psi}({\color{red}\check{u}-2i})\, .
\eeq
It can be easily expanded at weak coupling using the formulae for fermions given in appendix~\ref{fundamentaltransitions}, e.g.,
\beqa\label{weak-exp}
E_{\mathcal{F}_{12}}(u)\!\!\!&=&\!\!\!2+2 g^2 \left(H_{-i u}+H_{i u}\right)+\dots\nn\\
p_{\mathcal{F}_{12}}(u)\!\!\!&=&\!\!\!2 u+g^2 \left(-2 \pi  \coth (\pi  u)+\frac{2}{u-2 i}\right)+\dots\\
\mu_{\mathcal{F}_{12}}(u)\!\!\!&=&\!\!\! \frac{\pi g^2}{u \sinh(\pi u)}\!\Big[1+\nn\\
&&\!\!\!+g^2\! \left(\!\frac{3 u^2-6i u-4}{u^2 (u-2 i)^2}\!-\!\frac{2i
   \pi  \coth (\pi  u)}{u(u-2i)}\!+\!\pi ^2 \text{csch}^2(\pi  u)\!+\!\frac{\pi ^2}{3}\!-\!\left(H_{-i u}\right){}^2\!-\!\left(H_{i u}\right){}^2\!\right)\!+\!\dots \Big]\nn
\eeqa 
where $H_{x}\equiv \partial_x \log\Gamma(1+x)+\gamma_E$ with $\gamma_E$ the Euler-Mascheroni constant. We see from~(\ref{weak-exp}) that the integral in~(\ref{largesmall}) starts at one loop at weak coupling, in contrast with the ones in~(\ref{largelarge}) which we claimed are two-loop suppressed. This is something that should be stressed: Of the two contributions originating from the integral~(\ref{Wpsipsi}) it is the one involving a small fermion that gives the leading contribution at weak coupling.

This concludes the mathematical analysis of the fermion integrals. At this point all the contributions can be readily expanded at weak coupling and compared to the perturbative data. Prior to performing this analysis, we shall comment on the interpretation of our results.

\begin{figure}[t]
\centering
\def\svgwidth{16cm}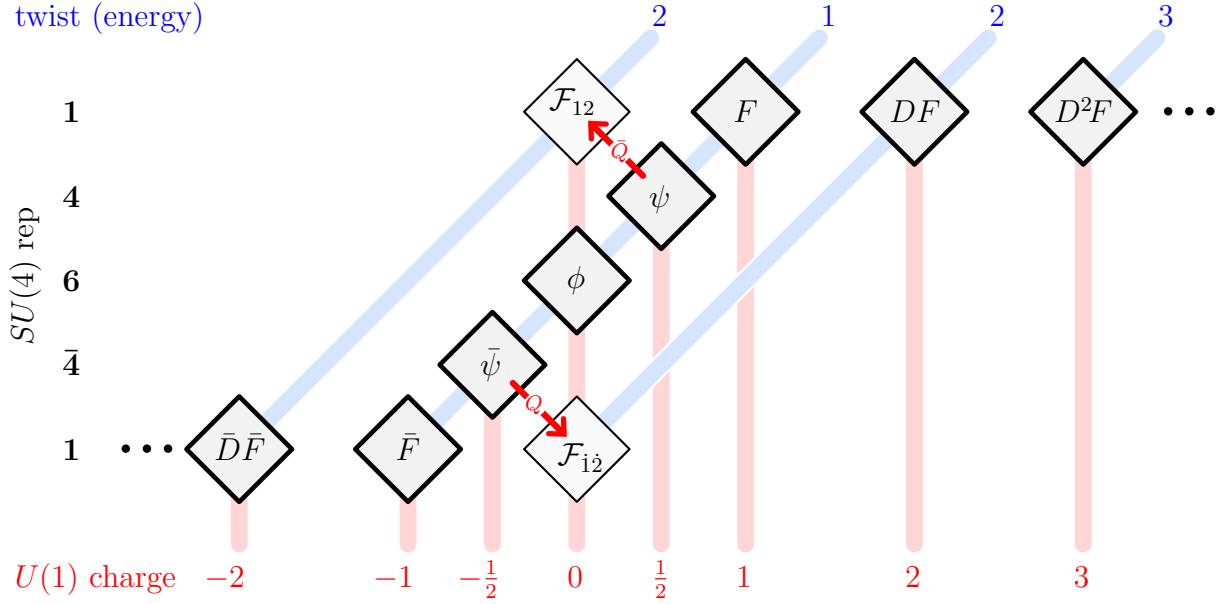
\caption{Extended table of weak-coupling excitations. Besides the fundamental excitations, sitting in the boldfaced squares, stand the effective excitations $\cF_{12}$ and $\cF_{\dot{1}\dot{2}}$. They were obtained by attaching a small fermion to a large one -- or equivalently by acting with a supercharge $\bar{Q}, Q$ on a twist-one fermion $\psi, \bar{\psi}$.  }
\la{Fundamental8}
\end{figure}

We found that the two-fermion contribution naturally decomposes into two terms. One of them is akin to the two-scalar or two-gluon contributions. It is the one given in (\ref{largelarge}). It involves two fermions integrated on the large sheet and seemingly covers the whole range of momenta for the fermions. We know however that this cannot be the full thing since the relevant integrals are performed in rapidity space and two copies (per fermion) of this space are needed to cover the set of all possible (real) momenta, see figure~\ref{fermionsintegration}. 
This is why we need this extra region called small sheet. We recall that the momentum in this sheet is very close to zero at weak coupling for any given rapidity, e.g., $p_{\psi}({\color{red} \check u})=O(g^2)$.  The contribution $\cW^{LS}_{\psi\bar\psi}$ in (\ref{largesmall}) corresponds then to a configuration where one fermion has finite momentum while the other one has vanishing momentum at weak coupling. A fermion with zero momentum is a supersymmetry generator \cite{AldayMaldacena} and as such can be thought of as acting on other excitations in the state. In our case it is a fermion acting on an anti-fermion or vice-versa. This action gives rise to a twist-two particle with the same quantum numbers as those of an excitation created by the component $\mathcal{F}_{12}$ or $\mathcal{F}_{\dot 1\dot 2}$ of the Faraday tensor, here written in the bi-spinor notations. This explains why the expression~(\ref{largesmall}) looks exactly like (twice) the contribution of a single-particle excitation with twist $2$. In the terminology of~\cite{Straps} it is the contribution associated to the $F_{+-} \propto \mathcal{F}_{12}+\mathcal{F}_{\dot{1}\dot{2}}$ component of the Faraday tensor.
 
It is important to stress that the excitation created by $F_{+-}$ is not a new fundamental excitation that should be added to the flux-tube spectrum reviewed in section~\ref{reviewSec}. Instead, as shown by our analysis, it emerges from (a combination of) these fundamental excitations and, therefore, to treat it as an extra fundamental excitation would be overcounting.
Furthermore, looking at~(\ref{weak-exp}) we notice that its dispersion relation is not purely real, in contrast with those for the fundamental flux-tube particles. This means that, in practice, it would not be easy to identify from scratch what is the domain of momenta one should integrate over when dealing with this excitation. It is worth emphasizing that this difficulty is bypassed by our approach which predicts both the precise weight of this excitation and the precise contour of integration. Nonetheless, the $F_{+-}$ boson behaves \textit{effectively} like any fundamental excitation at weak coupling, where the imaginary part in its dispersion relation is found to be small. This explains why this excitation appeared on an equal footing with the fundamental ones in the weak-coupling OPE analysis of~\cite{Bootstrapping,Straps,Wang}. We shall refer to these kind of excitations as composite or effective (or emergent) ones, to distinguish them from the fundamental ones that support the Hilbert space of the theory and underpin our analysis.
 
We further noticed that the contribution of two small fermions is zero, i.e., $\mathcal{W}^{SS}=0$. This fact can be given the following interpretation. This type of configuration can be associated at weak coupling to the covariant derivative $D_+ = \mathcal{D}_{\dot{2}\dot{2}}$, where `$+$' indicates the flux-tube time direction. Since in our two-particle example there are not enough excitations in the state to support the action of the derivative  $D_+$ we could not get anything else than zero. To reveal these kind of excitations one would have to start with more than just a fermion-pair in the state.
 It should become clear that the excitations $\cF_{12}, \cF_{\dot 1\dot2}, D_+,$etc., are just the tip of an iceberg of effective excitations with arbitrarily large twist. They have in common that they are all obtained by binding small fermions to almost everything else there could be in the state. The curious reader will find more about them in the discussion section. At twist two, the ones uncovered earlier are however all we need for proceeding.

We now come back to the weak coupling expansion of our expressions. Combining all the different contributions (with twist smaller or equal to two) we found
\beqa
\mathcal{W}&=&1+\underbrace{2\cos(\phi) \,\cW_F  }_\text{lightest single particles}+\underbrace{2\cos(2\phi) \,\cW_{DF} + 2\,\cW_{\mathcal{F}_{12}}}_\text{next-to-lightest single particles} \nn \\&+&\underbrace{\(\cW_{\phi\phi}+\cW^{LL}_{\psi{\bar \psi}}+\cW_{F\bar F} \) + 2\cos(2\phi) \cW_{FF} }_\text{two particles} + \mathcal{O}(e^{-3\tau})\, . \la{fullexp}
\eeqa 
To the leading order at weak coupling they read
\beqa\la{list}
\begin{array}{lll}
\mathcal{W}_F &\!\!\!\!\displaystyle =  g^2 \,e^{-\tau} \int\limits_{\mathbb{R}} \frac{du}{2\pi} \,\frac{-\pi}{ \cosh(\pi u) \(u^2+\frac{1}{4}\)} \,e^{2iu \sigma}  &\!\!\!\! + \,\,\mathcal{O}(g^4)\, , \\
\mathcal{W}_{DF} &\!\!\!\!\displaystyle =  g^2\, e^{-2\tau} \int\limits_{\mathbb{R}} \frac{du}{2\pi}\, \frac{\pi u}{ \sinh(\pi u) \(u^2+1\)} \,e^{2iu \sigma} &\!\!\!\!+\,\, \mathcal{O}(g^4)\, ,  \\
\mathcal{W}_{\mathcal{F}_{12}} &\!\!\!\!\displaystyle=  g^2\, e^{-2\tau} \int\limits_{\mathbb{R}+i0} \frac{du}{2\pi}\, \frac{\pi}{u \sinh(\pi u)}  \,e^{2iu \sigma} &\!\!\!\! + \,\, \mathcal{O}(g^4)\, ,  \\
\mathcal{W}_{F\bar F}&\!\!\!\!\displaystyle=g^4 \,e^{-2\tau} \int\limits_{\mathbb{R}} \frac{du}{2\pi} \,\int\limits_{\mathbb{R}} \frac{dv}{2\pi} \,\frac{\pi ^3 (\tanh (\pi  u)-\tanh (\pi  v))}{(u-v) 
   ((u-v)^2+1) \cosh(\pi u) \cosh(\pi v)} \,e^{2i(u+v)\sigma} &\!\!\!\! + \,\, \mathcal{O}(g^6)\, ,  \\
   \mathcal{W}_{\phi \phi}&\!\!\!\!\displaystyle=g^4 \,e^{-2\tau}  \int\limits_{\mathbb{R}} \frac{du}{2\pi} \,\int\limits_{\mathbb{R}} \frac{dv}{2\pi} \, \frac{3 \pi ^3 (u-v)  (\tanh (\pi  u)-\tanh (\pi 
   v))}{\left((u-v)^2+1\right) \left((u-v)^2+4\right)\cosh(\pi u) \cosh(\pi v)} \,e^{2i(u+v) \sigma} &\!\!\!\! + \,\, \mathcal{O}(g^6)\, ,  \\
   \mathcal{W}^{LL}_{\psi \bar \psi}&\!\!\!\!\displaystyle=g^4 \,e^{-2\tau}  \int\limits_{\mathbb{R}+i0} \frac{du}{2\pi} \,\int\limits_{\mathbb{R}+i0} \frac{dv}{2\pi} \,   \frac{4\pi ^3 (\text{cotanh}\, (\pi  v)-\text{cotanh}\, (\pi 
   u))}{\left((u-v)^2+4\right) (u-v)\sinh(\pi u) \sinh(\pi v)}\,e^{2i(u+v)\sigma}&\!\!\!\! + \,\,\mathcal{O}(g^6)\, ,  \\ 
   \mathcal{W}_{FF}&\!\!\!\!\displaystyle=g^8 \,e^{-2\tau}  \int\limits_{\mathbb{R}} \frac{du}{2\pi} \,\int\limits_{\mathbb{R}} \frac{dv}{2\pi} \,   \frac{\pi ^3 (u-v)   (\tanh (\pi  u)-\tanh (\pi  v))}{2\left(
   u^2+\ft{1}{4}\right)^2 \left(v^2+\ft{1}{4}\right)^2 \cosh(\pi u) \cosh(\pi v)} \,e^{2i(u+v)\sigma}&\!\!\!\! + \,\,\mathcal{O}(g^{10})\, .
   \end{array}
\eeqa
Higher-loop corrections are more bulky but trivially generated using formulae in Appendix~\ref{fundamentaltransitions}, see, e.g.,~appendix~E in~\cite{long} for a more detailed exposition of the procedure.
We easily verify the aforementioned feature that the contributions of the single-particle type start at one loop while multi-particle ones are more suppressed. In this respect, it is interesting to note that the two-gluon component $\mathcal{W}_{FF}$ is far more suppressed than all the other contributions, i.e., it only kicks in at four loops.

It remains to evaluate the above integrals over the rapidities. It is always possible to expand them in powers of $e^{\sigma}$, as explained below. We do not know, however, how to systematically obtain these integrals in a closed form, especially at high loop orders. One strategy is to identify the proper basis of functions in terms of which the integrals can be expressed and fix the unknown coefficients by comparison, see e.g.~\cite{long} for the twist-one contributions and, more generally, \cite{4Loops}. More ambitiously, one could try to compute these integrals directly along the lines of~\cite{Papathanasiou:2013uoa}.

Here, the goal is in comparing our predictions with the existing perturbative data for the hexagonal Wilson loop. This can be done quite convincingly by matching directly the two expressions either numerically or at the level of their series expansion around particular points. Below we shall focus on the point $\sigma = -\infty$ where the computation of our integrals boils down to extracting residues in the lower-half of the rapidity plane, e.g.,
\beq
\mathcal{W}_{\mathcal{F}_{12}} = -i\sum_{n=0}^{\infty} \underset{u=-in}{\operatorname{{\rm residue}}}\[\frac{\pi g^2}{u \sinh(\pi u)}e^{2iu \sigma-2\tau} + \mathcal{O}(g^4)\]\, .
\eeq
By truncating these sums, one can easily obtain any desired order in the small $e^\sigma$ expansion, as illustrated in table~1 (up at order $e^{2\sigma}$).

At one and two loop orders, we can easily see that our expansion is in perfect agreement with well established perturbative results. Namely, we can construct the ratio of polygons $\mathcal{W}$ using the one loop result -- given by the BDS ansatz \cite{BDS} -- and the two loop correction -- first computed in \cite{twoLoopsBefore} and  dramatically simplified in \cite{twoLoopsAfter}. Expanding this ratio for $\tau\to +\infty$ and then for $\sigma \to -\infty$ perfectly reproduces the result obtained from the OPE. This is already exciting since it is at two loops that most of the two-particle contributions kick in, see (\ref{list}). More precisely, only the contribution of two gluons of the same helicity, i.e., $\mathcal{W}_{FF}$, starts at a higher loop order and is therefore not yet probed by this comparison.

The story becomes even more interesting at three loops and higher. At these orders, a first principle computation of the Wilson loop using Feynman diagrams seems currently out of reach. Instead, we can rely on the alternative approach proposed by Dixon and collaborators for bootstrapping these objects at remarkably high number of loops~\cite{Lance,4Loops}. The basic idea of their method is to identify an appropriate basis of functions in terms of which the hexagon Wilson loop can be expressed in perturbation theory. Their main assumption is that these so-called \textit{hexagon functions} can be written in terms of multiple polylogarithms or, equivalently, in terms of a particular set of iterated integrals -- reminiscent of the recent symbol developments \cite{twoLoopsAfter} but upgraded to function level. The challenge is then to fix all, a priori unknown, coefficients in the corresponding ans\"atze for the perturbative result.

This is where the OPE appears helpful in providing valuable boundary data. One can expand the ans\"atze in terms of hexagon functions for {$\tau\sim \infty$ and $\sigma\sim -\infty$ and match it against our OPE predictions until all the coefficients have been determined}. Needless to say that fixing a finite number of coefficients necessitates no more than finitely many terms in this expansion. Once the ans\"atze have been totally fixed, all the (infinitely many) other terms in the expansion constitute non-trivial consistency checks which provide strong support for both approaches. Moreover, since the hexagon functions ans\"atze saturate the information at a given loop order, once known, they feed back the OPE program with arbitrarily many points of comparison at higher twists. These are the motors of the rich interplay between the two approaches. 
 
We conclude this section with some further details on this exciting interplay. We refer the reader to \cite{4Loops} for a more detailed exposition of these comparisons and in particular to table 1 therein for a nice summary.

\newpage
\begin{landscape}
$\vspace{-.9cm}$
\footnotesize
\beq
\mathcal{W}_\text{hex}= 1+e^{-\tau} \(e^{i\phi}+e^{-i\phi}\)  {\color{black}\mathcal{A}}+ e^{-2\tau} \(e^{2i\phi}+e^{-2i\phi}\)  {\color{black}\mathcal{B}}+ e^{-2\tau} {\color{black}  \mathcal{C}} +\mathcal{O}(e^{-3\tau})\nn 
\eeq
\scriptsize
\begin{eqnarray*}
\color{black}  
\!\!\!\!\!\!\!\!\!\!\!\!\!\!\!\!\!\!\!\!\!\!\!\!\!\!\!\!\!\!\!\!\!\!\!\!\mathcal{A}\!\!\!\!&\color{black}=&\color{black}  \!\!\!\!g^2\Big[e^{\sigma } (2 \sigma -1)+\dots\Big]+g^4 \Big[e^{\sigma } (4-4 \sigma ) \tau +e^{\sigma } \left(-\frac{2 \pi ^2 \sigma }{3}-4 \sigma +6\right)+\dots\Big]+g^6 \Big[ e^{\sigma } (4 \sigma -6) \tau ^2+e^{\sigma } \left(-4 \sigma ^2+\frac{8 \pi ^2 \sigma }{3}+24 \sigma -\frac{5 \pi ^2}{3}-36\right) \tau+e^{\sigma } \Big(-6 \sigma^2+\frac{22 \pi ^4 \sigma }{45}+\frac{5 \pi ^2 \sigma }{3}+36 \sigma \\
\!\!\!\!&\color{black}+&\color{black}\!\!\!\!4 \zeta (3)-\pi ^2-60\Big)+\dots\Big]+g^8 \Big[e^{\sigma } \left(\frac{16}{3}-\frac{8 \sigma }{3}\right)\tau ^3+e^{\sigma } \left(8 \sigma ^2-4 \pi ^2 \sigma -48 \sigma -8 \zeta (3)+\frac{14 \pi ^2}{3}+80\right)\tau^2+e^{\sigma }   \Big(-\frac{8 \sigma ^3}{3}+4 \pi ^2 \sigma ^2+48 \sigma ^2-\frac{12 \pi ^4 \sigma }{5}-\frac{52 \pi ^2 \sigma }{3}-240 \sigma -24 \zeta (3)+\frac{4 \pi
   ^4}{3}\\
      \!\!\!\!&\color{black}+&\color{black}\!\!\!\!\frac{52 \pi ^2}{3}+400\Big) \tau +e^{\sigma } \Big(-\frac{16 \sigma ^3}{3}+\frac{14 \pi ^2 \sigma ^2}{3}+80 \sigma ^2-\frac{146 \pi ^6 \sigma }{315}-\frac{4 \pi ^4 \sigma }{3}-\frac{52 \pi ^2 \sigma
   }{3}-400 \sigma -8 \sigma ^2 \zeta (3)-16 \sigma  \zeta (3)^2+24 \sigma  \zeta (3)-40 \zeta (5)-\frac{4 \pi ^2 \zeta (3)}{3}-48 \zeta (3)+\frac{71 \pi
   ^4}{90}+\frac{40 \pi ^2}{3}+700\Big)+\dots\Big] +\mathcal{O}(g^{10}) \\
  \color{black} \mathcal{B}\!\!\!\!&  \color{black}=&  \color{black}\!\!\!\!g^2\Big[ e^{2 \sigma } \left(-\sigma -\frac{1}{4}\right)+\dots \Big]+g^4 \Big[e^{2 \sigma } \left(3 \sigma -\frac{1}{2}\right)\, \tau{\color{black} +\,e^{2 \sigma } \left(2 \sigma^2+\frac{\pi ^2 \sigma }{3}+\frac{\sigma }{2}+\frac{\pi ^2}{6}-\frac{3}{8}\right) +\dots \Big] }+ g^6\Big[e^{2 \sigma } \left(-\frac{9 \sigma}{2}+\frac{21}{8}\right) \tau ^2{ \color{black} + e^{2 \sigma } \left(-\frac{7 \sigma^2}{2}- 2 \pi ^2 \sigma -\frac{9 \sigma }{2}-\frac{\pi ^2}{8}+\frac{27}{4}\right) \tau +
   e^{2 \sigma } \Big(-\frac{4}{3} \pi ^2 \sigma ^2}\\ 
         \!\!\!\!&  \color{black}-&{  \color{black}\!\!\!\!\frac{27 \sigma^2}{8}-\frac{11 \pi ^4 \sigma }{45}-\frac{13 \pi ^2 \sigma
   }{24} - \frac{3 \sigma }{4}-\frac{5 \zeta (3)}{2}-\frac{11 \pi ^4}{90}-\frac{\pi ^2}{16}+\frac{105}{16}\Big) +\dots  \Big] }\color{black}+ g^8 \Big[ 
      e^{2 \sigma } \left(\frac{9 \sigma }{2}-\frac{9}{2}\right) \tau ^3 \color{black} + e^{2 \sigma } \tau ^2 \Big(\frac{3 \sigma^2}{2}+\frac{9 \pi ^2 \sigma }{2}+\frac{35 \sigma}{2}+5 \zeta (3)-\frac{13 \pi ^2}{8}-\frac{113}{4}\Big)+e^{2 \sigma } \tau  \Big(-\frac{7 \sigma ^3}{2}+\frac{7 \pi ^2 \sigma ^2}{2} +\frac{29 \sigma
   ^2}{2} 
   \\          \!\!\!\!&  \color{black}+&  \color{black}\!\!\!\!\frac{9 \pi ^4 \sigma }{5}+\frac{13 \pi ^2 \sigma }{4}+\frac{91 \sigma }{4}+\frac{37 \zeta (3)}{2}+\frac{17 \pi ^4}{90}-\frac{13 \pi^2}{6}-\frac{629}{8}\Big)   \color{black} \boldsymbol{+ e^{2 \sigma } 
   {\color{black} \Big( -\frac{37 \sigma ^3}{6}+\frac{6 \pi ^4 \sigma ^2}{5}+\frac{19 \pi ^2 \sigma ^2}{8}+\frac{111 \sigma^2}{4}+\frac{73 \pi ^6 \sigma }{315}+\frac{47 \pi ^4 \sigma }{90}+\frac{\pi ^2 \sigma }{6}-\frac{11
   \sigma }{8}+5 \sigma ^2 \zeta (3)+8 \sigma  \zeta (3)^2-\frac{5 \sigma  \zeta (3)}{2}}} \\
         \!\!\!\!& \color{black}\boldsymbol{+}& \color{black}\!\!\!\!\!
   \boldsymbol{{\color{black} 
   25 \zeta (5)+4 \zeta (3)^2+\frac{5 \pi ^2 \zeta (3)}{6}+\frac{39 \zeta (3)}{2}+\frac{73 \pi ^6}{630}+\frac{9 \pi^4}{160}-\frac{\pi ^2}{24}-\frac{5751}{64} \Big)}+\dots \Big]} \color{black}+\mathcal{O}(g^{10})\\
  {\color{black} \mathcal{C}}\!\!\!\!\!&\color{black}=&\color{black}\!\!\!\!\!g^2\Big[4 \sigma -2 e^{2 \sigma }+\dots\Big]+g^4\Big[ 8 e^{2 \sigma } \sigma  \tau{\, \color{black}\boldsymbol{+\,e^{2 \sigma } \left(4 \sigma ^2+\frac{\pi ^2}{3}+\frac{7}{2}\right) -\frac{4 \pi ^2 \sigma }{3}+\dots\Big]}}+g^6\Big[e^{2 \sigma } \Big(-8 \sigma ^2-8 \sigma- \frac{2 \pi ^2}{3}+8\Big) \tau ^2 \color{black}\boldsymbol{+e^{2 \sigma } \tau  \left(-8 \sigma ^2-\frac{16 \pi ^2
   \sigma }{3}-6 \sigma +8 \zeta (3)-\frac{2 \pi ^2}{3}\right)+\frac{44 \pi ^4 \sigma }{45}}\\
   \!\!\!\!& \color{black} \boldsymbol+&\!\!\!\!\!    \color{black}\boldsymbol{e^{2 \sigma } \Big(-\frac{10}{3} \pi ^2 \sigma ^2-4 \sigma ^2-\frac{2 \pi ^2 \sigma }{3}+12 \sigma -8
   \sigma  \zeta (3)+8 \zeta (3)-\frac{3 \pi ^4}{10}+\frac{\pi ^2}{3}-\frac{141}{4}\Big) + \dots \Big]}
   \color{black} 
   +g^8 \Big[ e^{2 \sigma } \tau ^3 \left(\frac{32 \sigma ^3}{9}+\frac{32 \sigma ^2}{3}+\frac{8 \pi ^2 \sigma }{9}-\frac{32 \sigma }{3}-16 \zeta
   (3)+\frac{8 \pi ^2}{9}\right)  \color{black}\boldsymbol{+e^{2 \sigma } \tau ^2 \Big(-\frac{32 \sigma ^3}{3} }
   \\
      \!\!\!\!& \color{black}\boldsymbol{+}& \color{black}\!\!\!\!\!
   \boldsymbol{
   \frac{32 \pi ^2 \sigma ^2}{3}+28 \sigma ^2+\frac{16 \pi ^2 \sigma }{3}+16 \sigma
   -16 \sigma  \zeta (3)-32 \zeta (3)+\frac{14 \pi ^4}{15}-3 \pi ^2-\frac{87}{2}\Big)+e^{2 \sigma } \tau  \Big(-\frac{8}{9} \pi ^2-\frac{80 \sigma
   ^3}{3}+\frac{32 \pi ^2 \sigma ^2}{3}+48 \sigma ^2+\frac{226 \pi ^4 \sigma }{45}-\frac{16 \pi ^2 \sigma }{3}+22 \sigma +16 \sigma ^2 \zeta (3) 
   }\\
         \!\!\!\!& \color{black}\boldsymbol{-}& \color{black}\!\!\!\!\!\boldsymbol{
   64 \zeta (5)-\frac{16 \pi ^2 \zeta (3)}{3}-24 \zeta (3)+\frac{14 \pi ^4}{15}+4 \pi ^2+8\Big)+e^{2 \sigma } \Big(-\frac{8 \pi ^2 \sigma ^3}{9}-\frac{64 \sigma ^3}{3}+\frac{10 \pi ^4 \sigma ^2}{3}+5 \pi ^2 \sigma ^2+\frac{137 \sigma ^2}{2}+\frac{22 \pi ^4 \sigma }{45}-8 \pi ^2 \sigma -168 \sigma +16 \sigma ^3 \zeta (3)-32 \sigma ^2 \zeta (3)}\\
   \!\!\!\!& \color{black}\boldsymbol+&{ \color{black}\boldsymbol{\!\!\!\!\!64
   \sigma  \zeta (5) +\frac{16}{3} \pi ^2 \sigma  \zeta (3) +56 \sigma  \zeta (3)-64 \zeta (5)-\frac{16 \pi ^2 \zeta (3)}{3}-48 \zeta (3)+\frac{296 \pi ^6}{945}-\frac{\pi
   ^4}{20}+\frac{25 \pi ^2}{12}+\frac{3217}{8}\Big)-32 \sigma  \zeta (3)^2-\frac{292 \pi ^6 \sigma }{315}+\dots \Big] } }\color{black}+\mathcal{O}(g^{10})   
\end{eqnarray*}
\small
{\small Table 1: Hexagon WL $\mathcal{W}$ as $\tau \to \infty$, $\sigma \to -\infty$ and $g\to 0$.
The contribution $e^{-\tau \pm i \phi}\mathcal{A}$ corresponds to the propagation of a single gluon with helicity $\pm 1$ and was studied in \cite{short,long}. The contribution $e^{-2\tau \pm 2 i \phi}\mathcal{B}$ describes the propagation of states with total twist $2$ and total helicity $\pm 2$. These can either be a bound-state of two gluons or a state with a pair of gluons of the same helicity. The latter start at four loops and appear boldfaced. Finally, the contribution $e^{-2\tau}\mathcal{C}$ describes the propagation of states of twist $2$ and zero total helicity. These can be a singlet pair of scalars, fermions, or gluons with zero total $R$-charge. All these contributions start at two loops and appear boldfaced. The fermions pair also produces the $F_{+-}$ excitation that shows up already at one loop. The terms in $\mathcal{C}$ which are not boldfaced are captured by this effective excitation. }
\end{landscape}
\newpage

At three loops, the single-gluon contribution is already enough to fix the full ansatz up to a single constant. (Amongst all the parameters fixed in this way, two of them, referred to as $\alpha_1$ and $\alpha_2$ in the literature, played historically an important role. These are the only two parameters that survive at the symbol level and were the only remaining ones in~\cite{Lance} after imposing the constraint from the multi-Regge limit. They were previously fixed by means of the ${\bar Q}$-equation in~\cite{Qbar} and their values reproduced using the OPE in~\cite{short}.) Therefore, up to three loops, the two-particle contributions were matched with a single free parameter (which shows up in the so called beyond-the-symbol part of the three loop result).

At four loops, the single-gluon contribution fixes most of the constants but not all of them (even at the symbol level). Those leftovers can be found by matching the twist-two contribution proportional to $e^{\pm 2i \phi}$. Recall that this is a particularly simple part of the full twist-two result since it is entirely governed by gluons and their bound-states, see~(\ref{fullexp}). Hence, the more complicated term without any $\phi$ dependence -- which probes fermions, scalars, etc. -- is checked in this analysis without any free parameters.  Finally, we should stress that the four-loop match is especially interesting from the OPE viewpoint since it is at this loop order that the two-gluon contribution $\mathcal{W}_{FF}$ shows up for the first time, see~(\ref{list}). In sum, at four loops, we successfully probed, for the first time, \textit{all} the two-particle contributions to the hexagon Wilson loop.

\section{Strong coupling}\la{strongcouplingsec}
We can follow our finite coupling expressions for the single- and two-particle contribution all the way from weak to strong coupling where an emergent description in terms of classical strings should emerge.  In this section we highlight the main features of the strong coupling analysis and make contact with the minimal surface computations \cite{AldayMaldacena,AGM,AMSV}. 

We start with the simplest contributions of all, i.e., with the single-gluon contributions captured by~(\ref{hexagonSingle}). We recall that there are two such gluons, with positive and negative $U(1)$ charge respectively.
At strong coupling, these gluons behave as relativistic particles with mass $\sqrt{2}$, as depicted in figure \ref{masses}. They are then more conveniently parametrized in terms of an hyperbolic rapidity $\theta$, such that $E_{F} \pm p_{F}= \sqrt{2} \, e^{\pm \theta} + O(1/g)$. (This immediately follows from the strong coupling evaluation of their dispersion relation~\cite{BenDispPaper} using the map $u=2g \tanh(2\theta)$.) The gluon measure also drastically simplifies at strong coupling~\cite{short} where it simply evaluates to $\mu_{F}(u) =  -1+O(1/g)$. After performing the straightforward change of rapidities in the integral, the single-gluon contributions~(\ref{hexagonSingle}) are then found to be~\cite{short}
\beq
\mathcal{W}^\text{1-gluon} \simeq
-\frac{\sqrt{\lambda}}{2\pi}\, (e^{i\phi}+e^{-i\phi})  \int \frac{d\theta}{\pi\cosh^2(2\theta)}  e^{-\sqrt{2}\tau\cosh\theta+i \sqrt{2}\sigma\sinh\theta}\,,\la{singleStrong}
\eeq
when $g = \sqrt{\lambda}/(4\pi) \gg 1$. This result was already matched with the strong coupling minimal area computation of the hexagon Wilson loop \cite{AGM} in \cite{short}. For completeness, let us briefly review how the comparison goes.

First, we should recall that the string minimal area is given by a sum of several terms~\cite{OPEpaper}. However, as pointed out in \cite{short}, all of these terms cancel out when constructing the Wilson loop ratio $\mathcal{W}$, except for the most interesting one, the so-called Yang-Yang functional. It is therefore against the latter quantity, henceforth denoted $YY_c$, that we should compare all our predictions.

For generic kinematics, i.e., choice of cross ratios, the Yang-Yang functional is given in terms of the solution to the TBA equations for the minimal surface, whose explicit form is not known analytically. Nonetheless, one can straightforwardly derive closed expression at large $\tau$, or more generally at any order in the collinear limit expansion, as explained in \cite{OPEpaper}. One easily obtains, for instance, the first few terms of the development, that read
\beqa
\mathcal{W}^\text{string} \simeq \exp\(-\frac{\sqrt{\lambda}}{2\pi} YY_c\)=1- \frac{\sqrt{\lambda}}{2\pi}\, (e^{i\phi}+e^{-i\phi})  \int_{\mathbb{R}} \frac{d\theta}{\pi\cosh^2(2\theta)}  e^{-\sqrt{2}\tau\cosh\theta+i \sqrt{2}\sigma \sinh\theta} \nn\\
+\frac{\sqrt{\lambda}}{2\pi}   \int_{\mathbb{R}+i0} \frac{d\theta}{\pi\sinh^2(2\theta)}  e^{-2\tau \cosh\theta+2i \sigma \sinh\theta}+ \dots\,. \la{YYc}
\eeqa
We immediately recognize, in the first line, the contribution (\ref{singleStrong}) from the gluons, hence confirming the agreement with string theory.

String theory also provides us with a simple geometrical understanding for both terms in (\ref{YYc}). The first line describes the contribution of the two $\rm AdS$ string modes of mass~$\sqrt{2}$ which appear in the semi-classical quantization of the GKP string~\cite{arkady}. These modes can be viewed as the quantum fluctuations polarized along the two $\rm AdS_5$ directions that are orthogonal to the $\rm AdS_3$ subspace in which the classical string is moving. The last term in~(\ref{YYc}) pertains to the third $\rm AdS$ string mode, a particle of mass $2$ associated to fluctuations inside the $\rm AdS_3$ subspace. There are of course other string modes -- with polarizations in the sphere or along the fermionic directions -- but these are not relevant classically for the minimal surface of interest. The latter is living purely inside $\rm AdS$ and thus only the three AdS modes show up in~(\ref{YYc}).
At the end of this section, we shall point out, using our OPE series, that the sphere is not that irrelevant, \textit{even classically}.

For the moment, we note that we have a puzzle. We claimed earlier that the only single particle states contributing were those corresponding to the two gluonic excitations~$F$ and~$\bar{F}$. However, at strong coupling, we ought to find a third mode, with mass $2$, if we are to reproduce~(\ref{YYc}). The problem is that there is no fundamental excitation whose mass goes to $2$ at strong coupling. The main purpose of this section is to explain how this third mode emerges in the OPE context. We shall find it as an $SU(4)$-singlet compound state of two fermions and we will show that its contribution perfectly reproduces the last term in~(\ref{YYc}).

With this in mind, we turn our attention to the twist-two contribution (\ref{full2pt}) and its several terms. Given our discussion, we shall focus first on the two-fermion contribution $\mathcal{W}_{\psi{\bar \psi}}$.

\begin{figure}
\centering
\def\svgwidth{15cm}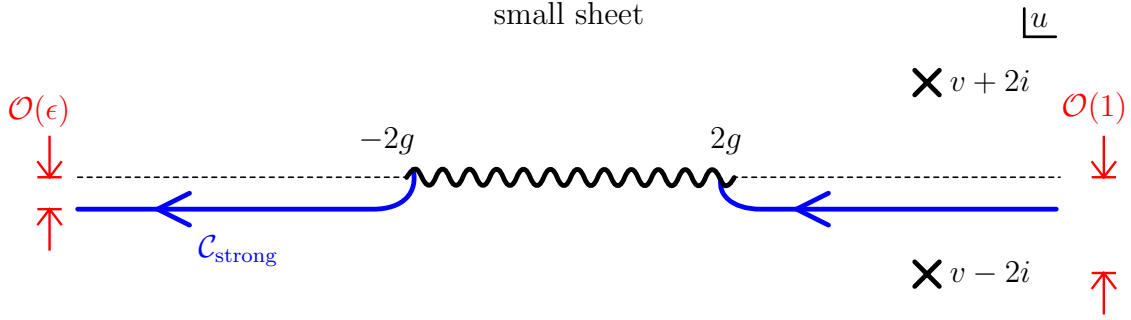
\caption{The relevant contribution in the fermion integrals at strong coupling comes from the sheet where fermions have momenta of order $\sim O(g^0)$. This is what we dubbed small sheet in the text. Fermions lying below the cut, on the so-called large sheet, have semiclassical energy of order $\sim O(g)$ and thus give rise to exponentially small contributions at strong coupling~\cite{BenDispPaper}. Here, we depicted the relevant part of the fermion contour of integration at strong coupling, denoted $\mathcal{C}_{\textrm{strong}}$. The crosses indicate the positions of the poles due to the presence of the extra fermion in the pair. These poles are at a distance of $\mathcal{O}(1)$ from the contour of integration.}
\la{Cstrong}
\end{figure}

The strong coupling analysis of this term differs slightly from the weak coupling one. At strong coupling, we do not want to use the partition of the contour of integration as presented in figure~\ref{fermionsintegration2}. Instead, we want to get back to the one drawn in~\ref{fermionsintegration}, for a reason that will become clear shortly. We thus divide the contour into two new pieces: one running along $(-\infty, -2g) \cup (2g, \infty)$ on the large sheet and the other one along $\mathcal{C}_{\textrm{strong}}\equiv(\infty-i\epsilon, 2g) \cup (-2g, -\infty-i\epsilon)$  on the small sheet, see figure~\ref{Cstrong}. The important point is that the fermionic excitation looks very different on each one of these branches, as illustrated in figure~\ref{FermionScaling}. In the large sheet, for instance, both the energy and momentum are huge. They scale as $\sqrt{\lambda}$ which is typical of a classical excitation. The fermions can be interpreted there as spikes travelling on the GKP string~\cite{Dorey:2010iy} and their contributions are completely negligible at strong coupling. In the small sheet, on the other hand, the fermion energy and momentum are of order $1$. More explicitly, one has~\cite{BenDispPaper}
\beq
E_\psi({\check u})\pm p_\psi({\check u})\simeq\({\bar u\pm1\over\bar u\mp1}\)^{1\over4}\equiv e^{\pm\theta} \qquad\text{where}\qquad \bar u =\frac{u}{2g}\, , \la{disPsi}
\eeq
which is the dispersion relation for a relativistic particle of mass $1$. We note that, in order to keep the energy and momentum fixed, one should consider very large rapidities, scaling as~$g$. The proper rapidity (at strong coupling) is thus the rescaled one, i.e., $\bar{u}$ in~(\ref{disPsi}), and when this one varies from $\pm1$ to $\pm\infty$, as in $\cC_\text{strong}$, all the momenta of the relativistic excitation are covered once. In sum, at strong coupling, only the $\cC_\text{strong}$ contribution matters and $u$ will \textit{always} live on the small sheet. 

\begin{figure}[t]
\centering
\includegraphics[scale=1]{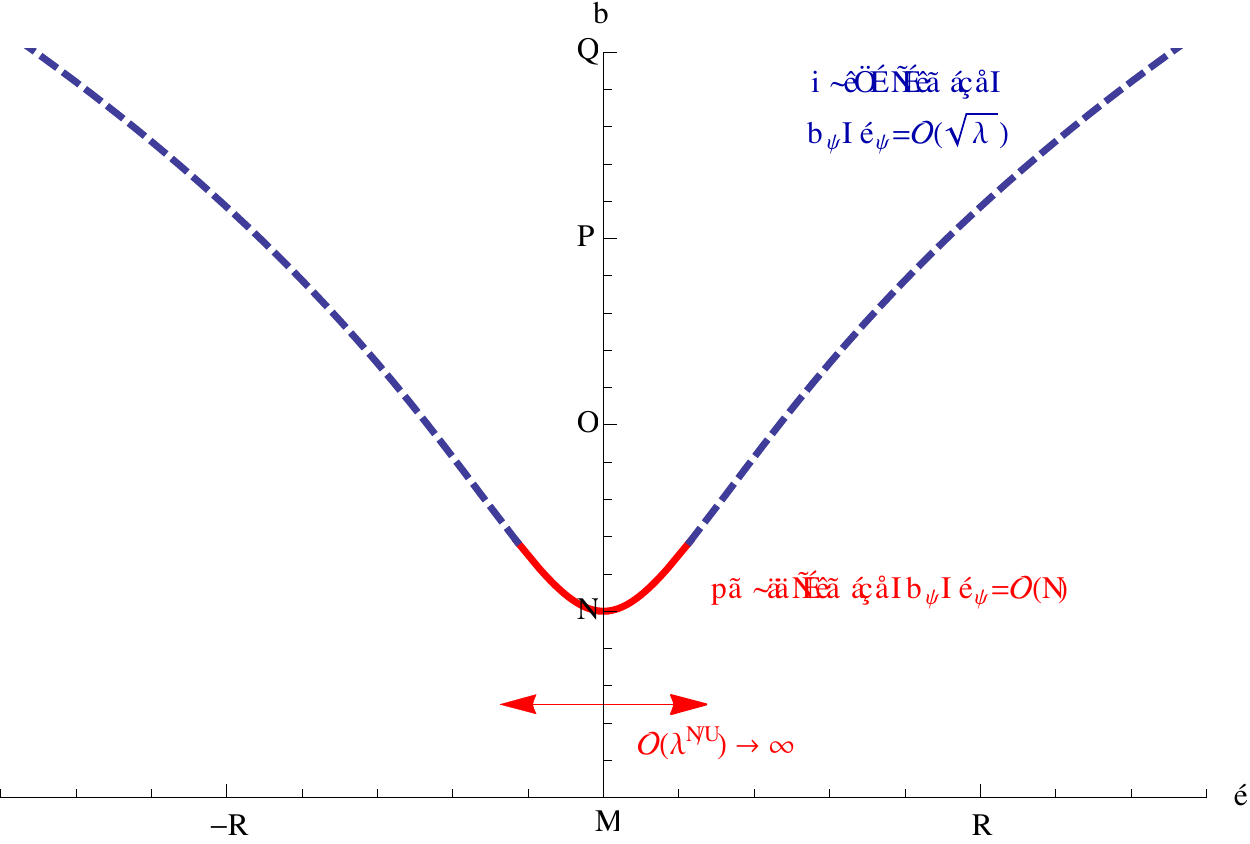}
\caption{The fermion dispersion relation at strong coupling. The solid part of the curve corresponds to a small fermion. It means that it is the region where the fermion rapidity is on the small sheet. The dashed part of the curve is associated to a large fermion. At strong coupling, the large fermion part is pushed to very high energies, signaling its classical interpretation \cite{Dorey:2010iy}. Hence, it drops out from the OPE sum, leaving us with the small fermion only.}\label{FermionScaling}
\end{figure}

Now we should recall that the integrand of the two-fermion contribution~(\ref{twofermions}) has poles for $u-v=\pm2i$. This leads to an interesting phenomenon. Namely, at strong coupling, the contour of integration in $u$ space gets pinched by these poles as we perform the rescaling of the rapidities. 
To properly handle the situation, we must deform the contour of integration into a safer region, as depicted in figure \ref{Cstrong}. Since in our case the full contour closes at $-i\infty$, we deform it toward the lower-half plane, below the point $u = v-2i$. Along the way, we extract the residue at this point. We refer to the corresponding contribution as the \textit{singular} part of the integral and denote it $\mathcal{W}_\text{sing}$, see figure \ref{Cstrong2}. The remaining piece, so-called \textit{regular} part $\mathcal{W}_\text{reg}$, is subleading compared to the latter one at strong coupling. This suppression directly results from the $1/(g^2(\bar{u}-\bar{v})^2)$ factor coming from the poles in~(\ref{twofermions}). 

The leading contribution is thus given by the residue at $v=u-2i$. It takes the form of the single integral
\beq
 \cW_\text{sing}  = -\int \limits_{{\cC_\text{strong}}} \!\!\!\frac{du}{2\pi}\,\frac{\mu_\psi({\check u}) \mu_\psi({{\check u}-2i})}{{P}_{\psi\bar{\psi}}({\check u}|{{{\check u}-2i}}){P}_{\psi\bar{\psi}}({{{\check u}-2i}}|{ \check u})} e^{-\tau(E_{\psi}(\check u)+E_{\psi}(\check u-2i))+i \sigma(p_{\psi}(\check u)+p_{\psi}(\check u-2i))}\, , \la{Wsing}
\eeq
which, with hindsight, we rewrite as 
\beq\la{Wsing2}
 \cW_\text{sing}  = \int \limits_{{\cC_\text{strong}}} \!\!\!\frac{du}{2\pi}\, {\mu}_{m=2}(\check u-i) e^{-\tau E_{m=2}(\check u-i)+i \sigma p_{m=2}(\check u-i)}\, .
\eeq
At strong coupling, we can expand the measure ${\mu}_{m=2}(\check u)$ using the finite-coupling expressions for fermions given in Appendix~\ref{fundamentaltransitions} (the details of this analysis will be presented in~\cite{to appear}). One simply finds at the end that $\mu_{m=2}(\check u) \simeq -1$. We also notice that, since at strong coupling the rapidities are very large, the imaginary shifts in~(\ref{Wsing},\ref{Wsing2}) can be dropped, in a first approximation. This way the energy $E_{m=2}(\check u)$ and momentum $p_{m=2}(\check u)$ become twice those of the mass $1$ fermion with same rapidity. In other words, they describe a relativistic mass $2$ particle. Re-writing then the integral~(\ref{Wsing2}) in terms of the hyperbolic rapidity $\theta$, defined in~(\ref{disPsi}), we conclude that
\beq
\cW_\text{sing}\simeq {\sqrt\lambda\over2\pi}\int\limits_{\mathbb{R}+i\epsilon}{d\theta\over\pi\sinh^2(2\theta)}e^{-2\tau\cosh\theta+2i\sigma\sinh\theta}\, ,
\eeq
to leading order at strong coupling. As anticipated above, we have reproduced the (missing) mass $2$ boson contribution to~(\ref{YYc}) from the two-fermion integrals.

\begin{figure}[t]
\centering
\def\svgwidth{15cm}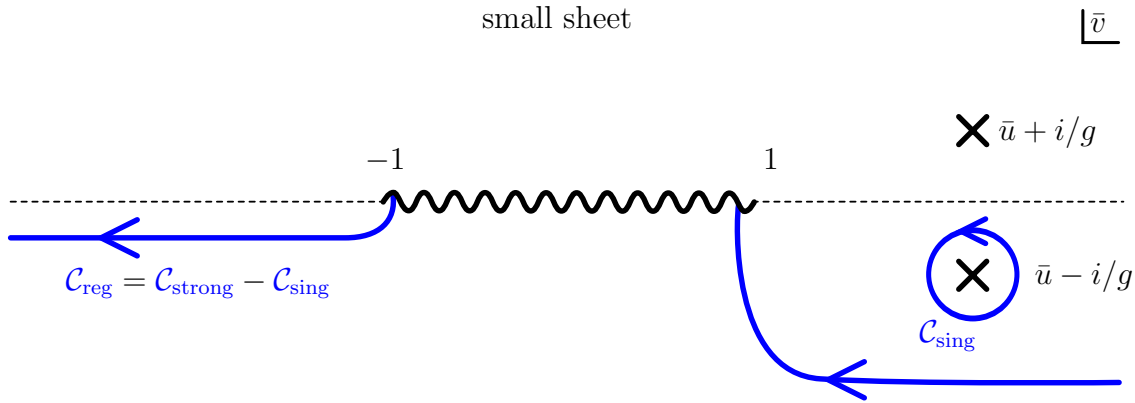
\caption{In the strong coupling limit, the integration contour of one of the two fermions gets pinched between two poles. We deform then the contour into a safer region in the lower-half plane, such as to avoid this pinching. This has the effect of extracting the residue of the pole at $\bar v= \bar u-i/g$. The two-fermion contribution breaks down into this residue and the leftover integral over the safer contour $\mathcal{C}_{\textrm{reg}}$. The latter is subleading at strong coupling compared to the former.}
\la{Cstrong2}
\end{figure}

A few comments about the nature of this excitation are in order. The mass 2 dispersion relation 
\beq\la{mass2-string}
E_{m=2}(p) \qquad \leftrightarrow \qquad  \left\{ \begin{array}{l}
\!E_{m=2}=E_{\psi}(\check u+i)+E_{\psi}(\check u-i) \\
p_{m=2}=p_{\psi}(\check u+i)+p_{\psi}(\check u-i) \end{array} \right.
\eeq
is manifestly real, without any imaginary part signalling an instability of sort. The absence of instability can be confirmed by energetic considerations. Namely, to corroborate that the mass 2 boson is perfectly stable, we should note that its energy is given by
\beq
E_{m=2}(p) -2E_{\psi}(\frac{p}{2}) = - \frac{p^4(4+p^2)^{3\over2}}{128g^2}+ O(1/g^3)\, ,\la{correction}
\eeq
at strong coupling. This directly follows from the fermion dispersion relation~\cite{BenDispPaper} and~(\ref{mass2-string}). What really matters here is the sign of the difference of energies in the left-hand side of~(\ref{correction}). This one is negative, implying that the mass $2$ boson is sitting below the two-fermion continuum in the spectrum. This conclusion actually holds at finite coupling, as illustrated in figure~\ref{Smiley}. The boson is thus clearly \textit{not} an unstable bound state, in the usual sense of the word. How should we think about it then?

The mass 2 boson started its existence at the $\psi\bar{\psi}$-threshold at strong coupling and later moved below it. However, this fact alone does not guarantee that the boson is a genuine asymptotic excitation. It could very well be that the boson migrated to the wrong side of the two-fermion cut, that is toward the unphysical sheet, as we decreased the coupling. Whether this is happening or not can be established by considering the Bethe ansatz equations for two-fermion states on the GKP string. The boson corresponds then to a pole in the singlet-channel scattering phase for the two fermions. A simple kinematical analysis reveals that this pole is indeed \textit{not} sitting in the right strip and thus cannot be associated to a genuine bound state. It corresponds, instead, to what Zamolodchikov called a virtual state in~\cite{Zamolodchikov:2013ama} (following the terminology used in potential scattering). This analysis and its conclusion also agree perfectly with what Zarembo and Zieme described as the dissolution of the mass~2 boson into the continuum of two-fermion states in~\cite{Zarembo:2011ag}. 

\begin{figure}[t]
\centering
\includegraphics[scale=.65]{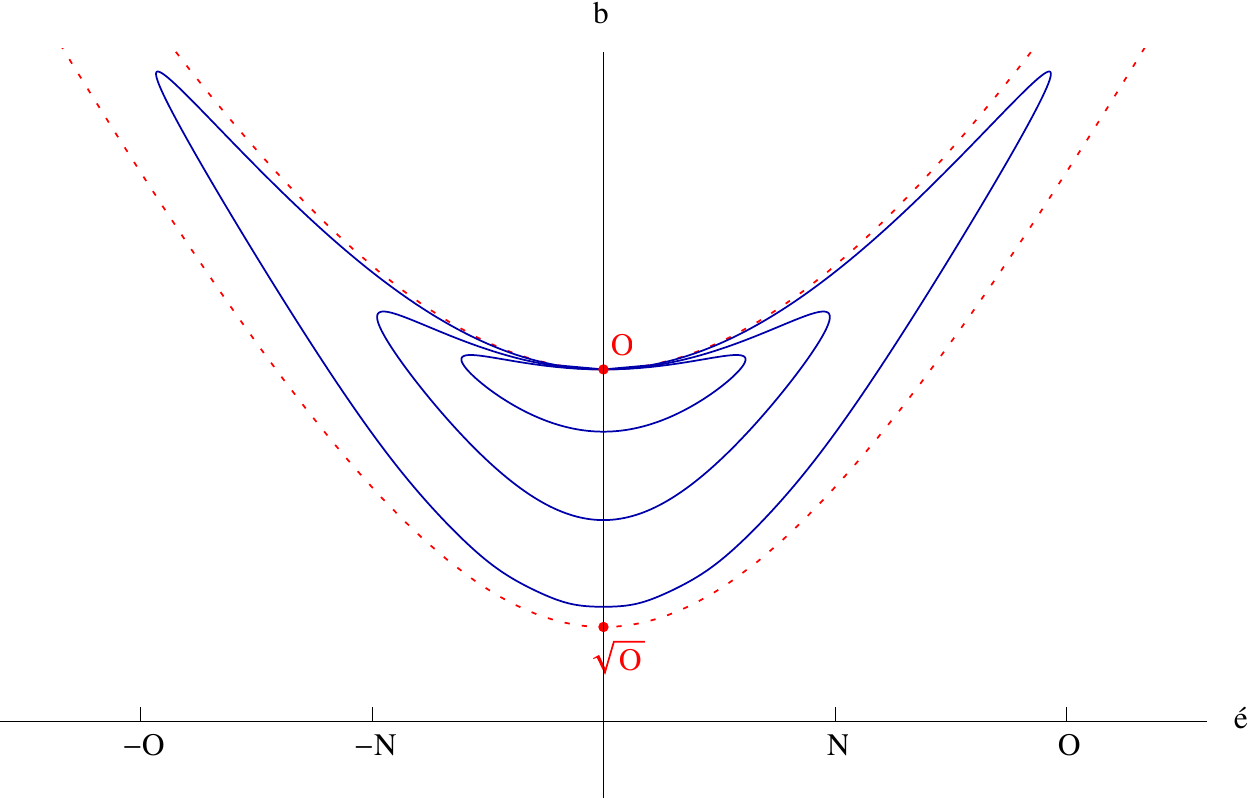}
\includegraphics[scale=.6]{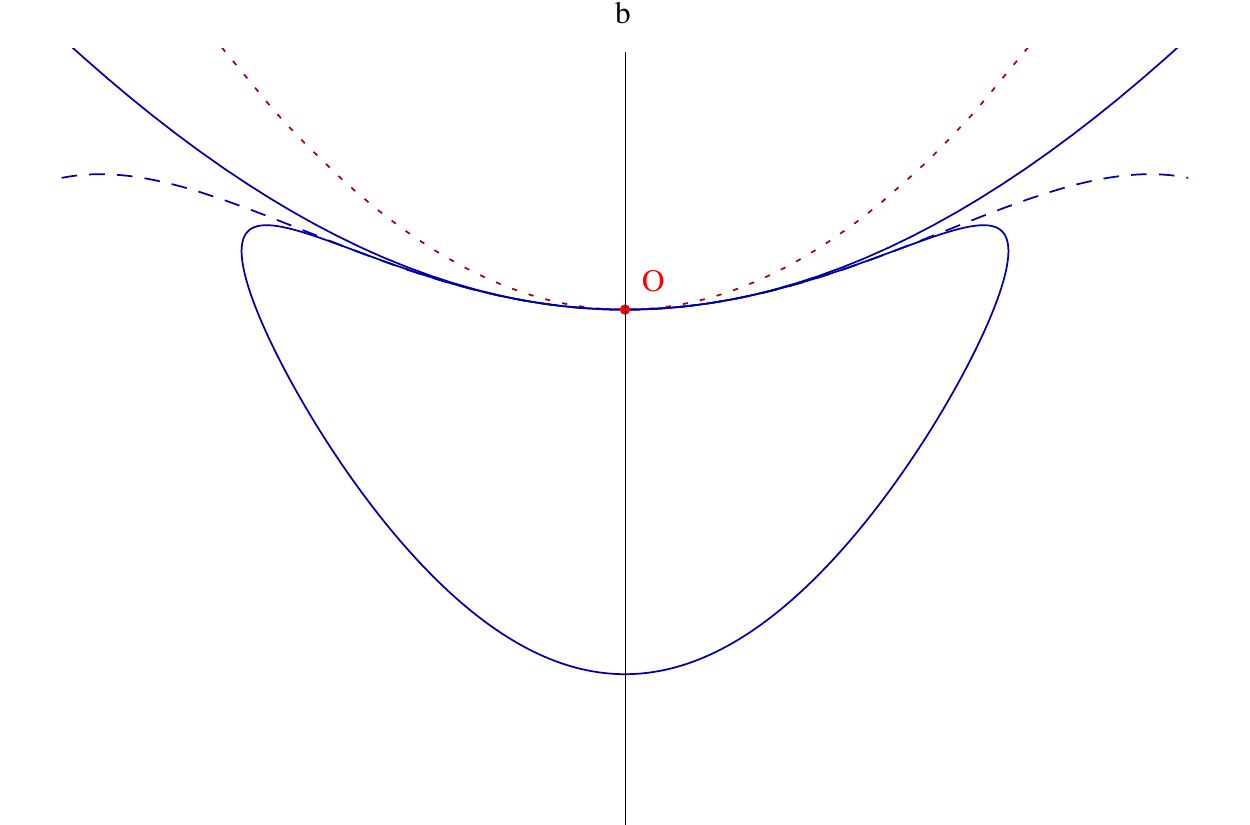}
\caption{Dispersion relation $E_{m=2}(p)$ for the mass $2$ boson from weak to strong coupling. At finite coupling $E_{m=2}(p)$ has two branches that join at some (coupling-dependent) values of the momentum into a smiley shape. On the left, we depicted this dispersion relation for the couplings $g=1,2,10$. The larger the coupling, the larger the smiley will be. As $g\to \infty$ the turning points go to infinity and the two branches effectively decouple. They approach the relativistic dispersion relation of a particle of mass $\sqrt{2}$ and $2$, respectively. On the right, we plotted the dispersion relation of the mass $2$ boson for $g=1$. The outer solid line depicts the energy $2E_{\psi}(p/2)$ of two fermions, moving together. This line is always above the smiley figure, corroborating that the mass $2$ boson is energetically stable. In the bottom dashed line, we represented $2E_{\psi}(p/2)- p^4(p^2+4)^{3/2}/128g^2$, which provides, in agreement with~(\ref{correction}), a better approximation to the upper branch. Finally, the top dotted line is the curve $E=\sqrt{4+p^2}$, towards which all the curves move as $g\to \infty$.}\label{Smiley}
\end{figure}

This achieves our discussion of the two-fermion term $\mathcal{W}_{\psi\bar\psi}$. We shall now briefly sketch the roles of the four remaining twist-two contributions in~(\ref{full2pt}). 

Three of them, namely $\mathcal{W}_{FF}, \mathcal{W}_{F\bar F}, \mathcal{W}_{DF}$, are gluonic and scale like $e^{-2\sqrt{2}\tau}$ in the collinear limit at strong coupling.
For the agreement with the string theory, they ought to match with subleading terms in the OPE of the Yang-Yang functional, or, more precisely, with the first terms in the dots in~(\ref{YYc}). The detailed analysis is beyond the scope of this paper and shall be presented elsewhere~\cite{toappear}. Here, we just want to draw the attention of the reader to the fact that the \textit{full} minimal surface area will be recovered only after including the OPE contributions of all these multi-particle states.
This contrasts with the pattern that we started to unveil at weak coupling, where we observed that multi-gluon contributions are delayed by higher powers of $g^2$. At strong coupling, all multi-gluon corrections should be taken at once to witness the emerge of the classical surface out of the OPE series.

Finally, we close this section with the case of the scalars contribution. This is the last twist-two correction in~(\ref{full2pt}), whose expression we recall here for convenience,
\beq\la{Twoscalars2}
\mathcal{W}_{\phi\phi}=\frac{1}{2} \times 6 \, \int \frac{du \,dv}{(2\pi)^2} \frac{\mu_\phi(u)\mu_\phi(v)\,e^{-\tau(E_\phi(u)+E_\phi(v))+i \sigma (p_\phi(u)+p_\phi(v))}}{g^4\((u-v)^2+4\)\((u-v)^2+1\)P_{\phi\phi}(u|v)P_{\phi\phi}(v|u)} \,.
\eeq
The crucial observation is that the scalars can be identified~\cite{AldayMaldacena} with the excitations of the (two-dimensional) $O(6)$ sigma model, at strong coupling. They become relativistic~\cite{Basso:2008tx,BenDispPaper},
\beq
E_{\phi}(u)\simeq m \cosh(\ft{\pi}{2} u) \,, \qquad p_{\phi}(u)\simeq m \sinh(\ft{\pi}{2} u)\, ,
\eeq
and their mass \cite{AldayMaldacena,Fioravanti:2008rv,Basso:2008tx},
\beq\la{mass}
m= e^{-\frac{\sqrt{\lambda}}{4}} \,{2^{1/4}\lambda^{1/8}/\Gamma(\ft{5}{4})}\(1+ \dots\) \,,
\eeq
exponentially small (as $\lambda \to \infty$). This, alone, leads us to expect that these very light excitations can trigger an interesting behaviour in the terms of the OPE series. In particular, given that they can made arbitrarily light, we should be obliged to re-sum all their contributions at large enough coupling. In other words, the truncation to the two-scalar contribution is almost never justified at strong coupling, as should become clear shortly. 

To demonstrate all that, we first note that the combination of pentagon transitions and measures, appearing in the scalars integrals~(\ref{Twoscalars2}), is \textit{not} suppressed at large coupling. Instead, one has
\beq\label{mu-P-O6}
\mu_{\phi}(u) \simeq g \frac{\sqrt{\pi} \Gamma(\ft{3}{4}) }{ \Gamma(\ft{1}{4})} \, , \qquad P_{\phi\phi}(u|v) \simeq \frac{ \Gamma(\ft{1}{4}-\ft{i}{4}u+\ft{i}{4}v)\Gamma(\ft{i}{4}u-\ft{i}{4}v)}{4g\Gamma(\ft{3}{4}-\ft{i}{4}u+\ft{i}{4}v)\Gamma(\ft{1}{2}+\ft{i}{4}u-\ft{i}{4}v)} \, , 
\eeq
as explained in appendix~\ref{scalar-manip}. They are such that the explicit $g$ dependence actually drops out of the ratio in~(\ref{Twoscalars2}). The result depends thus on the `bare' coupling constant $g$ only through the `physical' mass $m$, in line with the low-energy effective description in terms of the $O(6)$ model.

The relativistic invariance is also made manifest in~(\ref{mu-P-O6}), with the measure being a constant and the pentagon transition a function of the difference of rapidities. 
The immediate consequence of this boost symmetry is that the two-scalar contribution only depends on $\tau, \sigma$ and $m$, at strong coupling, through the relativistic invariant combination $z\equiv m \sqrt{\tau^2+\sigma^2}$. 

Plugging now the pentagon transition and measure~(\ref{mu-P-O6}) into the integrals~(\ref{Twoscalars2}) and performing the integration over the centre of mass rapidity $u+v$, we end up with the single integral
\beq
\mathcal{W}_{\phi\phi}(z) \simeq  \,3 \int\limits_{-\infty}^{+\infty} \frac{d\theta}{4 \pi ^2}\, \frac{\theta\tanh \left(\frac{\theta }{2}\right)  \Gamma \left(\frac{3}{4}\right)^2\Gamma \left(\frac{3}{4}-\frac{i \theta }{2 \pi }\right) \Gamma \left(\frac{3}{4}+\frac{i
   \theta }{2 \pi }\right) }{\,\left(\theta ^2+\pi ^2\right)\, \Gamma
   \left(\frac{1}{4}\right)^2 \Gamma \left(\frac{5}{4}-\frac{i \theta }{2 \pi }\right) \Gamma \left(\frac{5}{4}+\frac{i \theta }{2 \pi }\right)} \,K_0(2 z \cosh \left(\ft{\theta }{2}\right)) \,, \la{2ptS}
\eeq
with $K_0$ the modified Bessel function of the second kind and $\theta = \ft{\pi}{2}(u-v)$. Though it captures the two-scalar states only, this integral perfectly illustrates the type of contributions the scalars give at strong coupling. We expect, from the $O(6)$ model perspective, that all multi-scalar states will yield $O(\lambda^0)$ contributions of this kind, though with more integrals and rapidities involved. This is in line with the stringy intuition according to which the scalars, i.e., the sphere, should contribute towards the $\lambda$ independent prefactor in the partition function~$\cal{W}_{\textrm{string}}$. While this is morally correct, strictly speaking, the reality is a bit more complicated and interesting.

The catch is that the integral (\ref{2ptS}) barely converges for large coupling, since $z$ is exponentially small for $\tau,\sigma \sim 1$. The two-particle integral (\ref{2ptS}) displays, as a result, a logarithmically divergent behaviour $\propto \log z$ for $z\sim 0$, {which is visible in the yellow curve of figure~\ref{scalar plot}}. 
From the low-energy viewpoint, this logarithmic scaling has an UV origin, indicating that it should be generic to all the multi-scalar integrals.  The expectation, in fact, is that the integrals involving more scalars will diverge as higher powers of $\log z$. This points toward the need for their re-summation, obviously. Discerning the true $\lambda$ dependence of the sphere correction at strong coupling entails then uncovering the genuine small $z$ scaling of the sum of all multi-scalar corrections. If this sum enhances to a power law behaviour in $z$, as it is often the case in these circumstances, we would get a contribution from the scalars \textit{of the same order of magnitude} as the minimal surface area!
As surprising as this might be, we claim here that this is what is happening~\cite{to appear}. We note in this regard that this unexpected `semi-classical' correction must be a constant independent of the WL geometry, since $\log z=-\ft{1}{4}\sqrt\lambda+(\text{subleading})$ at strong coupling. 

Clearly, it is not an easy task to re-sum all the scalars and check all these claims. (One difficulty is related to the matrix parts of the multi-scalar transitions. As the number of scalars increases, the $SU(4)$ index structure of the transitions becomes more and more cluttered and it is a fascinating problem to uncover their pattern.) We gave here all the evidence we could, based on the two-particle information, and shall present a more complete study in a future publication~\cite{to appear}. Our above analysis already illustrates, however, how rich is the scalar sector at strong coupling.

\begin{figure}[t]
\centering
\includegraphics[scale=1]{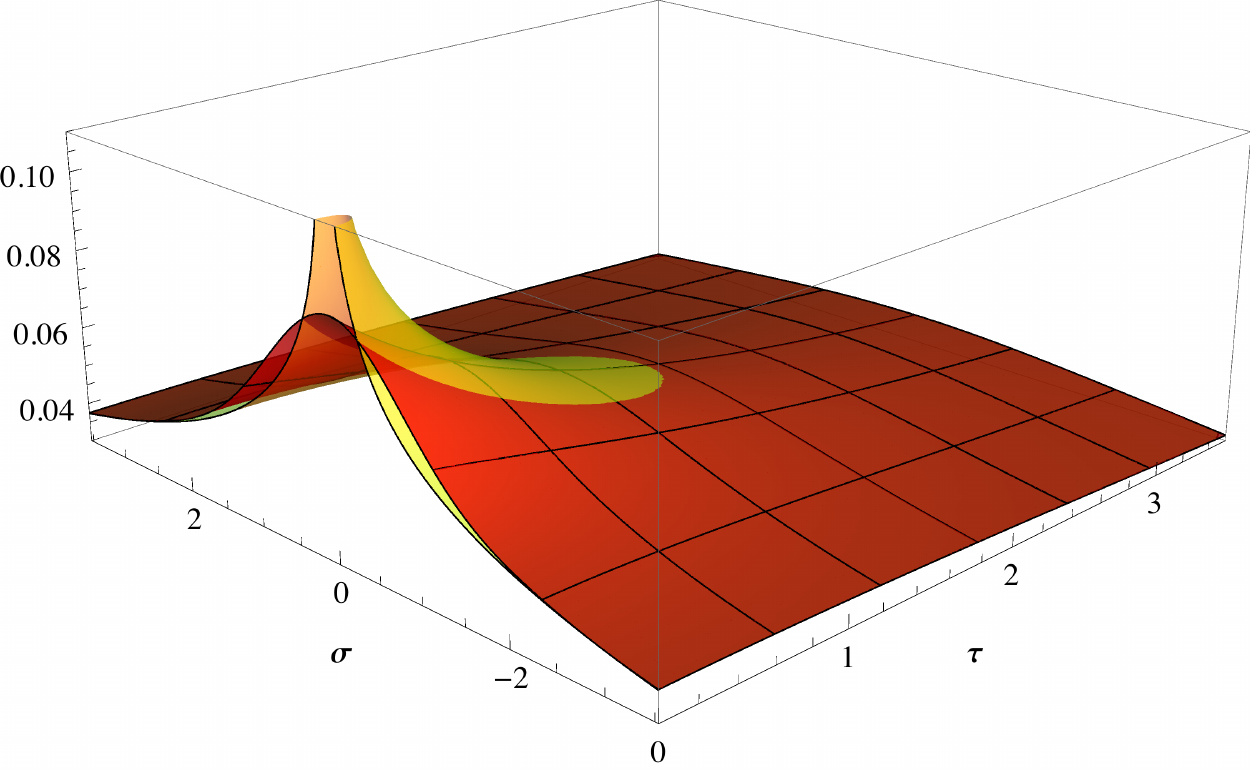}
\caption{{The two-scalar contribution $\mathcal{W}_{\phi\phi}$ is plotted here in red for $g=2$ (i.e., for large 't Hooft coupling $\lambda \simeq 630$). The yellow curve corresponds to the asymptotic formula~ (\ref{2ptS}) evaluated at the same value of the coupling, i.e., for a mass $m\simeq 0.0055$ (related to $g=2$ through~(\ref{mass})). We see that the yellow and red plots are well on top of each other except for a small region close to $\tau,\sigma \to 0$. In the latter (UV) region the strong coupling formula develops a logarithmic singularity as discussed in the main text.}
}\label{scalar plot}
\end{figure}

\section{Discussion}\la{discussion-sect}

Scattering amplitudes in planar $\mathcal{N}=4$ SYM theory are dual to null polygonal Wilson loops and as such
are subject to
the  OPE. This remarkable circumstance allows us to represent the scattering amplitude in the form of the OPE series (\ref{hexagon2}) which can be thought of as a partition function of sort for the gas of flux-tube excitations. This picture is manifestly similar to the form factor expansion for correlators in massive two-dimensional quantum field theories and, therefore, is very well-suited for setting up the integrable boostrap of the gauge/string theory.
Its most valuable feature is that it is completely non-perturbative, being valid all the way from weak to strong coupling.  
In this paper, following~\cite{short,long}, we have shown how the OPE sum reproduces, at weak coupling, the most advanced perturbative field theory computations and how, as we crank up the coupling, a dual description in terms of strings in higher dimensional space-time emerges from it. In our opinion, this provides us with a rather palpable and beautiful realization of the holographic duality.

This paper was focused on the two-particle contributions. Its main goal was to pave the way for the daunting prospect of solving the OPE many-body problem and spelling out the \textit{full} OPE of the
the hexagonal Wilson loop.
Adopting now the conventional philosophy about integrable systems -- namely that the two-particle contributions unravel most, if not all, of the many-body physics -- we shall end this paper with a broad discussion about multi particles and their contributions. A more serious study will be presented in \cite{toappear}.

\subsubsection*{Finite Coupling}

In the OPE approach one sums over all the flux tube eigenstates $\psi$.
These are simply given as multi-particle 
states constructed out of the few fundamental flux tube excitations. The latter comprise the lightest excitations, which are eight fermions, six scalars and two gluons, plus the heavier bound-states formed out of gluons, see figure~\ref{fundamentalExcitations}. These excitations are gapped, stable, and expected to span the complete spectrum of flux-tube eigenstates.
Hence, states made out of these fundamental excitations are all we need in order to write down the full OPE series at any finite value of the coupling.
Moreover, the heavier a multi-particle state is the smaller its contribution is (close enough to the collinear limit).
As so, for many numerical purposes,
on can merely truncate the OPE sum to its first few terms. Clearly, the Wilson loop is begging to be put on a computer and plotted all the way from weak to strong coupling. 
{That this can be done in a completely non-perturbative fashion was illustrated in figures~\ref{gluonsPlot},~\ref{fermionsPlot} and~\ref{scalar plot} where several contributions to the OPE are plotted at finite values of the coupling. It remains to add all of them together.}

In this paper, we already acquired most of the ingredients for moving forward with this exciting endeavour. Indeed, as already exemplified in~\cite{short} for the gluonic excitations, multi-particle contributions are often no more complicated than the two-particle ones. There is, however, an extra complication to overcome  when dealing with scalars and fermions. The point is that scalars and fermions carry R-charge indices. As a result, their pentagon transitions are tensors with regards to incoming and outgoing indices. Precisely, they are believed to factorize~\cite{long} into a scalar part, whose form is (almost) universal, and a much simpler tensorial part, to which we colloquially referred as the \textit{matrix part} in~\cite{long}. The latter does not depend on the coupling constant when written in terms of rapidities. After combining the pentagon transitions together and contracting their indices, as was done for instance in sections~\ref{scalar-section} and~\ref{fermionssection}, these matrix parts will give rise to rational factors in the OPE integrands. For illustration, the rational functions of the rapidities appearing explicitly in (\ref{Twoscalars}) and (\ref{twofermions}) stand as the simplest such factors. An efficient analytic understanding of these matrix parts, or more exactly of their contractions, is the main new conceptual ingredient in the way to plotting the amplitude at any finite coupling. This part is beyond the scope of this paper and will be examined elsewhere. It is tempting to muse that these rational factors, because of their algebraic origin, are somehow connected to the perimeter and/or off-shell nested Bethe ansatz. The latter was recently revived in \cite{Frassek:2013xza} and argued to be relevant to the study of the spectrally deformed integrands for scattering amplitudes~\cite{spectralstuff} while the former was designed to offer an handle on the tensorial part of form factors in certain integrable theories, see~\cite{Babujian:2006md} and references therein.

\subsubsection*{Weak Coupling}

As we have seen in this paper an interesting phenomenon occurs when expanding the OPE series at weak coupling. 
It is tied to the strange behaviour of the fermionic excitations. These ones naturally split into two types, dubbed {\it small} and {\it large} fermions, with rather distinct  features at weak coupling. Large fermions carry momenta of order $O(1)$ and behave like any other single-particle excitation. Small fermions on the other hand have vanishingly small momenta in perturbation theory and play quite a special role.

Let us first recall what makes these two classes of excitations so much different at weak coupling. We begin with the generic behaviour of the flux-tube excitations to which the large fermions conform, as any of the fundamental excitations depicted in figure~\ref{fundamentalExcitations}. It has to do with the remarkable fact that the creation or annihilation of multi-particle states
is more and more loop suppressed as the number of particles grows. This, on its own, is not surprising and admits a direct Feynman diagrammatic explanation. What is perhaps more curious and appears to defy the diagrammatic intuition is how much suppressed some contributions happen to be. For instance, the contribution of a multi-particle state consisting of $N$ gluons with same helicity only kicks in at $N^2$ loops, that is, appears roughly $N$ times more suppressed than naively expected. Though a bit extreme, this case illustrates the predominant characteristic of this generic class of excitations, namely that at any given loop order the amount of such particles in the contributing states  is bounded by a fairly small number. This aspect, solely, makes the re-summation of the OPE contributions of these excitations at weak coupling elementary if not superfluous.

The small fermions deviate significantly from the above behaviour and span a new direction in regards of the weak-coupling re-summation.
The main point is that zero-momentum fermions act as supersymmetry generators \cite{AldayMaldacena}. Therefore, adding small fermions to any given state does not cost an additional power of the coupling. Figuratively fermions of this kind are cheap, and we can have any number of them already at one-loop order.
The first time they reveal their originality is at the two-fermion level, with one thereof being large and the other one small. This is the case encountered in this paper. The small fermion acted then as a SUSY generator on the large one, converting it into an higher-twist excitation which we qualified as composite or effective and which we represented as the Faraday tensor component $F_{+-}$.
We have seen how precisely this happened with the integration over the small-fermion rapidity localizing completely around the pole induced by its partner, hence forming a string or composite object ready to be identified with the $F_{+-}$ excitation.

\begin{figure}[t]
\centering
\def\svgwidth{16cm}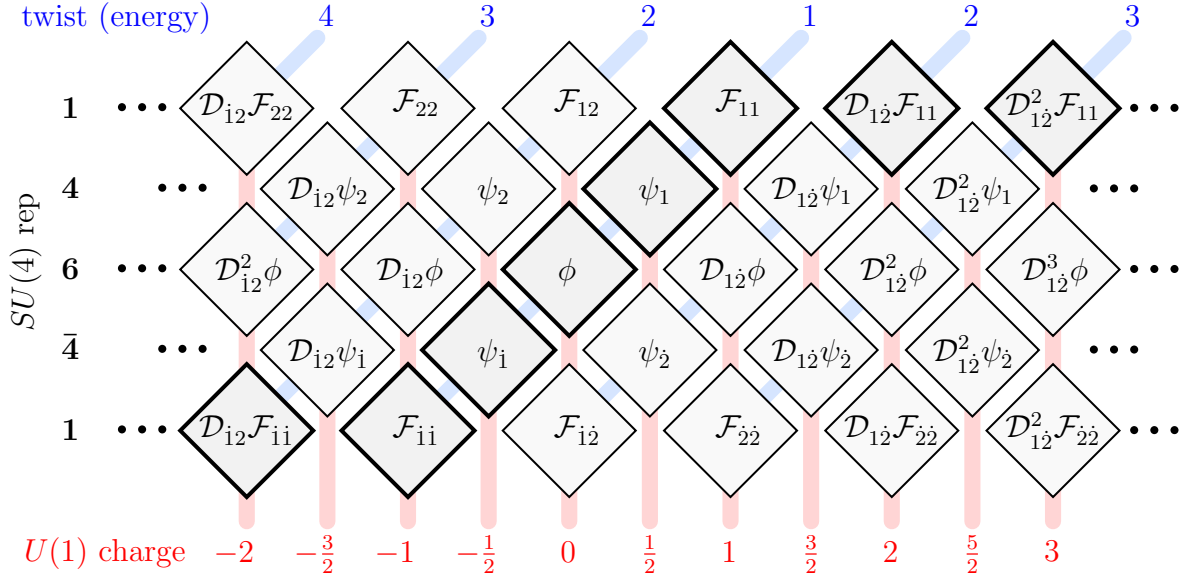
\caption{Completed table of weak-coupling excitations. The fundamental excitations sit in the boldfaced squares.
All the other fields in this table correspond in our terminology to composite or effective excitations, which emerge as long-lived resonances at weak coupling. In the double-spinor notation, see e.g.~\cite{Beisert:2003jj}, the flux-tube symmetries are generated by $\cD_{1\dot2}=D_z$ (equivalently $\cD_{\dot12}=D_{\bar z}$) for rotation in the transverse plane, $\cD_{1\dot1}= D_-$ and $\cD_{2\dot2}= D_+$ for translations along the flux-tube space and time directions. The dictionary also includes $\cF_{11}= F_{-z}$, $\cF_{\dot1\dot1}= F_{-\bar z}$, $\cF_{22}= F_{+\bar z}$, $\cF_{\dot2\dot2}= F_{+z}$, $\cF_{12}=F_{+-}+F_{z\bar z}, \cF_{\dot1\dot2}=F_{+-}-F_{z\bar z}$, and so on.
 In the table we plotted only those excitations that do not involve the action of $D_+$ derivatives.
The complete set of weak coupling excitations would be straightforwardly obtained by adding a third direction to the picture for including these derivatives. These ones were called descendants in \cite{Bootstrapping,Straps,Wang}.
} \la{Fundamental}
\end{figure}

It is not hard to figure out what will happen when more fermions will be included.
Since small fermions correspond to supersymmetry generators, which can be chosen as the spinor components $Q_2$ and $Q_{\dot 2}$, they will combine with the $SU(4)$ generators of the flux tube to yield an enhanced $SU(1,1|4)$ symmetry at weak coupling. The spectrum is then expected to organize itself into $SU(1,1|4)$ multiplets of effective particles, as pointed out in~\cite{BenDispPaper}. By restricting to the primaries of the $SU(1,1)$ subgroup, we obtain the array of excitations in figure \ref{Fundamental}, whose representatives are uniquely fixed up to the use of the field equations. One can move diagonally in this figure by adding more (at most $\mathcal{N}=4$) small fermions as indicated by the (red) arrows in figure~\ref{Fundamental8}. 
The full $SU(1,1|4)$ multiplet is obtained by adding the $SU(1,1)$ descendants, that is, by acting with any number of twist-two derivatives $D_+=\{Q_2,Q_{\dot 2}\}$ on the fields in figure~\ref{Fundamental}. These derivatives are nothing but ($SU(4)$ singlet) pairs of small fermions as described in the main text.

A more complete description of the multi-particle excitations at weak coupling is then the following. At any given loop order, we have a \textit{maximum} number of fundamental excitations and \textit{arbitrarily many} small fermions (which we can integrate out exactly). 
Equivalently, we can write our OPE partition function as a sum over multi-particle states built out of the effective excitations in figure \ref{Fundamental} plus their $SU(1,1)$ descendants. In this language, the total number of excitations is bounded at any loop order but they are chosen out of a much larger group of possible excitations as compared to the fundamental ones depicted in figure \ref{fundamentalExcitations}. This was the point of view implicitly adopted in the earlier OPE papers \cite{OPEpaper,Bootstrapping,Straps} where at one loop one had at most one effective particle, at two loops at most two, etc. It is a bit ironic at the end that the perturbative regime, with its associated zoo of effective excitations, appears more intricate than its finite coupling counterpart. 

Involved as it might be, the weak coupling OPE for the hexagonal Wilson loop has been extensively checked against perturbative results. So far, this scrutiny has yielded nothing but perfect agreement between the two, see  \cite{long,Lance} and \cite{4Loops} for notable recent examples at high loop orders. As we move beyond six edges, dual to six gluons, the available perturbative data becomes unfortunately less abundant. Promising ongoing progress in revealing the structure of loop amplitudes with higher multiplicity make us of the so called Cluster Coordinates and Cluster Polylogarithms \cite{Golden:2013xva,GoldenTalk}. Once combined with a minimal OPE input, these may allow one to bootstrap $n$-gon Wilson loop at higher loop orders, along the lines of \cite{Lance,4Loops}, see \cite{SHe} as well for related work in 2d kinematics. It seems also propitious to look for connections between the OPE program and recent perturbative approaches such as the Integrand \cite{ArkaniHamed:2010kv}, the Positive Grassmanian \cite{ArkaniHamed:2012nw}, or the Amplituhedron \cite{Arkani-Hamed:2013jha}. The latter, which their avalanche of symmetries, could shed light on hidden simplicities of the OPE series, for instance. This might require first upgrading them from the integrand to amplitude level, possibly along of the lines of \cite{Lipstein:2013xra}.

\subsubsection*{Strong Coupling}

As we move to higher loops at weak coupling we need to include more and more particles. Not surprisingly, at strong coupling, we need to sum over infinitely many of them. 
What appears quite remarkable is that, in this limit, one can re-sum exactly the OPE series and reproduce the (exponential of the) minimal area of a string in~$AdS_5$. From what we have learned in this paper, as well as in \cite{short}, we can already anticipate a bit on our forthcoming analysis~\cite{to appear}  and see how this comes about. 

The string minimal surface area admits an OPE expansion which involves three different modes. These }are in a one-to-one correspondence with the $AdS_5$ fluctuations of the Gubser-Klebanov-Polyakov string, see~\cite{Frolov:2002av,AldayMaldacena}.
These three excitations
should thus emerge out of our OPE series at strong coupling. Two of them do it in a rather straightforward way, as already observed in~\cite{short}. These are the two gluonic excitations with mass $\sqrt{2}$.
The third $AdS$ mode has mass $2$ and is more elusive. For instance, 
there is no fundamental excitation whose mass approaches $2$ at strong coupling. In this paper we clarified this mini puzzle by showing how this mode arises as a composite excitation made out of two fermions with mass $1$ each. 

We further showed, following~\cite{short}, that the OPE contributions associated to these three modes perfectly reproduce the leading collinear behaviour of the strong coupling Y-system~\cite{OPEpaper}. 
Though rewarding, this agreement is not yet the end of the story at strong coupling. We know that the full minimal surface area  encompasses much more than just these three single-particle excitations. It indeed re-sums the contributions of \textit{all} multi-particle states made out of these excitations and, importantly enough, of their bound states as well. This can be seen either by analyzing the string theory answer, given by the Yang-Yang functional, or, for what matters more here, by staring at the OPE decomposition at strong coupling.

That  the mass $\sqrt{2}$ gluons can form bound states is well established from the flux-tube spectral analysis \cite{BenDispPaper} and it is natural to see these bound-states playing a role at strong coupling. More exotic are the contributions of certain bound-states of the mass $2$ modes. Both the latter and the former are predicted by the stringy Yang-Yang functional that treats them all in a rather symmetric way. This is not obviously the case in the OPE series. Given what we learned in this paper, one could expect however bound-states of mass $2$ bosons to emerge as composite states of $2n$ fermions. We claim this is precisely what happens. Fermions can form effective excitation of mass $2n$ for all integer $n$, at strong coupling. These effective excitations pop up from our decomposition through the same kind of integration contour pinching mechanism as the one uncovered in this paper for $n=1$.

{In sum, at strong coupling, we need to sum over all possible states with any number of gluons and their bound-states plus any even number of fermionic excitations. The latter then give rise to bound states of the mass 2 excitation. All the other OPE contributions are expected to be subleading (with a minor caveat for the scalars' ones).  
We stress again that this qualitative picture is quite clear both from the OPE decomposition and from the stringy Yang-Yang functional. What is more exciting, and will be explained in detail in \cite{to appear}, is to observe the precise \textit{quantitative} match between the many terms of the OPE sum and the corresponding ones in the IR expansion of the stringy minimal surface area.

Finally, things are expected to be even more interesting as we go beyond the leading strong coupling analysis. Past this point, all flux tube excitations become relevant. Their physics is a fascinating problem awaiting to be explored. The scalars are a notable example. They are very light at strong coupling and therefore need to be re-summed, as already discussed in the main text. Surprisingly, their resummation even leads to an additional enhancement of the leading order result \cite{toappear}. This is just one out of many formidable avenues for the future.

\subsubsection*{Closing remarks}

As should be clear from the previous discussions, the re-summation of the full OPE expansion presents different challenges at weak and strong coupling. The weak coupling re-summation entails a truncation, at each loop order, to a maximum number of excitations, which are taken out of the enlarged table of effective excitations depicted in figure \ref{Fundamental}. At strong coupling we need to take into account an arbitrary number of flux tube excitations. At leading order, gluons and composite states of fermions suffice to completely recover the classical minimal area result, while all other excitations start showing up as stringy quantum corrections.

We do not know whether some sort of master object, re-summing the full OPE expansion at finite coupling, exists. What we \textit{do} know is that 
if such an object exists, its interpolation from weak to strong coupling must be quite non-trivial as it ought to transmute into very different incarnations at each end. This object might also not exist at all and our sums already be the best we can aim at at finite coupling. After all, our expressions are akin to the form factor expansions for correlators in integrable models and the common lore about them is that the sums are typically the best one can do. Planar $\mathcal{N}=4$ SYM is however an exceptional theory in many regards and one can be optimistic that further magical simplifications still awaits to be uncovered, as exemplified recently by the remarkable reformulation of the spectral problem equations of~\cite{Pmu}.

One extra reason why it would be fascinating to re-sum the OPE at finite coupling  is that some interesting kinematical regions, such as the so-called {\it multi-Regge} kinematics~\cite{multi-Regge} for instance, stand outside of the radius of convergence of the near collinear expansion. To access these regions one must re-sum completely (or at least partially) the OPE series. As far as the multi-Regge kinematics is concerned, the procedure is currently well-understood at strong coupling only, where the leading asymptotics was found to be captured by the excited-state version of the full TBA equations \cite{Bartels:2013dja,Bartels:2013dja2}. It would be wonderful if a similar story were present at finite coupling as well.

\section*{Acknowledgements}

We thank Lance Dixon, James Drummond, Matt von Hippel and Jeff Pennington for several discussions and exchanges pertaining the three loop analysis and Lance Dixon, James Drummond, Claude Duhr and Jeff Pennington for discussions and correspondence related to the four loop checks and for insightful comments on the draft. We thank Andrei Belitsky, Yasuyuki Hatsuda, Juan Maldacena, Joao Penedones and Kostya Zarembo for enlightening discussions. B.B, A.S and P.V thank IAS and ICTP-SAIFR for hospitality. B.B thank PI for hospitality. Research at the Perimeter Institute is supported in part by the Government of Canada through NSERC and by the Province of Ontario through MRI. A.S was supported in part by U.S. Department of Energy grant DE- SC0009988.

\appendix
\section{S-matrices and anomalous bootstrap for fermions}\la{paradoxSec}

In this appendix we present the expressions for the scattering matrix of two fermions in both the physical and mirror channels. They are the main building blocks entering the composition of our conjectures for the fermion transitions.

\begin{figure}
\centering
\def\svgwidth{10cm}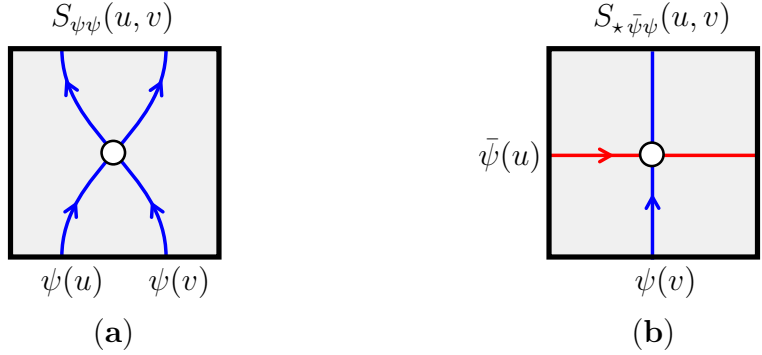
\caption{Fermion S-matrices. In ({\bf a}) we measure the physical S-matrix between two fermions by creating and annihilating a fermion pair at the bottom and top of a square Wilson loop. We would get instead the mirror S-matrix if one of the fermion was created/annihilated on the left/right edge of the Wilson loop. This is what is depicted in ({\bf b}) for a mirror anti-fermion scattering against a real fermion. Rotating this picture by $90^{\circ}$ would change the meaning of who is mirrored and who is real. The whole S-matrices also include $SU(4)$ indices for the ingoing and outgoing excitations that are not represented here.}
\la{MirorSmatrix}
\end{figure}

We shall first report the physical S-matrix which was extracted from the analysis of the spectral equations for the fermionic excitations of the Gubser-Klebanov-Polyakov string~\cite{GKP} using the underlying integrable spin-chain description~\cite{BS}. The end result follows directly from the generalization of the study of~\cite{withAdam} to the fermionic sector along the lines of~\cite{BenDispPaper} and has some overlap with the alternative treatment proposed in~\cite{Fioravanti:2013eia}.

The difficulties inherent to the fermion case shall arise when we will be considering the scattering of two fermions that live in two different channels of the square. The process that we will have in mind is pictured in figure~\ref{MirorSmatrix}.b. There we see an anti-fermion propagating horizontally and crossing at some point the trajectory of a fermion evolving vertically. It is equivalent to stating that the former has imaginary energy and momentum in the frame where the latter has both being real. In these circumstances we say that the anti-fermion is mirror rotated with respect to the fermion. Who is real and who is mirrored depends on how we look at the square in figure~\ref{MirorSmatrix}.b and the ``amplitude'' we associate to this process should not depend on it. To make the case of the mirror S-matrix even more clear we depicted in figure~\ref{MirorSmatrix}.a the process corresponding to a real scattering among two fermions, for comparison.

We will argue that there is no continuous transformation that rotates a fermion from one channel to the other, in contrast to the scalar and gluon cases for which such maps were easily constructed in~\cite{MoreDispPaper}. Instead we will find that fermions transform anomalously under a mirror rotation. Most of this appendix is dedicated to explaining how to derive the fermion mirror S-matrix using this anomalous map. We will end with a perturbative check of our result and a summary of the obtained S-matrices.

\subsection{The physical S-matrix}\la{Phys-S}

We begin with the exposition of the S-matrix between two fermions prepared in the same channel. We can think of the S-matrix as an operator that permutes the two excitations. This viewpoint has become conventional in 2d integrable models and relies on the fact that one can associate a natural and unambiguous ordering between particles. Namely the two-fermion state
\beq
\left|\psi_{A}(u)\psi_{B}(v)\right>
\eeq
can be seen either as an in-state of two fermions, carrying rapidities $u,v$ and $SU(4)$ charges $A,B = 1, \ldots , 4$, or as an equivalent out-state, depending on whether the rapidities are ordered or anti-ordered. In these terms the S-matrix, which by definition relates unitarily the two basis, can be written as
\beq
\left|\psi_{A}(u){\psi}_{B}(v)\right> = -S_{\psi\psi}(u, v)_{AB}^{CD}\left|{\psi}_{D}(v)\psi_{C}(u)\right>\, .
\eeq
The overall minus sign is conventional for fermions and would already be present in a free theory, which would correspond to $S = 1$ in our notations. Unitarity implies then that
\beq
S_{\psi\psi}(u, v)_{AB}^{CD}S_{\psi\psi}(v, u)_{DC}^{FE} = \delta_{A}^{E}\delta_{B}^{F}\, .
\eeq

In order to fully characterize the fermion scattering we should also consider processes where the two incoming fermions carry different $U(1)$ charges. In this case we scatter distinguishable particles and need two matrices to account for both transmission and reflection processes,
\beq\label{gen-form-S}
\left|\psi_{A}(u)\bar{\psi}^{B}(v)\right> = -S_{\bar{\psi}\psi}(u,v)_{AD}^{CB} \left|\bar{\psi}^{D}(v)\psi_{C}(u)\right> + \tilde{S}_{\bar{\psi}\psi}(u,v)_{AC}^{DB} \left|\psi_{D}(v)\bar{\psi}^{C}(u)\right>\, .
\eeq
Please note that we use the same letter for the $SU(4)$ index of fermion and anti-fermion, which in both cases can take the four possible values $1, 2, 3, 4$. Our convention is that the index is down for the fermion which belongs to the fundamental representation, i.e., $\textbf{4}$, and up for its anti-partner which is in the complex conjugate (and inequivalent) $\bar{\textbf{4}}$ irrep. 
The unitarity relations corresponding to the scattering~(\ref{gen-form-S}) could easily be obtained from the analysis of~\cite{Berg:1977dp}. Below we shall present the ones that apply to our specific theory, see~(\ref{unitarity-fin}).

From the analysis of the Bethe ansatz equations for fermionic excitations of the GKP string, one can deduce that the fermion S-matrix is consistent with a reflectionless $U(4)$ invariant scattering. This means that $\tilde{S} = 0$ on the one hand and that the two remaining matrices are proportional to the fundamental $SU(4)$ $R$-matrices of~\cite{Berg:1977dp} on the other hand. This way we can write
\beq\label{S-matrix-ferm}
S_{\psi\psi}(u, v)_{AB}^{CD} = S_{\psi\psi}(u, v) R_{\textbf{44}}(u-v)_{AB}^{CD}\, , \qquad S_{\psi\bar{\psi}}(u, v)_{AD}^{CB} = S_{\psi\bar{\psi}}(u, v)R_{\textbf{4}\bar{\textbf{4}}}(u-v)_{AD}^{CB}\, ,
\eeq
where
\beq\label{SRmatrices}
R_{\textbf{44}}(w)_{AB}^{CD} = \frac{w}{w-i}\delta_{A}^{C}\delta_{B}^{D}-\frac{i}{w-i}\delta_{A}^{D}\delta_{B}^{C}\, ,\qquad\text{and}\qquad
R_{\textbf{4}\bar{\textbf{4}}}(w)_{AD}^{CB} = \delta_{A}^{C}\delta_{D}^{B}+\frac{i}{w-2i}\delta^{B}_{A}\delta^{C}_{D}\, .
\eeq
In these circumstances the unitarity relations take the simplest possible form
\beq\label{unitarity-fin}
S_{\psi\psi}(u, v)S_{\psi\psi}(v, u) = S_{\psi\bar{\psi}}(u, v)S_{\bar{\psi}\psi}(v, u) = 1\, ,
\eeq
with $S_{\psi\bar{\psi}}(u, v) = S_{\bar{\psi}\psi}(u, v)$ by charge conjugation. Notice that the formulae~(\ref{S-matrix-ferm}, \ref{SRmatrices}) encode all information about the scattering of fermions in all possible $SU(4)$ channels. For instance,  one easily derives from them the expression for the scattering phase in the singlet channel,
\beq
{\bar S}_{\textrm{singlet}}(u,v)  = \frac{1}{4}\delta^{A}_{B}\delta_{C}^{D}S_{\psi\bar{\psi}}(u, v)_{AD}^{CB} = \frac{u-v+2i}{u-v-2i} \, S_{\psi\bar{\psi}}(u, v)\, ,
\eeq
or the one for two $\bf{4}$'s in the symmetric representation,
\beq
S_{\textrm{sym}}(u, v) = S_{\psi\psi}(u, v)_{11}^{11} = S_{\psi \psi}(u,v)\, .
\eeq

Having in hand the matrix form~(\ref{S-matrix-ferm}, \ref{SRmatrices}) it remains to determine the two scalar factors $S_{\psi\psi}(u, v)$ and $S_{\psi\bar{\psi}}(u, v)$. This is where the intricate dynamics of the flux tube theory is sitting, including in particular all the dependence on the coupling constant. Here again the necessary information can be extracted from the spin chain analysis. What we found from this study is that the two scalar factors are not independent. In fact, they fulfill the curious equality
\beq
S_{\psi \psi}(u,v)  = S_{\psi \bar \psi}(u,v) \la{SequalBarS}\, .
\eeq
This is a great simplification that leaves us with $S_{\psi \psi}(u,v)$ only. As far as the latter quantity is concerned we can not do better than providing a representation in the form of~\cite{withAdam}. In this representation the main information is encoded into a set of functions $f_{1, 2, 3, 4}(u, v)$ that depend on the two rapidities $u, v$ of the scattered particles and the coupling constant. The recipe for computing these functions is universal and presented in the appendix~\ref{summary-S-matrices}.

\subsection{Tension with mirror symmetry}

Given the knowledge of the (finite-coupling) physical S-matrix, the challenge is to derive the expression for the mirror S-matrix. It shall become clear shortly that this step is nontrivial for fermions. In preparation for it we shall first remind the reader of the difficulties~\cite{MoreDispPaper,long} that one faces in attempting to define a mirror or crossing rotation for fermions.

We first recall that a mirror transformation, a.k.a. double-Wick rotation~\cite{Zamolodchikov:1989cf}, is a map~$\gamma$, from the rapidity plane to itself, which for any rapidity $u$ gives us back another rapidity denoted for short as $u^{\gamma}$. When applied to a given excitation $X$, this map should have the property of swapping energy and momentum up to a rescaling by the imaginary unit. This translates mathematically into
\beq\label{mirror-map}
E_{X}(u^{\gamma}) = ip_{X}(u)\, , \qquad p_{X}(u^{\gamma}) = iE_{X}(u)\, ,
\eeq
for any rapidity $u$. The details of the map might of course depends on the excitation $X$ to which it applies. More importantly, when writing~(\ref{mirror-map}), we assumed implicitly that the physics is invariant under mirror rotation, which is reflected in the fact that the energy and momentum are just swapped under the rotation. This should be the case for the flux tube theory under study~\cite{OPEpaper} where it is understood as a symmetry of the square Wilson loop under rotation by $45^{\circ}$.

An elementary illustration of a mirror rotation of the type~(\ref{mirror-map}) is found for a relativistic particle $E(\theta) \pm p(\theta) = m e^{\pm \theta}$ where it takes the simple form $\gamma:\theta \rightarrow \theta^{\gamma} = \theta+\frac{i\pi}{2}$. An example of a theory with a mirror rotation but no mirror symmetry is given by the theory of magnons over the BMN vacuum~\cite{Ambjorn:2005wa}. 

We note finally that it is straightforward to define a crossing transformation that exchanges particle and anti-particle by applying the mirror rotation twice.

\begin{figure}[t]
\centering
\includegraphics[scale=0.6]{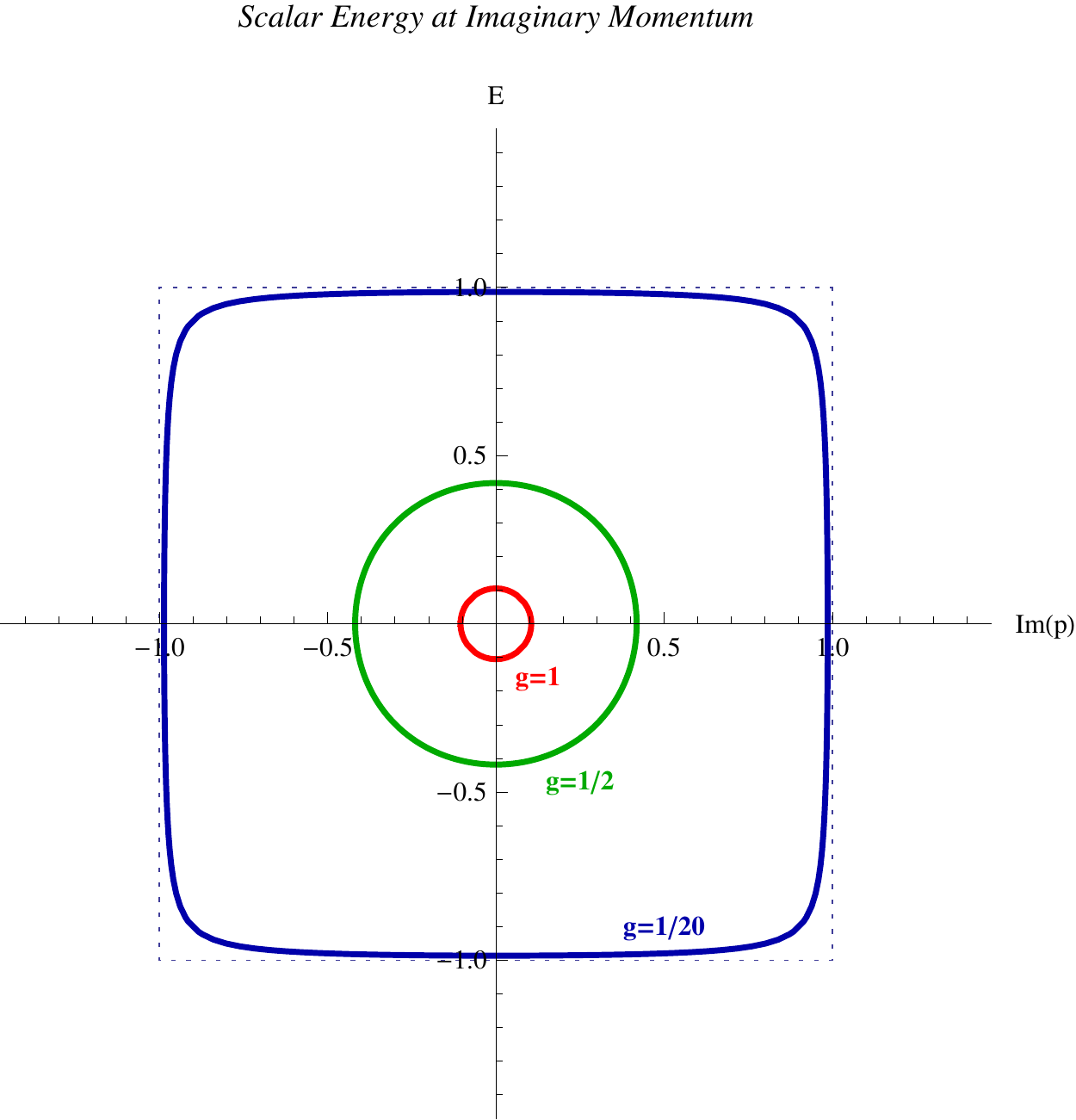}
\includegraphics[scale=0.6]{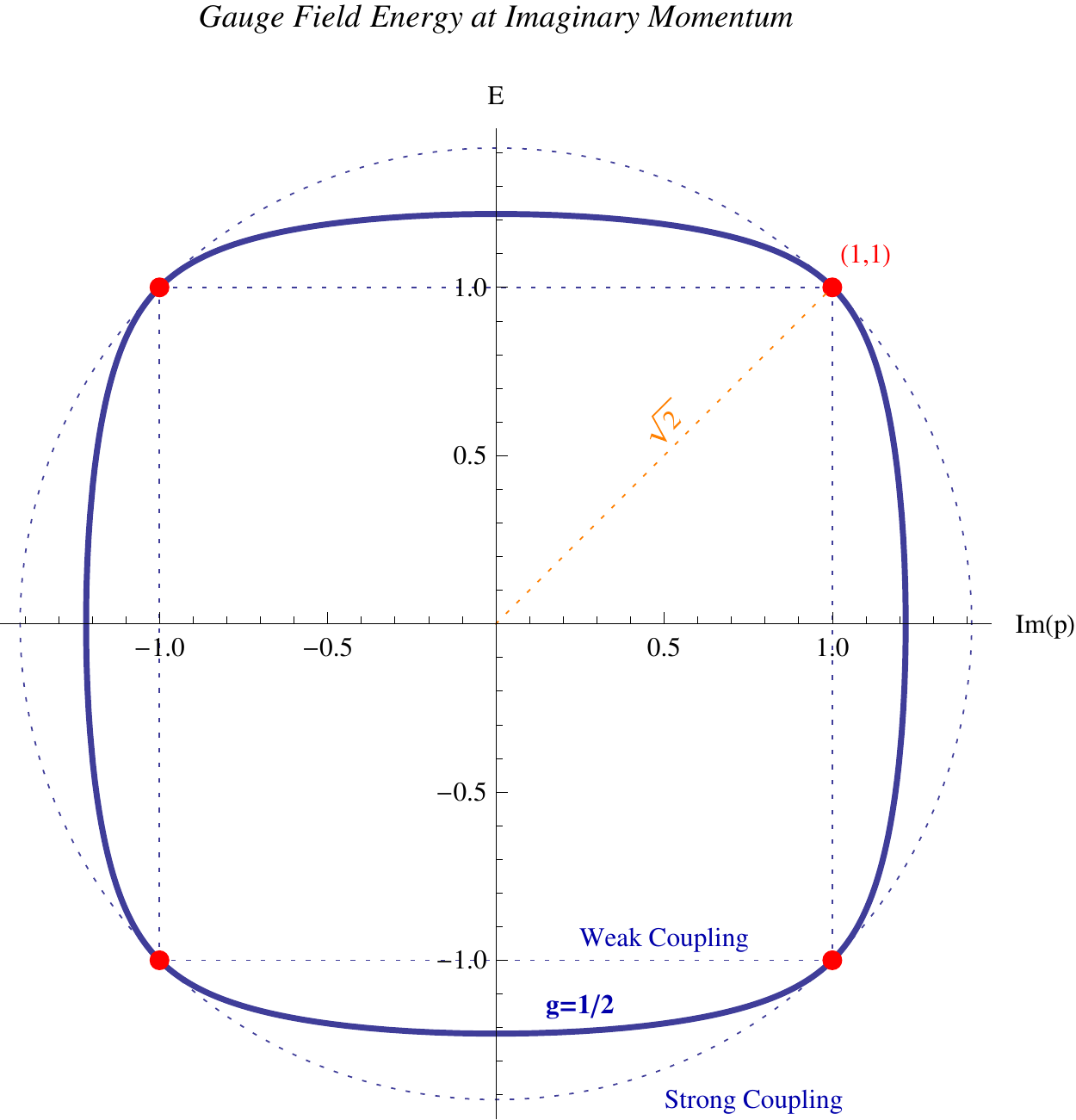}
\caption{Scalar and gauge field energies for purely imaginary momentum at finite coupling. The scalar and gluon curves are both given by a square centred around $(E, p) = (0, 0)$ at weak coupling. At strong coupling the excitations become relativistic and their curves circular with an exponentially small radius for the scalar and a radius equal to $\sqrt{2}$ for a gluon. We verify that the gluon curve passes through the Goldstone point $(E, p) = (1, i)$ for any value of the coupling, in agreement with~\cite{AldayMaldacena, MoreDispPaper}. The scalar and gauge field curves are mirror symmetric, meaning that they are invariant under reflection w.r.t. the diagonal $E = -ip$ (or equivalently $E=ip$ by parity). The particles are at rest in the real or crossed kinematics when $p = 0$ and in the mirror or crossed mirror kinematics when $E=0$, respectively.}\label{scalar curves}
\end{figure}

The goal of this appendix is to provide evidence for the claim that there is no map $\gamma$ fulfilling~(\ref{mirror-map}) for a fermion, i.e., for $X = \psi$, despite the fact that such maps were found for all the other fundamental excitations~\cite{MoreDispPaper}, i.e., for $X = \phi, F,$ and bound states. We shall do it by exhibiting a manifestation of an existing tension between a (naive form of) mirror rotation for fermions and the expected mirror symmetry of the flux-tube background.

It is often the case that the mirror rotation appears as a specialization of a continuous transformation. This is so for scalars and gluons where the real and mirror kinematics are analytically related to one another~\cite{MoreDispPaper}. A way of finding such an analytic continuation is to start with an excitation at rest and look for a direction along which the momentum becomes purely imaginary and the energy gets decreased. Following this path up to the point where the energy vanishes and the momentum extremizes gives a mean to identifying the mirror rotation, assuming such a point exists. For instance, in the relativistic example mentioned before, one would start at rest with $\theta = 0$ and move along the imaginary axis in the upper-half plane until one reaches $\theta = \frac{i\pi}{2}$ where the energy is zero and the momentum is maximal. In this way we would rederive the familiar $\theta\rightarrow \theta +\frac{i\pi}{2}$ mirror transformation for a relativistic particle, which of course is just a particular instance of an euclidean rotation.

\begin{figure}[t]
\centering
\includegraphics[scale=0.35]{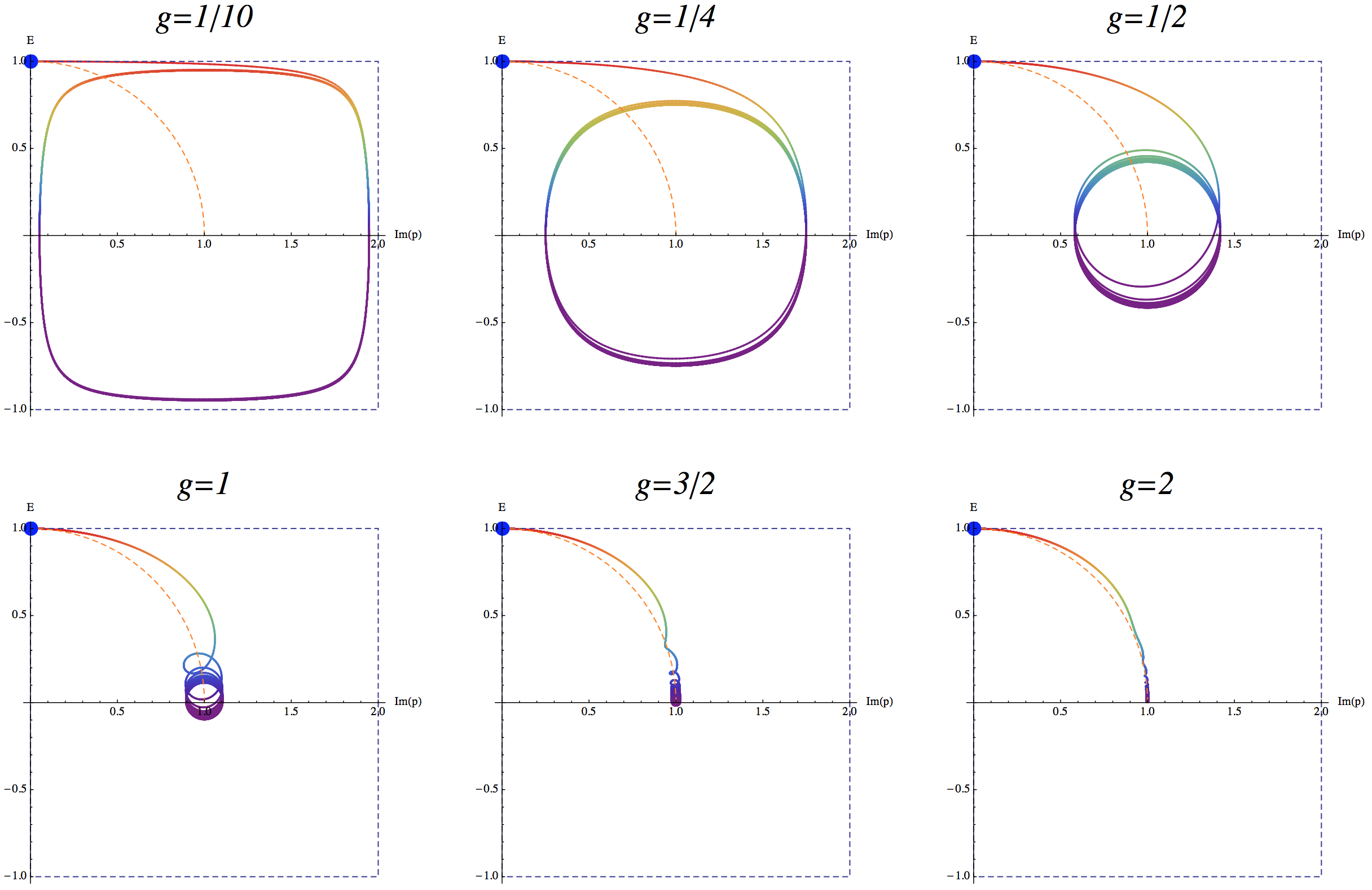}
\caption{Plots of the fermion energy for purely imaginary momentum at various values of the coupling $g$. The curves are parameterized by the Bethe rapidity $u$ that takes purely imaginary values ranging between $u=\mp i\infty$. The point $u=-i\infty$ is regular and corresponds to a fermion at rest, i.e., $(E, p) = (1, 0)$.  As we move away from this point, by increasing the imaginary value of $u$, the energy first decreases and then starts oscillating in a rather complicated way. As we move toward the (singular) point $u=i\infty$ we get closer and closer to a limiting cycle that is given by the scalar energy centred around $(E, p) = (0, i)$, i.e., $E(p) \sim E_{\textrm{scalar}}(p-i)$. At zero coupling the fermion energy exactly coincides with its limiting cycle which is just a square. At strong coupling the limiting cycle and the associated oscillations are exponentially suppressed and the energy is well described by the relativistic law $E = \sqrt{1+p^2}$ represented by the dashed line.}\label{fermioncurves}
\end{figure}

The above strategy for identifying a mirror map works very well for the scalar and gluon excitations of the colour flux tube. (This is actually the way the mirror rotations $\gamma:u\to u^{\gamma}$ for these two excitations were originally found in~\cite{MoreDispPaper}.) The results of this procedure are depicted in figure~\ref{scalar curves} where we plotted side-by-side the exact scalar and gluon dispersion relation for purely imaginary values of the rapidity, equivalently momentum. This figure makes manifest the mirror symmetry of the flux tube; it is reflected by the symmetry of the dispersion relation $E(p)$ with respect to the diagonals $E = \pm ip$.  

When applied to the fermion dispersion relation, the aforementioned analysis produces a very different outcome, as illustrated in figure~\ref{fermioncurves} for several values of the coupling constant. In this case as well we can start at rest, which corresponds to the point $u=-i\infty$ on the small sheet for a fermion, and move along the imaginary rapidity axis. When we reach the point $u=0$ we are sitting on the cut that connects the small and large sheet. Nothing dramatic happens though at this point or in its surroundings. The energy is smaller than the fermion mass but not yet vanishing nor evidencing that something already went wrong. We should therefore continue the process further by stepping inside the large sheet. We are now lying on this sheet and heading upward along the imaginary $u$-axis. This path entails crossing the infinite sequence of cuts that populates the large fermion sheet (see~\cite{BenDispPaper} for details on the analytic structure of the fermion dispersion relation in the rapidity plane). There is no problem passing smoothly through all of these cuts at finite coupling. The presence of this infinite tower of cuts has nonetheless the effect of adding an oscillatory component to the energy and momentum of the fermion. This on its own would not be problematic if this behaviour was purely periodic. This is for instance what happens for the scalar dispersion relation that oscillates periodically between real, mirror, crossed, crossed-mirror, ... kinematics as we increase the imaginary part of the rapidity. (This is also the behaviour of a relativistic dispersion relation $E = m \cos{\phi}$ for $\phi = -i\theta$.) This is not the case for the fermion dispersion relation. Instead we find the rather complicated behaviour depicted in figure~\ref{fermioncurves}. From this picture we see that the oscillatory behaviour sets in at some point (which depends on the strength of the coupling) and that it keeps going for ever afterward. Asymptotically, at the very end of our path, i.e., when we get closer and closer to $u=i\infty$, the oscillatory component of the fermion dispersion relation becomes more and more periodic. It is possible to show indeed that, in this limit, the fermion dispersion relation is controlled by the scalar one, which as we already mentioned is purely periodic,
\beq\label{Osc-beh}
E_{\psi}(u) \sim E_{\phi}(u-\ft{i}{2})\, , \qquad p_{\psi}(u)-i \sim p_{\phi}(u-\ft{i}{2})\, .
\eeq   
This means that the fermion dispersion relation eventually circles (infinitely many times) around the point $(E, p) = (0, i)$ with a radius $\sim m_{\phi}$, where $m_{\phi}$ is the scalar mass. At strong coupling, this mass is exponentially small and the fermion dispersion relation looks mirror symmetric to a very good approximation (precisely to all orders in the strong coupling expansion~\cite{MoreDispPaper}).  At finite coupling, there is however no recognizable mirror symmetric pattern in the curves displayed in figure~\ref{fermioncurves}. Neither do we actually reach the point $(E, p) = (0, i)$ that corresponds to the mirror image of a fermion at rest. In a nutshell, what seems to obstruct the realization of a continuous rotation of a fermion from the real to the mirror kinematics is the oscillatory component of the dispersion relation which looks asymptotically like~(\ref{Osc-beh}). We will see in the next subsection that this is not the only obstacle for such a continuous interpolation between the two kinematics. The extra argument given below will however show us the way out from this paradoxical situation.

\subsection{The anomalous mirror rotation of fermions}\la{fermion-bootstrap}

As it is often the case with puzzling situation, the resolution of our problem turns out to be elementary. Essentially, the reason we could not find a decent interpolation between the real and mirror kinematics for a fermion is that there cannot be any. The crux of the proof is the geometrical argument~\cite{long} according to which a mirror rotation should flip the sign of the $U(1)$ charge of the excitation. For instance, when we rotate a gluon $F$ from one edge of the Wilson loop to the neighbouring one we end up with an $\bar{F}$ and not an $F$. If we now try to apply this general lesson to the fermion we immediately run into a contradiction. Namely, any continuation of a fermion form one edge to another should preserve the $R$-symmetry and simultaneously flips the $U(1)$ charge. The problem is that for a twist-one fermion these two quantum numbers are correlated to one another: if we fix the $SU(4)$ representation then we have no freedom to choose the $U(1)$ charge anymore, and vice-versa. For symmetry reason, there cannot be any smooth interpolation between twist-one fermions living in two different kinematics. This is the explanation for the obstruction met earlier.

It might happen however that an interpolation exists between a twist-one fermion and an higher-twist component of the Weyl field. This would be a mirror map between a fundamental fermion and a composite (or effective) one, in our terminology. This is what we shall advocate in this appendix. Staring at the table of excitations~\ref{Fundamental}, we realize that there are not that many options for what this effective excitation can be. If we insist on the $R$-charge conservation and $U(1)$ charge flipping, the outcome can only be the twist-two component of the Fermi field, or possibly any of its $\mathfrak{sl}(2)$ descendants. The $\mathfrak{sl}(2)$ primary being the most likely candidate, we are led to consider an anomalous mirror transformation $\omega$ acting like
\beq\label{omega}
\omega : \left|\psi_{A}(u)\right>_{\textrm{bottom edge}} \to \left|\psi_{A}(u^{\omega})\right>_{\textrm{bottom edge}} = \left|\Psi_{A}(u)\right>_{\textrm{left edge}}\, ,
\eeq
where $\Psi$ denote the twist-two component with opposite $U(1)$ charge but same $R$-charge as~$\psi$. The transformation is depicted in figure~\ref{mirrorfermion}. We denote it by $\omega$  and call it anomalous mirror map to distinguish it from the naive (and, as we understand now, forbidden) mirror rotation $\gamma$ that would have mapped a fermion back to itself, see~(\ref{mirror-map}).  

\begin{figure}
\centering
\def\svgwidth{14cm}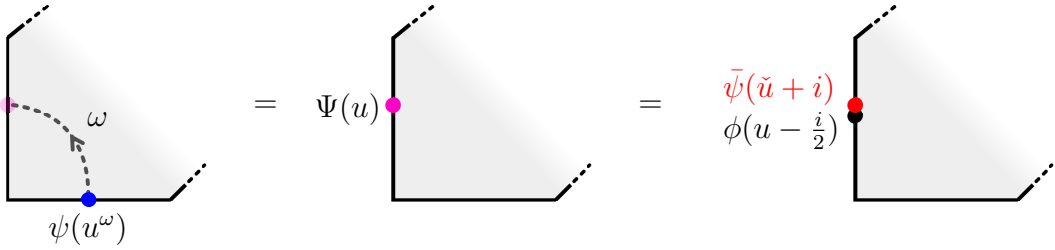
\vspace{.4cm}
\caption{The anomalous rotation of a fermion. Under the anomalous mirror rotation $\omega$ the fundamental fermion $\psi$ maps into the effective excitation $\Psi$. The latter object is a composite excitation made out of a scalar $\phi(u-\ft{i}{2})$ and a small anti-fermion $\bar{\psi}(\check{u}+i)$, projected into the appropriate $SU(4)$ representation. The $\omega$ map is a consistent continuous transformation that preserves the $R$-charge and flips the $U(1)$ charge.}
\la{mirrorfermion}
\end{figure}

As a side remark, let us point out that the composite fermion $\Psi$ did not appear in our analysis of the two-particle contributions to the hexagonal Wilson loop because, despite having twist two, it carries $R$-charge. It must therefore be accompanied by at least one more fermionic excitation to become visible in the OPE of any bosonic Wilson loop. This has the effect of delaying its appearance to the twist-three and higher contributions in our case. 

The reason why the map~(\ref{omega}) is useful for constructing processes involving fermions in mirror channels is that the effective excitation $\Psi$ is not that far from being a twist-one fermion. The former excitation is indeed akin to the twist-two excitations $\cF_{12}$ and $\cF_{\dot{1}\dot{2}}$ that we encountered in Section~\ref{WeakSec} and that we know are made out of fermions. We understood why this is so in Section~\ref{discussion-sect}: e.g., the twist-two excitation $\cF_{12}$ falls in the $\mathfrak{sl}(2|4)$ multiplet of the twist-one fermion $\psi$ and thus can be obtained by binding small fermions to it (in the present case attaching a single small anti-fermion to $\psi$ is enough). The same logic applies for the composite excitation $\Psi$. Looking more carefully for the fermion $\Psi = \psi_{2}$ in table~\ref{Fundamental}, we see that it belongs to the $\mathfrak{sl}(2|4)$ multiplet of the twist-one scalar $\phi$. This means that we can view the fermion $\Psi$ as a descendant of $\phi$ and therefore that we can obtain the former by binding appropriately small fermions to the latter. In fact, as we will confirm below, the fermion $\Psi$ is a string (i.e., bound-state) of a scalar $\phi$ and a small fermion $\bar{\psi}$ lying at a distance $\ft{3i}{2}$ from one another in rapidity space. (The fact that the distance here is $\ft{3i}{2}$ while it was $2i$ for the $\psi\bar{\psi}$ pairs corresponding to $\cF_{12}$ or $\cF_{\dot{1}\dot{2}}$ has to do with the $SU(4)$ symmetry.)  In $SU(4)$ covariant notations we can write
\beq\label{meaning-Psi}
\Psi_{A}(u) = \frac{1}{\sqrt{6}}\rho^{i}_{AB}\bar{\psi}^{B}(\check{u}+i)\phi_{i}(u-\ft{i}{2})\, ,
\eeq
where $\rho^{i}_{AB}$ ($i=1, \ldots, 6, A, B = 1, \ldots, 4$)  stand for the matrices that project the tensor product $\bar{\textbf{4}}\otimes \textbf{6}$ to its irreducible $\textbf{4}$ component, see the following subsection for more details. This is the main result of this subsection. Combined with~(\ref{omega}) it implies that
\beq
\left|\psi_{A}(u^{\omega})\right>_{\textrm{bottom edge}} = \frac{1}{\sqrt{6}}\rho^{i}_{AB}\left|\bar{\psi}^{B}(\check{u}+i)\phi_{i}(u-\ft{i}{2})\right>_{\textrm{left edge}}\, .
\eeq
It means at the end of the day that we can transport a fermion from one edge to another if we take care of removing the scalar component in the process. This shall give rise to a bootstrap that relates fermions and scalars to one another. Though it is of a more complicated nature than the one for the other twist-one excitations, it is easily implemented and solved for the mirror fermion S-matrix. This is what we shall prove in the next subsection.

Having understood the physics we now have to deal with the mathematics. The first thing that has to be done is to identify the map $\omega$ and verify that it produces the effect we want. After all, we did not give yet evidence that such a map actually exists. This is what we shall establish now. The starting point are the identities
\beq\label{toward-mirro-ferm}
E_{\psi}(u) = E_{\phi}(u-\ft{i}{2})+ip_{\psi}(\check{u})\, , \qquad p_{\psi}(u) = p_{\phi}(u-\ft{i}{2})+iE_{\psi}(\check{u})\, ,
\eeq
that relate energy and momentum of a fermion to the ones of a scalar. These identities were derived in~\cite{MoreDispPaper} and can be checked in perturbation theory at weak coupling using formulae in Appendix~\ref{pert-exp-app} for instance. What we are going to do next does not commute with perturbation theory though. The main observation is that the equations~(\ref{toward-mirro-ferm}) would be of the expected mirror type if $E_{\phi}, p_{\phi}$ on their right-hand sides were replaced by $ip_{\phi}, iE_{\phi}$. This is fortunately simple to fix since this replacement amounts to performing a mirror transformation on the scalar. We know that this transformation is obtained by shifting the rapidity by the imaginary unit $u\to u+i$. The shift should however be understood as an analytical continuation through the strip $-2g < \Re e\, u < 2g$ in the upper-half rapidity plane. Denoting such a shift by $\omega$ we can write
\beq\label{final-rule}
E_{\psi}(u^{\omega}) = ip_{\phi}(u-\ft{i}{2})+ip_{\psi}(\check{u}+i)\, , \qquad p_{\psi}(u^{\omega}) = iE_{\phi}(u-\ft{i}{2})+iE_{\psi}(\check{u}+i)\, .
\eeq
It has exactly the expected form given the string interpretation of the fermion $\Psi$ presented before,
\beq
E_{\Psi}(u) = E_{\phi}(u-\ft{i}{2}) + E_{\psi}(\check{u}+i)\, , \qquad p_{\Psi}(u) = p_{\phi}(u-\ft{i}{2}) + p_{\psi}(\check{u}+i)\, .
\eeq
We conclude therefore that the sought $\omega$ map that transforms a fermion $\psi$ into the fermion $\Psi$ on the mirror edge is nothing else that the analytical continuation $u^{\omega} = u+i$ through the cut connecting $-2g+i$ and $2g+i$ in the upper-half plane of the large fermion sheet. The precise path is depicted in figure~\ref{uomega}.

\begin{figure}[t]
\centering
\def\svgwidth{8cm}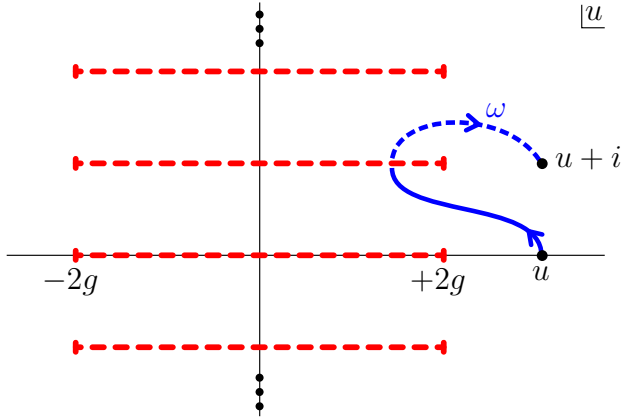
\caption{The $\omega$ map for a fermion. Starting from a rapidity $u$ on the large sheet we analytically continue to $u^{\omega} = u+i$ by going through the first cut in the upper-half plane. We should take care of avoiding going through the middle cut (on the real line) which would just bring us to the small sheet which is free of any additional cut.}\label{uomega}
\end{figure}

Equipped with the above mathematical rule it is now straightforward to perform the $\omega$ transformation on the fermion S-matrix. Prior to doing it we would like to add one more comment. There exists another version of our map that is obtained by continuing the rapidity to the lower-half plane instead. It is nothing else that the mirror image of the one pictured in figure~\ref{uomega} with respect to the real axis and we shall denote it by $-\omega$. Its effect on the dispersion relation is given by
\beq
E_{\psi}(u^{-\omega}) = -ip_{\phi}(u+\ft{i}{2})-ip_{\psi}(\check{u}-i)\, , \qquad p_{\psi}(u^{-\omega}) = -iE_{\phi}(u+\ft{i}{2})-iE_{\psi}(\check{u}-i)\, ,
\eeq
which, not surprisingly, is the complex conjugate of~(\ref{final-rule}). The interpretation of this map is that it transports the fermion $\psi$ to the rightmost edge, which makes it akin to an inverse mirror rotation instead. What we obtain at the end has again the interpretation of being a string made out of a scalar and a small fermion standing at a distance $\ft{3i}{2}$ from each other. However this string is not exactly the same as the one we introduced before. Instead the former is an upside down version of the latter,
\beq
\left|\psi_{A}(u^{-\omega})\right>_{\textrm{bottom edge}} = \frac{1}{\sqrt{6}}\rho^{i}_{BA}\left|\bar{\psi}^{B}(\check{u}-i)\phi_{i}(u+\ft{i}{2})\right>_{\textrm{right edge}}\, .
\eeq
The reason is that the fermion $\Psi$ does not have a real dispersion relation. (This feature is shared among all the effective excitations depicted in table~\ref{Fundamental}.) It exists therefore in two distinct forms that are complex conjugate of one another. Which form is the relevant one depends on the context. In the following we shall only consider the effect of the direct $\omega$ map, which means that it is the representation~(\ref{meaning-Psi}) that will matter. 

Finally, let us add that everything we said so far has an obvious counterpart for an anti-fermion by charge conjugation.

\subsection{The mirror S-matrix}

\begin{figure}[t]
\centering
\def\svgwidth{14cm}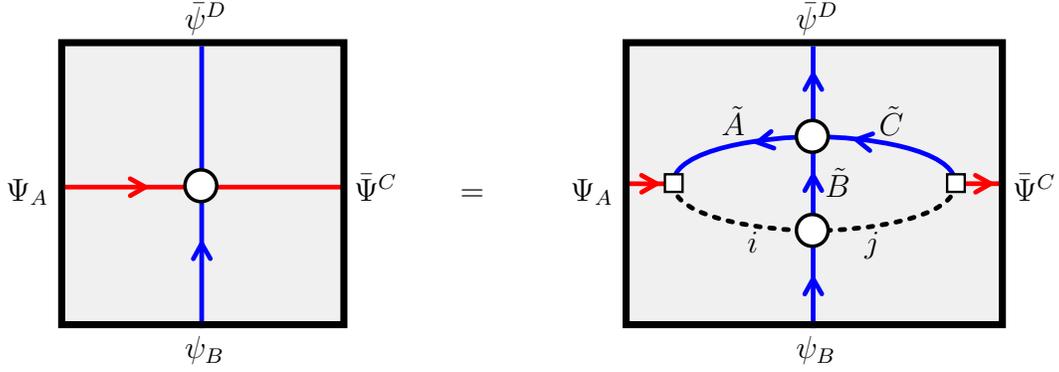
\caption{Fusion of mirror S-matrices. We pictured on the left-hand side the mirror S-matrix between the composite fermion $\Psi$ and the fundamental one $\psi$. On the right-hand side, the same process is realized by decomposing $\Psi$ into a string (or dipole) of a scalar $\phi$ and an anti-fermion $\bar{\psi}$. Our conventions here are that $\Psi, \psi$ or $\bar{\psi}$, and $\phi$, are associated to red, blue, and dashed, lines, respectively. The square dots represent $\rho$ matrices while the round ones are (mirror) S-matrices.}\label{fusion-fermion}
\end{figure}

The strategy for extracting the mirror S-matrix for fermions can now be spelled out. The idea is to consider the effect of the $\omega$ map on the physical S-matrix and interpret it along the lines of the previous subsection as giving us the mirror S-matrix between the twist-two and twist-one fermions. We thus write down
\beq\label{def-Psipsi}
S_{\psi\psi}(u^{\omega}, v)_{AB}^{CD} = S_{\star\Psi\psi}(u, v)_{AB}^{CD}\, ,
\eeq
where the lower script $\star$ is used throughout the paper to indicate a mirror process, e.g., here $\star \Psi$ means that the excitation $\Psi$ lives on the edge located to the left of $\psi$. We then use our understanding that the fermion $\Psi$ has the meaning of the string~(\ref{meaning-Psi}).  This implies that the mirror S-matrix between $\Psi$ and $\psi$ can be represented by the fusion in figure~\ref{fusion-fermion}. Mathematically it reads
\beq\label{main-fusion}
 S_{\star\Psi\psi}(u, v)_{AB}^{CD} =  \frac{1}{6}\rho^{i}_{A\tilde{A}}S_{\star\bar{\psi}\psi}(\check{u}+i, v)_{\tilde{C}\tilde{B}}^{\tilde{A}D}S_{\star\phi\psi}(u-\ft{i}{2}, v)_{iB}^{j\tilde{B}}\rho^{\tilde{C}C}_{j}\, ,
\eeq
where $S_{\star\bar{\psi}\psi}$ and $S_{\star \phi \psi}$ are the mirror S-matrices for anti-fermion-fermion and scalar-fermion scatterings, respectively. As before, the $\rho$ matrices, that we shall review shortly, project the product of the latter two matrices to the appropriate $SU(4)$ representation; they literally fuse the pair $\bar{\psi}-\phi$ into $\Psi$ at the $SU(4)$ level. The relevance of the equation~(\ref{main-fusion}) lies in the fact that it is telling us \textit{how precisely} we can access to the sought-after mirror S-matrix $S_{\star\bar{\psi}\psi}$ from the knowledge of the matrices $S_{\star \phi \psi}$ and $S_{\star\Psi\psi}$. This is what is demonstrated below.

The matrix $S_{\star\Psi\psi}$ is given by the analytic continuation~(\ref{def-Psipsi}). It remains thus to construct $S_{\star \phi \psi}$ prior to make use of~(\ref{main-fusion}). This step is fortunately straightforward. At first, the S-matrix between scalar and fermion is defined through
\beq
\left|\phi_{i}(u)\psi_{A}(v)\right> = S_{\phi\psi}(u, v)_{iA}^{jB}\left|\psi_{B}(v)\phi_{j}(u)\right>\, ,
\eeq
and has a form that can be read out directly from the Bethe ansatz equations. Not surprisingly, it is proportional to the $R$-matrix between the $\textbf{6}$ and $\textbf{4}$ of $SU(4)$,
\beq\label{SandR}
 S_{\phi\psi}(u, v)_{iA}^{jB} =  S_{\phi\psi}(u, v) R_{\textbf{64}}(u-v)_{iA}^{jB}\, , \qquad R_{\textbf{64}}(w)_{iA}^{jB} = \delta_{i}^{j}\delta_{A}^{B} + \frac{i}{2w-3i}\rho_{iAC}\rho^{jCB}\, .
\eeq
The $\rho$-matrices entering the $R$-matrix~(\ref{SandR}) are the same as the ones met earlier and are defined as the off-diagonal components of the 6D Dirac $\gamma$-matrices in the Weyl representation. They satisfy the Clifford algebra
\beq
\rho_{iAC}\rho^{jCB} + \rho_{jAC}\rho^{iCB} = 2\delta_{i}^{j}\delta_{A}^{B}\, , \qquad \textrm{where} \qquad \rho^{iAB} = -\rho^{iBA} = -\rho^*_{iAB}\, .
\eeq
We can further use the freedom of raising or lowering the vector indices $i, j$ by means of the Kr\"onecker delta, whenever it is needed or natural to do so. An explicit representation for these matrices can be found in the appendices of ref.~\cite{Giombi:2009gd}. The scalar factor $S_{\phi\psi}(u, v)$ in~(\ref{SandR}) is, like any other element of the flux-tube S-matrix, a complicated function of the coupling which can be constructed by adapting the analysis of~\cite{withAdam} to our case. Its main property, besides the unitarity relation $S_{\phi\psi}(u, v)S_{\phi\psi}(-u, -v) =1$, is the crossing identity
\beq\label{cross-scalar-fermion}
S_{\phi\psi}(u^{2\gamma}, v) =  \frac{u-v+\ft{i}{2}}{u-v+\ft{3i}{2}}S_{\phi\psi}(-u, -v)\, , 
\eeq
where $\gamma$ denotes the mirror map for a scalar~\cite{MoreDispPaper}, i.e., $u^{\gamma} = u+i$ with the shift involving crossing the first cut in the upper-half plane if present. The mirror S-matrix $S_{\star\phi\psi}$ appearing in~(\ref{main-fusion}) is simply defined by
\beq\label{def-mirro-S-scalar-fermion}
S_{\star \phi\psi}(u, v)_{iA}^{jB} = S_{\star\phi\psi}(u, v)R_{\textbf{64}}(u+i-v)_{iA}^{jB}\, , \qquad S_{\star\phi\psi}(u, v)=S_{\phi\psi}(u^{\gamma}, v)\, ,
\eeq
and is thus known once $S_{\phi\psi}$ is known.

It is now possible to employ the fusion equation~(\ref{main-fusion}). One should first observe that, thanks to the $SU(4)$ symmetry, the mirror S-matrix $S_{\star\bar{\psi}\psi}$ admits two channels only, i.e., can be decomposed as
\beq
S_{\star\bar{\psi}\psi}(u, v)_{CB}^{AD} = S^{I}_{\star\bar{\psi}\psi}(u, v)\delta_{C}^{A}\delta_{B}^{D} + S^{II}_{\star\bar{\psi}\psi}(u, v)\delta_{C}^{D}\delta_{B}^{A}\,.
\eeq
Plugging this form inside~(\ref{main-fusion}), as well as the known expressions for $S_{\Psi\psi}$ and $S_{\phi\psi}$, see Eqs.~(\ref{def-Psipsi},\ref{S-matrix-ferm},\ref{SRmatrices}) and~(\ref{SandR}) respectively, and performing the $\rho$-matrix algebra, one would immediately find that the ratio $S^{I}/S^{II}$ is fixed and expressed as a rational function of the difference of rapidities. In fact, it is such that the mirror S-matrix becomes proportional to the $R$-matrix
\beq\label{extra-R}
R_{\bar{\textbf{4}}\textbf{4}}(w)_{CB}^{AD} = R_{\textbf{4}\bar{\textbf{4}}}(w)_{BC}^{DA}\, ,
\eeq
with $R_{\textbf{4}\bar{\textbf{4}}}$ as defined in Eq.~(\ref{SRmatrices}) and $w=u+i-v$ in the current case. We can thus write
\beq\la{Sstar}
S_{\star\bar{\psi}\psi}(u, v)_{CB}^{AD} = S_{\star\bar{\psi}\psi}(u, v) R_{\bar{\textbf{4}}\textbf{4}}(u+i-v)_{CB}^{AD}\, , 
\eeq
with $S_{\star\bar{\psi}\psi}(u, v)$ a scalar factor. This one is also fixed by the relation~(\ref{main-fusion}) which imposes that
\beq\label{scalarfactor-fusion}
S_{\psi\psi}(u^{\omega}, v) = S_{\star \bar{\psi}\psi}(\check{u}+i, v)S_{\star \phi\psi}(u-\ft{i}{2}, v)\, .
\eeq
One can verify a posteriori that the equations~(\ref{Sstar},\ref{scalarfactor-fusion}) are consistent with~(\ref{main-fusion}) thanks to the identity
\beq
R_{\textbf{4}\textbf{4}}(w)_{AB}^{CD} = \frac{1}{6}\rho^{i}_{A\tilde{A}}R_{\bar{\textbf{4}}\textbf{4}}(w+i)^{\tilde{A}D}_{\tilde{C}\tilde{B}}R_{\textbf{6}\textbf{4}}(w-\ft{i}{2})_{iB}^{j\tilde{B}}\rho_{j}^{\tilde{C}C} \, ,
\eeq
among the $R$-matrices.

Equation~(\ref{scalarfactor-fusion}) provides us with a sharp definition of the scalar factor $S_{\star\bar{\psi}\psi}$ and thereby of the mirror S-matrix among fermions via Eq.~(\ref{Sstar}). Though this factor is given above for $u$ lying in the small-fermion sheet, as indicated by the check mark on top of $u$ in the right-hand side of~(\ref{scalarfactor-fusion}), it is valid for any rapidity through analytic continuation. One thing we can easily do, in order to simplify~(\ref{scalarfactor-fusion}), is to undo the $\omega$ transformation in the left-hand side of~(\ref{scalarfactor-fusion}), see figure~\ref{uomega}. To this end, we first cross the cut stretching between $u = \pm 2g$, which is present in both $S_{\psi\psi}(u^{\omega}, v)$ and $S_{\star \phi\psi}(u-\ft{i}{2}, v)$ in~(\ref{scalarfactor-fusion}), and then shift $u\rightarrow u-i$. This way we derive the equivalent relation
\beq\label{Smirror-scalar-factor}
S_{\star \bar{\psi}\psi}(\check{u}, v) = S_{\psi\psi}(u, v)/S_{\phi\psi}(u-\ft{i}{2}, v)\, .
\eeq
Note that this operation has transformed $S_{\star \phi\psi}$ into $S_{\phi\psi}$ because the aforementioned cut is a mirror cut for the scalar-fermion S-matrix, i.e., a cut connecting the real and mirror kinematics of a scalar. The equation~(\ref{Smirror-scalar-factor}) has the advantage of expressing the mirror factor $S_{\star \bar{\psi}\psi}$ in terms of real scattering phases solely, which, as already emphasized a couple of times in this paper, are directly accessible from the study of the asymptotic Bethe ansatz equations. 
It is also possible to obtain a representation for the mirror factor that applies when both the two rapidities are in the large-fermion sheet. One way of writing it is
\beq\label{large-representation-mirror-factor}
S_{\star \bar{\psi}\psi}(u, v) = \frac{u-v+i}{u-v}S_{\psi\psi}(\check{u}, v)S_{\star \phi\psi}(u+\ft{i}{2}, v)\, .
\eeq
It follows from the fact that on the other side of the cut, which we ought to cross to implement the small-to-large-sheet transition $u\leftrightarrow \check{u}$, we can use that $S_{\phi\psi}(u-\ft{i}{2}, v)$ in~(\ref{Smirror-scalar-factor}) reads
\beq
S_{\phi\psi}(u^{-\gamma}+\ft{i}{2}, v) = \frac{u-v}{(u-v+i)S_{\phi\psi}(u^{\gamma}+\ft{i}{2}, v)}  = \frac{u-v}{(u-v+i)S_{\star\phi\psi}(u+\ft{i}{2}, v)}\, ,
\eeq
with the middle equality being equivalent to the crossing relation~(\ref{cross-scalar-fermion}) and the last one to the definition in~(\ref{def-mirro-S-scalar-fermion}). Finally, it is important to stress that
\beq\label{sym-Sstar}
S_{\star \bar{\psi}\psi}(u, v) = S_{\star \bar{\psi}\psi}(v, u) \,.
\eeq
This important property is not patent in the representation~(\ref{large-representation-mirror-factor}) but is manifest in the explicit expression given in appendix~\ref{summary-S-matrices}.

We could have proceeded with the extraction of the mirror S-matrix $S_{\star\psi\psi}$ as well. We leave the algebra to the motivated readers and simply report here the final expression. We found that
\beq\label{S-mirror-two-psi}
S_{\star\psi\psi}(u, v)_{AB}^{CD} = S_{\star\psi\psi}(u, v)R_{\textbf{44}}(u-v+i)_{AB}^{CD}\, ,
\eeq
where the $R$-matrix was given in~(\ref{SRmatrices}) and the scalar factor is related to the one obtained before by
\beq\label{2Sstar}
S_{\star\psi\psi}(u, v) = \frac{u-v}{u-v+i}S_{\star\bar{\psi}\psi}(u, v) \, .
\eeq
The two mirror scalar factors are thus the same, up to a simple rational function of the difference of rapidities. This was actually predictable and as a consistency check we can verify that the relation~(\ref{2Sstar}) is precisely the one respecting the mirror symmetry. By rotating the square in figure~\ref{MirorSmatrix}.b one can indeed map the S-matrices~(\ref{Sstar}) and~(\ref{S-mirror-two-psi}) into one another, such that the relation
\beq
S_{\star\bar{\psi}\psi}(u, v)_{CB}^{AD} = S_{\star\psi\psi}(v, u)_{BC}^{DA}
\eeq
should be observed. This, at the end, immediately follows from~(\ref{2Sstar},\ref{sym-Sstar}), given the obvious relation
\beq
R_{\bar{\textbf{4}}\textbf{4}}(w+i)_{CB}^{AD} = \frac{w}{w-i}R_{\textbf{4}\textbf{4}}(-w+i)_{BC}^{DA}\, ,
\eeq
among the $R$-matrices~(\ref{SRmatrices},\ref{extra-R}).

The equations~(\ref{Sstar},\ref{scalarfactor-fusion},\ref{Smirror-scalar-factor},\ref{large-representation-mirror-factor},\ref{S-mirror-two-psi},\ref{2Sstar}) are the main results of this subsection. They were derived by implementing the bootstrap equation~(\ref{main-fusion}), depicted in figure~\ref{fusion-fermion}, which resulted from the anomalous mirror rotation~(\ref{omega},\ref{meaning-Psi}) of a fermion.

\subsection{Direct test of the mirror S-matrix}

The end result for the mirror S-matrix $S_{\star\bar\psi\psi}(u,v)$ between two fermions is summarized in appendix~\ref{summary-S-matrices}. Its derivation from the physical S-matrix depicted in figure \ref{MirorSmatrix}.a was a rather involved procedure whose rightness relies entirely on whether our interpretation for the $\omega$ transformation introduced in~\ref{fermion-bootstrap} is correct. To support further this picture, 
we shall now confront our finding against a direct perturbative result at weak coupling.

The weak coupling expressions for the mirror S-matrix $S_{\star\bar\psi\psi}(u,v)$ is given in~(\ref{weak-S-app}).
We would like to match it against the direct computation of the square Wilson loop in figure~\ref{MirorSmatrix}.b. To do so, we should insert the four fermion states~$\{\bar\psi(u),\psi(v),\psi(-u),\bar\psi(-v)\}$ on the four consecutive edges of the square, as shown in figure \ref{MirorSmatrix}.b. The mirror S-matrix is then obtained by evaluating the corresponding dressed Wilson loop expectation value (and dividing the result by the bare square Wilson loop). We choose the two fermions $\psi(u)$ and $\psi(v)$ to carry different R-charge so that we probe the scalar factor $S_{\star\bar\psi\psi}$ directly, which is a particular component of the complete S-matrix~(\ref{Sstar}). Since the two fermionic lines, each stretching between a $\psi$ and an $\bar\psi$, must cross each other, the expectation value of the dressed Wilson loop starts at one loop (in the planar limit) in agreement with (\ref{weak-S-app}). 

Instead of computing the associated one-loop Feynman diagrams directly, we shall extract the result from an already known scattering amplitude using the flattening method of~\cite{long}. The idea is to engineer the insertions of the fermions on the four edges of a square by deforming a suitably chosen components of a super amplitudes \cite{superloopskinner,superloopsimon}, as we now explain. 

\begin{figure}[t]
\centering
\def\svgwidth{10cm}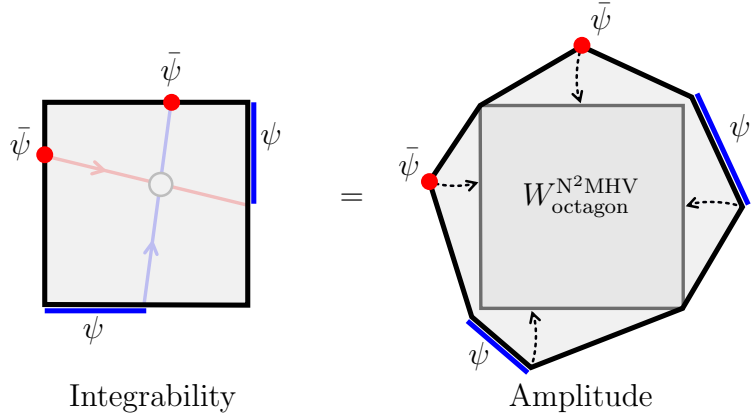
\caption{Computation of the mirror fermion S-matrix from an N$^2$MHV eight-gluon amplitude. The multi-collinear limit or flattening of the dual octagon super WL produces a square with one fermion insertion on each of its sides.}
\la{FourFermions}
\end{figure}
The flattening method for extracting OPE data out of known amplitudes was explained in details in \cite{long}. Therefore we will be brief here. The total helicity weight of our four fermions is $8=1_\psi+1_\psi+3_{\bar\psi}+3_{\bar\psi}$. We should thus consider an N$^2$MHV amplitude. For each insertion of a fermion, we replace the corresponding edge of the square by a ``bump" made of two edges, so that in total we end up with an octagon. We then flatten the bumps by taking their collinear limits with respect to the symmetries of the square, see figure \ref{FourFermions}. 

Now, different components of the N$^2$MHV super loop correspond to different way of decorating the octagon with insertions. With this respect, fermionic insertions along the super loop come in two different flavours: fermions $\psi$ appear integrated along edges while their conjugate insertions $\bar\psi$ are produced at cusps, see \cite{superloopsimon}. We are thus interested in the configuration illustrated in figure \ref{FourFermions}. After extracting the result in position space we shall Fourier transform it into momentum space and compare it with the integrability prediction~(\ref{weak-S-app}). The fact that the two fermions $\psi$ happen to be integrated up to the position of the cusp while the two conjugate fermions $\bar\psi$ are just inserted at the cusp requires introducing a (simple) form factor as we will do. The difference between integrating a fermion to the right or the left of the insertion point will translate into an $i\epsilon$ prescription for handling the poles at zero momenta. The form factors as well as the $i\epsilon$ prescription will be further precised below. 

To specify a component of the N$^2$MHV octagon we need to assign eight $\eta$'s to its edges. There is more than one choice of such assignment that yield, among other insertions, a fermion on an edge or cusp. For the fermion insertion to be well regularized, we will avoid integrating it all the way to a cusp of the square, see figure \ref{FourFermions}. Moreover, our choice of component is made such that at leading order in perturbation theory only the fermion insertion contributes. One possible choice, together with a corresponding assignment of eight twistors is given in figure \ref{octagonN2MHV}.
\begin{figure}[h]
\centering
\def\svgwidth{18cm}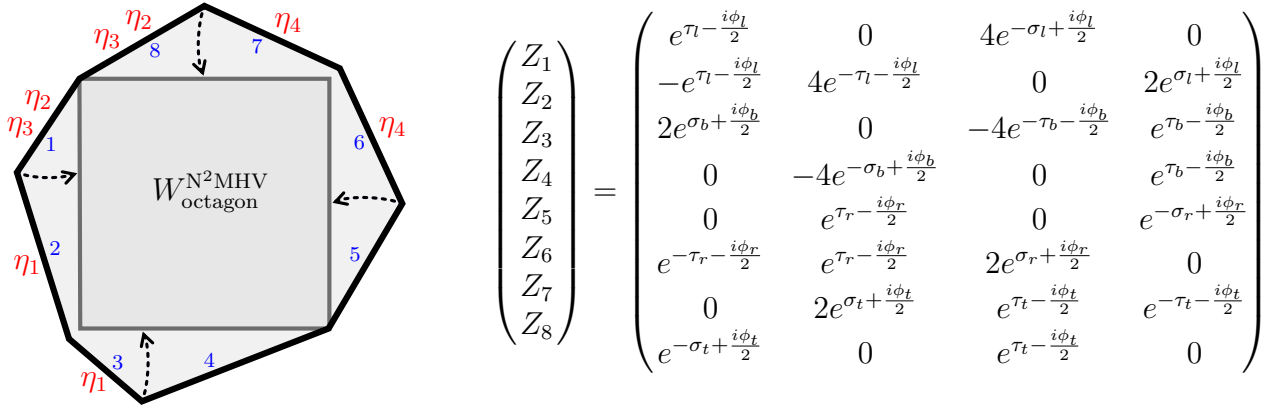
\caption{On the left we specify the component of the N$^2$MHV octagon of interest. On the right are the eight momentum twistors used to parametrize the family of octagons. There are four pairs of twistors corresponding to the four bumps on top of the square. On each of theses bumps we act with the three symmetries of the square $\{\tau_i,\sigma_i,\phi_i\}$ with $i\in\{l,b,r,t\}$.}
\la{octagonN2MHV}
\end{figure}

To evaluate this N$^2$MHV octagon at one loop we use the package introduced by Bourjaily, Caron-Huot and Trnka in~\cite{Bourjaily:2013mma}, set the \verb"Zs" therein to be given by the ones in figure \ref{octagonN2MHV} and run the command\\ \\
\verb"evaluate@superComponent[{2,3},{1},{1},{},{},{4},{4},{2,3}]@ratioIntegral[8,2]". \\ \\
Finally, we \verb"Series" expand the result at large $\tau_l,\tau_b,\tau_r$ and $\tau_t$. The details of how exactly we have chosen to deform the octagon edges away from the square, as given by the twistors assignments in figure \ref{octagonN2MHV}, only effect the subleading contributions in the collinear limit. To leading order we find
\beq\la{octagonN2MHVOPE}
\cR_8^{\text{N}^2\text{MHV}}\xrightarrow[\tau_i\to\infty]{}{e^{\tfrac{i}{2}(\phi_b-\phi_t+\phi_r-\phi_l)}\over 16\,e^{\tau_l+\tau_b+\tau_r+\tau_t}}\[{2y^2\(\text{Li}_2\left(-x^2\right)+\log (x y) \log \left(x^2+1\right)\)\over y^2-x^2}+(y\leftrightarrow x)\]
\eeq 
where $x=e^{-\sigma_l-\sigma_r}$ and $y=e^{-\sigma_t-\sigma_b}$ are the relative insertion points of the pairs of conjugate fermions. 

Next, we would like to Fourier transform the result (\ref{octagonN2MHVOPE}). It is equivalent, simpler and more instructive to Fourier transform the integrability prediction
\beqa\la{expect0}
\int{du\over2\pi}\int{dv\over2\pi}x^{-2iu}y^{-2iv}{x(u)\over g^2}{x(v)\over g^2}\,\mu_\psi(u)\mu_\psi(v)\,S_{\star\bar\psi\psi}(u,-v) \,,
\eeqa
instead. Here $\mu_\psi(u)$, $\mu_\psi(v)$ are the measure factors for integrating the rapidity of a fermion (\ref{measureweak}) while $x(u)/g^2$ and $x(v)/g^2$ are the form factors alluded to above. In our convention for the fermonic measure and transition, all fermions are integrated. However, in the case at hand the two $\bar\psi$'s on the top and left edges are not integrated and therefore two form factors need to be incorporated for those two insertions.  We already considered similar form factors for the gluonic excitations in \cite{long} -- see section 4.5 there -- and the fermion form factors here should in principle be justified along the same lines as there. Here, however, since we are only interested in the leading order weak coupling result, it is simpler to argue for them. After all, to go from an integrated to a non-integrated fermion we simply apply a derivative. In momentum space, it amounts to multiplying the result by $p(u)=2u+\cO(g^2)$ and indeed $x(u)\simeq u$ at weak coupling. The additional factor of $1/g^2$ for non integrated fermion is built into the super loop of \cite{superloopskinner,superloopsimon}. In the right hand side of (\ref{expect}) we also included the $i0$ prescription for preforming the momentum integrals, see discussion around (\ref{Cusp-behav}). 

In sum, to leading order at weak coupling our prediction reads
\beqa\la{expect}
\int du\int dv\, x^{-2iu}y^{-2iv}{\pi g^2\sinh(\pi(u+v))\over4 \sinh^2(\pi u+i0)\sinh^2(\pi v+i0)(u+v)}+\cO(g^4)
\eeqa
Preforming the Fourier integrals in (\ref{expect}) is now straightforward. We simply expand the integral at small positive $y$ and large $x$ by picking the residues in the upper (lower) $v$ ($u$) plane. 

In this way, we find a perfect match with the same expansions of the square bracket in~(\ref{octagonN2MHVOPE}) up to an overall numerical factor of $-16$. In fact, finding an agreement up to an overall constant is what we should expected at this point. 

This is because $\cR_8^{\text{N}^2\text{MHV}}$ in (\ref{octagonN2MHVOPE}) carries helicity weight. This means that if we rescale the projective twistors in figure \ref{octagonN2MHV}, we rescale the corresponding result in the right hand side of (\ref{octagonN2MHVOPE}). Still, with some ingenuity, it is possible to match the overall normalization of the mirror S-matrix as well. 

\begin{figure}[t]
\centering
\def\svgwidth{12cm}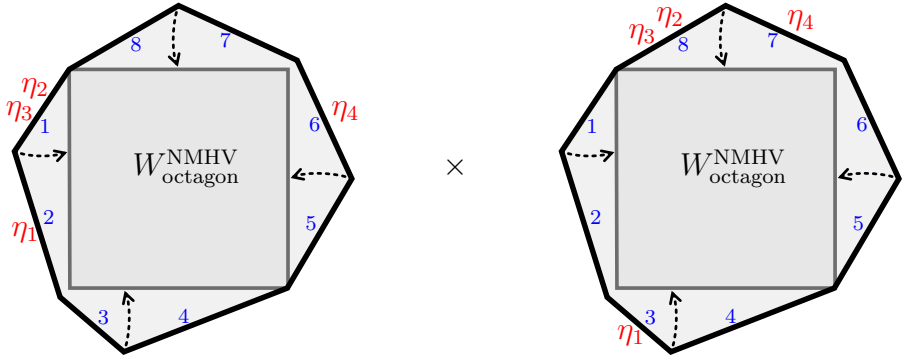
\caption{To construct an helicity weight free quantity from $\cR_8^{\text{N}^2\text{MHV}}$ in figure \ref{octagonN2MHV}, we divide it by the product of two NMHV components of the same octagon Wilson loop. These are plotted in this figure.}
\la{octagonNMHV}
\end{figure}
To do so we should construct an helicity weight free quantity. We can achieve this by normalizing $\cR_8^{\text{N}^2\text{MHV}}$ in (\ref{octagonN2MHVOPE}) with the free propagation of the fermions in the two channels of the square. These are given by the large $\tau$ limit of the two NMHV amplitudes in figure~\ref{octagonNMHV}. They start their lives at tree level. With our choice of twistors they behave, respectively, as
\beq
\cR_{\eta_1^2\eta_2\eta_6}^{\text{NMHV}}\xrightarrow[\tau_i\to\infty]{}-{1\over8}{e^{\tfrac{i}{2}(\phi_r-\phi_l)}\over e^{\tau_l+\tau_r}}{x^2\over1+x^2}\ ,\qquad\cR_{\eta_3\eta_7\eta_8^2}^{\text{NMHV}}\xrightarrow[\tau_i\to\infty]{}-{1\over2}{e^{\tfrac{i}{2}(\phi_b-\phi_t)}\over e^{\tau_b+\tau_t}}{y^2\over1+y^2}
\eeq
On the other hand our prediction for these would be
\beq
\int{du\over2\pi}x^{-2iu}{x(u)\over g^2}\,\mu_\psi(u)=\int du{x^{-2iu}\over2\sinh(\pi u+i0)}+\cO(g^2)=-i{x^2\over1+x^2}+\cO(g^2)
\eeq
and the same with $x\leftrightarrow y$ for the other channel. The overall factor of $-{1\over16} =\[{1\over8}\times{1\over2}\]/\[i\times i\]$ is nothing but the numerical constant encountered before! 

At the end of the day, it is quite remarkable that the flux-tube (mirror) S-matrix, which plays a pivotal role in our construction~\cite{short}, can be given a direct and elegant physical meaning. It is equal to the collinear limit of a suitably chosen (super) Wilson loop. It would be interesting to see if something similar can be done for the spin chain S-matrix on top of the BMN vacuum~\cite{Beisert} -- that is, to see whether this S-matrix can be defined directly in terms of certain observables of the ${\cal N}=4$ SYM theory.   

\subsection{Summary of the S-matrices}\label{summary-S-matrices}

In this subsection we summarize the expressions for the various scattering phases and mirror factors encountered in this paper.

The flux-tube scattering factors all follow the same universal pattern. For any pair $(a, b)$ of excitations, they can be written in the form
\beq\label{S-Smirror-app}
S_{ab}(u, v) = s_{ab}(u, v)e^{i\Phi_{ab}(u, v)}\, , \qquad S_{\star ab}(u, v) = s_{\star ab}(u, v)e^{-\Phi_{\star ab}(u, v)}\, ,
\eeq
where $s_{ab}, s_{\star ab}$ are explicitly known functions of the rapidities $u, v$ and coupling constant $g$. They typically involve integrals of the Bessel functions of the first kind (see Eq.~(\ref{easy-part-real},\ref{easy-part-mirror}) below for illustration), which can be evaluated analytically at weak/strong coupling or numerically for intermediate values of the coupling. The most involved, and also interesting, part of the flux-tube dynamics is encoded into the functions $\Phi_{ab}(u, v)$ and $\Phi_{\star ab}(u, v)$. Their expressions can be obtained by solving certain infinite systems of linear equations, that can be viewed as inhomogeneous versions~\cite{BenDispPaper} of the BES equation~\cite{BES} for the cusp anomalous dimension. The interested reader is referred to~\cite{withAdam} for the exposition of the analysis in the scalar case ($a=b=\phi$), and to~\cite{Fioravanti:2013eia} for an alternative treatment. In this appendix, we shall restrict ourselves to presenting the final expressions, for the few cases relevant to our study.

Prior to specialize to certain flavours of excitations, there is more we can say about the structure of the terms in~(\ref{S-Smirror-app}). First, we can always write
\beq\label{dyn-factor}
\Phi_{ab}(u, v) = -2f_{1\,ab}(u, v)+2f_{2\,ab}(u, v)\, , \qquad \Phi_{\star ab}(u, v) = -2f_{3\,ab}(u,v) +2f_{4\,ab}(u,v)\, ,
\eeq
in terms of functions $f_{1, \ldots , 4\,ab}(u, v)$. Roughly speaking, these functions saturate the information contained in the solution to the inhomogeneous BES equation. They form a closed algebra under mirror transformation, as explained in~\cite{withAdam} for the scalar case. Second, they have definite parity
\beq\label{basic-prop-1}
f_{1,3\,ab}(-u, v) = -f_{1,3\,ab}(u, v)\, , \qquad f_{2,4\,ab}(-u, v) = f_{2,4\,ab}(u, v)\, ,
\eeq
and simple permutation property,
\beq\label{basic-prop-2}
f_{1\,ab}(u, v) = f_{2\,ba}(v, u)\, , \qquad f_{3\,ab}(u, v) = f_{3\,ba}(v, u)\, , \qquad f_{4\,ab}(u, v) = f_{4\,ba}(v, u)\, .
\eeq
These equalities guarantee, in particular, that the identities
\beq
\Phi_{ab}(u, v) = -\Phi_{ab}(-u, -v) = -\Phi_{ba}(v, u)\, , \qquad \Phi_{\star ab}(u, v) = \Phi_{\star ab}(-u, -v) = \Phi_{\star ba}(v, u)\, ,
\eeq
hold true, at any coupling and for whatever rapidities. They are related to the unitarity of the S-matrix, in the first case, and to the fact that the mirror factor is almost a symmetric function, in the second case.

Finally, a nice property of the functions $f_{1, \ldots 4\,ab}(u, v)$ is that they can be written in a relatively closed form. For instance, using the matrix notation of~\cite{long}, which was inspired by the analysis of the BES equation given in~\cite{Benna:2006nd}, we have
\beq
\begin{aligned}\label{def-f-functions}
&f_{1\,ab}(u, v) = 2\tilde{\kappa}_{a}(u)^{t}\cdot \mathbb{Q}\cdot \mathbb{M}\cdot \kappa_{b}(v)\, ,\\
&f_{2\,ab}(u, v) = 2\kappa_{a}(u)^{t}\cdot \mathbb{Q}\cdot \mathbb{M}\cdot \tilde{\kappa}_{b}(v)\, ,\\
&f_{3\,ab}(u, v) = 2\tilde{\kappa}_{a}(u)^{t}\cdot \mathbb{Q}\cdot \mathbb{M}\cdot \tilde{\kappa}_{b}(v)\, ,\\
&f_{4\,ab}(u, v) = 2\kappa_{a}(u)^{t}\cdot \mathbb{Q}\cdot \mathbb{M}\cdot \kappa_{b}(v)\, ,
\end{aligned}
\eeq
where $\kappa_{a}(u), \tilde{\kappa}_{a}(u)$ are two infinite dimensional (column) vectors, $\kappa_{a}(u)^{t} \equiv (\kappa_{a}(u)_1, \kappa_{a}(u)_2, \ldots )$ and $\tilde{\kappa}_{a}(u)^{t}$ their transposed, and $\mathbb{Q}, \mathbb{M}$ two infinite dimensional square matrices. The two vectors $\kappa_{a}(u)$ and $\tilde{\kappa}_{a}(u)$ have different parity,
\beq\label{parity-prop-k}
\kappa_{a}(-u) = \kappa_{a}(u)\, , \qquad \tilde{\kappa}_{a}(-u) = -\tilde{\kappa}_{a}(u)\, ,
\eeq
and define, respectively, the two series of source terms for the parity even/odd inhomogeneous versions of the BES equation derived in~\cite{BenDispPaper}. (More precisely, the vector $\kappa_{a}(u)$ here collects the source terms for the various excitations given in~\cite{BenDispPaper} while for the tilde version one should first flip the sign of the odd components, i.e., $\tilde{\kappa}_{a}(u)_{n}$ here equals $(-1)^n\tilde{\kappa}_{a}(u)_{n}$ in~\cite{BenDispPaper}.) 
These two vectors are known explicitly and can be easily computed at weak and strong coupling. Their expressions are recalled in appendix~\ref{FCPs} for all the cases relevant to our analysis. The square matrix $\mathbb{Q} \cdot \mathbb{M}$ is more \textit{universal}: it is independent of both the flavours and rapidities of the excitations, i.e., of the indices $a,b$ and rapidities $u, v$, which are only carried by the vectors in~(\ref{def-f-functions}). In fact, it only depends of the coupling constant, but not through the matrix $\mathbb{Q}$ which is a simple diagonal matrix with entries $\mathbb{Q}_{ij}=\delta_{ij}(-1)^{i+1}i$. The main block is thus the matrix $\mathbb{M}$ that is given as the inverse of a coupling dependent matrix:
\beq
\mathbb{M}=\[\mathbb{I}+\mathbb{K}\]^{-1} \, , \qquad \mathbb{K}_{ij}=2j(-1)^{j(i+1)} \int\limits_{0}^\infty \frac{dt}{t} \frac{J_i(2gt)J_j(2gt)}{e^t-1} \la{Mdef} \,,
\eeq
where $\mathbb{I}$ is the identity matrix and $J_i(z) = O(z^i)$ the Bessel functions of the first kind. This matrix can be easily inverted at weak coupling,
\beq
\mathbb{M} = \mathbb{I} -\mathbb{K} +\mathbb{K}^2 -\ldots\, ,
\eeq
thanks to the scaling $\mathbb{K}_{ij} = O(g^{i+j})$. This, together with a similar scaling for the inhomogenous terms, make the formulae~(\ref{def-f-functions}) very convenient for perturbation theory. As an example, in all cases, one finds that the functions~(\ref{def-f-functions}) are explicitly suppressed at weak coupling, with $f_{1,\ldots 4\,ab} = O(g^2)$. This, in turn, implies that the leading weak coupling expressions for the S-matrix~(\ref{S-Smirror-app}) are entirely controlled by the simple prefactors $s_{ab}$ and $s_{\star ab}$. One can also easily verify that the product $\mathbb{Q} \cdot \mathbb{M}$ defines a symmetric matrix. This, together with~(\ref{parity-prop-k}), immediately explain the properties listed in~(\ref{basic-prop-1},\ref{basic-prop-2}). 

As an additional comment, let us stress that the infinite matrix~(\ref{Mdef}) simply arises from inverting the kernel of the BES equations, in the representation given in~\cite{BenDispPaper}. The bilinear representation~(\ref{def-f-functions}) makes then clear why the functions $f_{1, \ldots, 4\,ab}(u, v)$ appear as the main building blocks for the scattering factors. They essentially combine the two ingredients, i.e., the inhomogenous terms carrying the flavour information and the universal kernel~(\ref{Mdef}) encoding the flux-tube dynamics, in all practical ways.

Now, what remains to be done for fixing the S-matrices~(\ref{S-Smirror-app}) is to list the expressions for the prefactors $s_{ab}, s_{\star ab}$. For fermion-fermion and scalar-fermion scattering, we found
\beq
\begin{aligned}\label{easy-part-real}
&s_{\psi\psi}(u, v) = \exp\[2i\int\frac{dt}{t}\frac{J_0(2gt)\sin{(ut)}-J_0(2gt)\sin{(vt)}-\sin{((u-v)t)}}{e^{t}-1}\]\, , \\
&s_{\phi\psi}(u, v) = \exp\[2i\int\frac{dt}{t}\frac{J_0(2gt)e^{t/2}\sin{(ut)}-J_0(2gt)\sin{(vt)}-e^{t/2}\sin{((u-v)t)}}{e^{t}-1}\]\, ,
\end{aligned}
\eeq
and
\beqa
&&\!\!\!s_{\star \bar{\psi}\psi}(u, v) = -\frac{g^2}{x(u)x(v)}\times\nn\\
&&\qquad \,\, \exp\[\int\frac{dt}{t}\frac{\cos{((u-v)t)}-J_0(2gt)\cos{(ut)}-J_0(2gt)\cos{(vt)}+J_0(2gt)^2}{e^{t}-1}\]\, ,\nn \\
&&\!\!\!s_{\star\phi\psi}(u, v) = -\frac{g^2}{(u-v+\ft{i}{2})x(v)}\times\label{easy-part-mirror}\\
&&\qquad \,\, \exp\[\int\frac{dt}{t}\frac{e^{t/2}\cos{((u-v)t)}-J_0(2gt)e^{t/2}\cos{(ut)}-J_0(2gt)\cos{(vt)}+J_0(2gt)^2}{e^{t}-1}\]\, .\nn
\eeqa
Their form is reminiscent of the results obtained for scalars and gluons in~\cite{withAdam,long}. We also notice that the prefactor $s_{\star\bar{\psi}\psi}$ is symmetric under the permutation $u\leftrightarrow v$. When combined with the second equation in~(\ref{basic-prop-2}), it implies that the same property is observed by the mirror S-matrix $S_{\star \bar{\psi}\psi}$, as previously claimed in~(\ref{sym-Sstar}).

As said earlier, the expressions~(\ref{easy-part-real},\ref{easy-part-mirror}) can be directly used to determine the S-matrix~(\ref{S-Smirror-app}) to leading order at weak coupling. After a bit of algebra (see~\cite{withAdam} for illustration), we get
\beq
\begin{aligned}\label{weak-S-app}
&S_{\psi\psi}(u, v) \simeq s_{\psi\psi}(u, v) = \frac{\Gamma(1-iu)\Gamma(1+iv)\Gamma(1+iu-iv)}{\Gamma(1+iu)\Gamma(1-iv)\Gamma(1+iv-iu)} + O(g^2) \, , \\
&S_{\star\bar{\psi}\psi}(u, v) \simeq s_{\star\bar{\psi}\psi}(u, v) =  -\frac{\pi g^2\sinh{(\pi(u-v))}}{(u-v)\cosh{(\pi u)}\sinh{(\pi v)}} + O(g^4)\, , \\
&S_{\phi\psi}(u, v) \simeq s_{\phi\psi}(u, v) = \frac{\Gamma(\ft{1}{2}-iu)\Gamma(1+iv)\Gamma(\ft{1}{2}+iu-iv)}{\Gamma(\ft{1}{2}+iu)\Gamma(1-iv)\Gamma(\ft{1}{2}+iv-iu)} + O(g^2) \, , \\
&S_{\star\phi\psi}(u, v) \simeq s_{\star\phi\psi}(u, v) =  -\frac{\pi g^2\cosh{(\pi(u-v))}}{(u-v+\ft{i}{2})\cosh{(\pi u)}\sinh{(\pi v)}} + O(g^4)\, .
\end{aligned}
\eeq
The expression for $S_{\psi\psi}$ above agrees with the expected scattering phase for conformal spin~$1$ excitations~\cite{Belitsky:2006en} and with the explicit result of~\cite{Fioravanti:2013eia} -- note though that our sign convention might differ with those used in these papers.

Finally, for completeness we also give the expressions obtained after the analytic continuation to the small fermion. We have
\beq
s_{\psi\psi}(\check{u}, v) = s_{\psi\psi}(\check{u}, \check{v}) = s_{\phi\psi}(u, \check{v}) = 1\, , 
\eeq
and
\beq
s_{\star\bar{\psi}\psi}(\check{u}, v) = \frac{x(u)}{u-v}\, , \qquad s_{\star \bar{\psi}\psi}(\check{u}, \check{v}) = 1\, , \qquad s_{\star\phi\psi}(u, \check{v}) = -\frac{x(v)}{u-v+\ft{i}{2}}\, .
\eeq
These equations illustrate the general lesson according to which, when dealing with fermions, the simplest expressions are always found on the small sheet. This is made especially manifest here for the scattering phases involving a small fermion, which are completely captured by the universal factor~(\ref{dyn-factor}). It implies notably that all these scattering phases evaluate to~$1$ to leading order at weak coupling,
\beq
S_{\psi b}(\check{u}, v) = 1+O(g^2)\,,
\eeq
for any excitation $b$. This simplicity is tied to the fact that fermions have momenta $\sim g^2$ on the small sheet at weak coupling. The expression for a large fermion, which simply follows from analytic continuation in the rapidity, is significantly more involved.
In the first line of~(\ref{weak-S-app}), we see that it is of order $O(1)$ for these fermions with momenta $\sim u,v = O(1)$, as it was the case for other excitations like scalars or gauge fields (see~\cite{Peng,withAdam,long,Fioravanti:2013eia}).

\section{Manipulating the transitions}\la{appenpixB}

In this appendix we present the algebra that allows one to obtain the expressions for the pentagon transitions and form factors used in this paper. The manipulations we shall perform below rely mostly on the general properties of the flux-tube S-matrix, like crossing and unitarity for instance, which were given in~\cite{short,withAdam,long} and appendix~\ref{paradoxSec}.

\subsection{Gluons} \la{manipG}
To begin with, we shall demonstrate that the gluonic pentagon transitions~\cite{short,long} can be casted into the form~(\ref{trans-formMain},\ref{finFermions}) advocated in this paper. The algebra is straightforward.
For the sake of clarity we recall that the transitions are given by~\cite{short,long} 
\beq\label{recall-long}
\begin{aligned}
P_{F F}(u|v)^2 = \frac{f_{FF}(u,v)}{(u-v)(u-v-i)} \frac{S_{F F}(u,v)}{S_{F F}(u^{\gamma},v)} \, , \\
P_{F \bar F}(u|v)^2 = \frac{(u-v)(u-v-i)}{f_{FF}(u,v)} \frac{S_{FF}(u,v)}{S_{FF}(u^\gamma,v)}\, ,
\end{aligned}
\eeq
where the function $f_{FF}(u, v)$ is given in terms of Zhukowsky variables in~(\ref{f-typeMain}) and slightly differs from the similar function introduced in~\cite{short,long} by the overall factor $1/g^2$. To present these transitions in the form used in this paper we introduce the mirror S-matrices
\beq\label{def-mirror-S-gluons}
S_{\star FF}(u, v) \equiv S_{F\bar{F}}(u^{\gamma}, v)\, , \qquad  S_{\star F\bar{F}}(u, v) \equiv S_{FF}(u^{\gamma}, v)\, .
\eeq
These definitions incorporate the fact that the two gluonic excitations exchange their role under the mirror map $\gamma:u\rightarrow u^{\gamma}$, see appendix~D of~\cite{long}. The two scattering phases, and their mirror partners, are not independent and satisfy the relation~\cite{long}
\beq\label{ratio-S-mat-gluons}
\frac{S_{FF}(u, v)}{S_{F\bar{F}}(u, v)} = \frac{S_{\star F\bar{F}}(u, v)}{S_{\star FF}(u, v)} = \frac{u-v+i}{u-v-i}\, .
\eeq
With its help it is immediate to derive the representations~(\ref{trans-formMain},\ref{finFermions}) from~(\ref{recall-long},\ref{def-mirror-S-gluons}).

The focus of this paper is on the gluonic form factors~(\ref{tosimp}) which control the creation and annihilation of pairs of gluons on the pentagon Wilson loop. According to~(\ref{tosimp}) they can be obtained from the direct transitions~(\ref{recall-long}) by suitable sequences of mirror rotations, applied to one of the two gluon rapidities. For illustration, to obtain the first form factor in~(\ref{tosimp}), i.e., $P_{FF}(0|u, v)$, we can perform the crossing transformation $u \to u^{2\gamma}$ on the helicity-violating transition $P_{F\bar{F}}(u|v)$ in~(\ref{recall-long}). This is done as follows. By definition, the crossing transformation of a gluon amounts to carrying its rapidity $u$ along a closed path in the rapidity plane that crosses twice the Zhukowksy cuts associated to $x^{\pm} = x(u\pm \ft{i}{2})$, see figure 4 in~\cite{MoreDispPaper}. In other words both Zhukowsky variables $x^{\pm}$ map back to themselves at the end of the process, since $x^{\pm} \to g^2/x^\pm \to x^\pm$. It follows from it that only the ratio of S-matrices in (\ref{recall-long}) transforms non-tivially under crossing, as the function $f_{FF}(u, v)$ and the rational function of the rapidities in~(\ref{recall-long}) are entirely built out of invariant elements. The transformation of the former ratio only involves knowledge of the crossing and unitarity relations for the S-matrix which were both given in~\cite{short,long} and summarized here as $S_{FF}(u^{2\gamma},v) = S_{F\bar{F}}(v,u)=1/S_{F\bar{F}}(u,v)$. We conclude that 
\beq
P_{F \bar F}(u^{2\gamma}|v)^2 = \frac{g^2 (u-v)(u-v-i)}{f_{FF}(u,v)} \frac{S_{F\bar{F}}(u^{\gamma},v)}{S_{F\bar{F}}(u,v)} \, , \la{step1}
\eeq
which we recognize as being the same as $1/P_{FF}^2(u,v)$ thanks to~(\ref{ratio-S-mat-gluons},\ref{def-mirror-S-gluons}). This establishes the representation for $P_{FF}(0|u,v)$ given in~(\ref{creationFromTransition}) up to an overall sign. To fix the latter ambiguity one can observe with~\cite{short} that the gluonic pentagon transitions all become equal to $1$ at strong coupling and thus are invariant under any combination of mirror rotations in this limit. This singles out the branch choice to be as given in~(\ref{creationFromTransition}). Similarly, we could simplify the second relation in (\ref{tosimp}) by noticing that it amounts to performing an inverse crossing transformation $v\rightarrow v^{-2\gamma}$ on the pentagon transition $P_{F\bar{F}}(v^{-\gamma}|u)$. The latter is equivalent to $P_{FF}(u|v)$ by construction~\cite{short,long} and hence the entire problem of deriving the second relation in~(\ref{creationFromTransition}) from~(\ref{tosimp}) follows the same elementary steps as before. Finally, the annihilation amplitudes can also be obtained along these lines or, equivalently, using~(\ref{chain}).

\subsection{Scalars}\label{scalar-manip}

The analysis that leads to the scalar form factor~(\ref{scalarCase}) is very similar to the one performed previously for the gluons. The starting point is the scalar pentagon transition~\cite{long} which can be squared to the expression quoted in~(\ref{trans-formMain}). The form factor~(\ref{scalarCase}) follows directly from it after using the crossing transformations~\cite{withAdam,long}
\beq\label{crossing-scalar-S}
S_{\phi\phi}(u^{2\gamma}, v)S_{\phi\phi}(u, v) = \frac{u-v}{u-v+2i}\, , \qquad S_{\star\phi\phi}(u^{2\gamma}, v)S_{\star\phi\phi}(u, v)  = \frac{u-v+i}{u-v+3i}\, ,
\eeq 
with $S_{\star\phi\phi}(u, v) \equiv S_{\phi\phi}(u^{\gamma}, v)$ and $u^{\gamma} = u+i$ the scalar mirror rotation~\cite{MoreDispPaper}. Indeed, applying~(\ref{crossing-scalar-S}) to~(\ref{trans-formMain}), we find
\beq
P_{\phi\phi}(u^{2\gamma}|v)^2  = \frac{(u-v)S_{\star\phi\phi}(u, v)}{g^2(u-v+i)(u-v+2i)^2S_{\phi\phi}(u, v)}\, ,
\eeq
which is easily seen to be the square of the equality~(\ref{scalarCase}). The sign ambiguity can be fixed by going to strong coupling, as we will now demonstrate.

At strong coupling and for rapidities of order $O(g^0)$, the scalar pentagon transition takes the simple form given in~(\ref{mu-P-O6}). To derive this result, we can use the fact that the scalar excitations are controlled in this regime by the $O(6)$ sigma model~\cite{AldayMaldacena}. This implies in particular that the scalar S-matrix becomes identical to the one found by Zamolodhikov and Zamolodhikov~\cite{ZamolodchikovS},
\beq
S_{\phi\phi}(u, v)  = S_{O(6)}(\xi) + \ldots\, ,
\eeq
where $\xi = i(u-v)/4$ and
\beq\label{SO6-app}
S_{O(6)}(\xi) = -\frac{\Gamma(1+\xi)\Gamma(\ft{1}{2}-\xi)\Gamma(\ft{3}{4}+\xi)\Gamma(\ft{1}{4}-\xi)}{\Gamma(1-\xi)\Gamma(\ft{1}{2}+\xi)\Gamma(\ft{3}{4}-\xi)\Gamma(\ft{1}{4}+\xi)}\, .
\eeq
That this follows from the general formula for the scalar flux-tube S-matrix was established in~\cite{withAdam}, following the earlier study~\cite{Basso:2008tx}. The mirror expression is immediately obtained by applying $u\rightarrow u+i$ to the above equation, that is
\beq\label{SmO6-app}
S_{\star\phi\phi}(u, v) \simeq S_{O(6)}(\xi -\frac{1}{4}) = \frac{\xi}{\xi-\ft{1}{4}}\frac{\Gamma(\ft{1}{2}+\xi)\Gamma(\ft{1}{2}-\xi)\Gamma(\ft{3}{4}+\xi)\Gamma(\ft{3}{4}-\xi)}{\Gamma(1+\xi)\Gamma(1-\xi)\Gamma(\ft{1}{4}+\xi)\Gamma(\ft{1}{4}-\xi)}\, .
\eeq
After plugging the $O(6)$ values~(\ref{SO6-app}) and~(\ref{SmO6-app}) into the scalar pentagon transition~(\ref{trans-formMain}), one easily verifies that the expression~(\ref{mu-P-O6}) should hold true, up to an overall sign. To check that the sign in~(\ref{mu-P-O6}) is the correct one, we should recall that the pentagon transition $P_{\phi\phi}(u|v)$ also determines the square measure $\mu_{\phi}(u) = \lim_{v\rightarrow u} 1/(i(v-u)P_{\phi\phi}(v|u))$. The sign in~(\ref{mu-P-O6}) is as it should be for the measure $\mu_{\phi}(u)$ to be positive definite, which ought to be the case for any real rapidity $u$ and coupling $g$ within our normalizations~\cite{long}. Having established the strong coupling form~(\ref{mu-P-O6}) of the scalar transition, one can use it to test the crossing relation~(\ref{scalarCase}) and verify that in this case as well the sign ambiguity was properly addressed.

\subsection{Fermions} \la{conjectureFermions}

In section~\ref{fermionssection} we presented two conjectures for the fermion pentagon transitions (\ref{WFerm}) that we repeat here for convenience 
\beq
P_{\psi \psi}(u|v)^2 =\frac{f_{\psi\psi}(u,v)S_{\psi\psi}(u, v)}{(u-v)(u-v+i)S_{\star \psi\psi}(u, v)}\, , \qquad P_{\psi\bar \psi}(u|v)^2 =  \frac{S_{\bar \psi \psi}(u, v)}{f_{\psi\psi}(u,v)S_{\star \bar \psi \psi}(u, v)}\, . \la{WFerm2}
\eeq
Here we will show that they satisfy the fundamental relations to the S-matrix (see~(\ref{fund-rel-fermions}) and discussion after~(\ref{ansatzC}))
\beq\la{fundamental}
P_{\psi\psi}(u|v) = -S_{\psi\psi}(u, v)P_{\psi\psi}(v|u)\, , \qquad P_{\psi\bar\psi}(u|v) = -S_{\psi\bar\psi}(u, v)P_{\psi\bar\psi}(v|u)\, ,
\eeq
where $S_{\psi\psi}(u, v)$ and $S_{\psi\bar\psi}(u, v)$ are the fermion scattering phases defined in appendix~\ref{paradoxSec}. We will only consider the square of these relations, since, as explained in the main text,  the minus signs in these equations follow from unitarity of the S-matrices and the existence of the square measure pole in $P_{\psi\psi}$.

The two fundamental relations in (\ref{fundamental}) are not independent. Instead, we have that 
\beq\la{twoPtooneP}
P_{\psi\psi}(u|v)^2 = P_{\psi\bar\psi}(u|v)^2 \times{f_{\psi\psi}(u,v)^2\over(u-v)^2}\, ,
\eeq
where $f_{\psi\psi}(u,v)$ is the symmetric function (\ref{f-fermions}), as follows directly from the relations between S-matrices given in~(\ref{SequalBarS}) and~(\ref{2Sstar}). 
Hence it is enough to prove the relation $P_{\psi\bar\psi}(u|v)^2 =S_{\psi\bar\psi}(u, v)^2P_{\bar\psi\psi}(v|u)^2$ to establish simultaneously both equations in~(\ref{fundamental}). The latter relation is easily derived and follows from the fact that all factors entering $P_{\psi\bar\psi}(u|v)^2$ in (\ref{WFerm2}) are symmetric under the permutation $u\leftrightarrow v$, apart from $S_{\psi\psi}(u,v)$ which is required to fulfill $S_{\psi\psi}(u,v)S_{\psi\psi}(v,u)=1$ by unitarity. 

Finally, we easily verify that the pentagon transition $P_{\psi\psi}(u|v)$ has a pole at coinciding rapidities, as required by the existence of a square limit~\cite{short}. This pole is correctly embodied in our ansatz and is manifest in~(\ref{twoPtooneP}) since both $S_{\star \bar{\psi}\psi}(u, v)$ and $f_{\psi\psi}(u, v)$ are regular at $u=v$. Given that $S_{\bar{\psi}\psi}(u, u) = 1$, we could derive from it the representation
\beq
\frac{1}{\mu_\psi(u)^2} = - \frac{f_{\psi\psi}(u, u)}{S_{\star \bar{\psi} \psi}(u, u)}\, .
\eeq
This, together with our convention that the measure is positive for a large fermion, uniquely fixes $\mu_{\psi}(u)$. Note that with this choice of normalization, the measure is negative for a small fermion, i.e., $\mu_{\psi}(\check{u}) < 0$, because the sign of $\mu_{\psi}dp_{\psi}$ should be preserved by continuity and $dp_{\psi}/du$ changes sign when changing sheet.

\newpage

\section{Conjectures for all fundamental transitions}\la{fundamentaltransitions}
In this appendix we summarize our conjectures for all the transitions used in the text, that is for
\beq
P_{FF}(u|v) \,, \qquad P_{F\bar F}(u|v)\,, \qquad P_{\phi\phi}(u|v) \,, \qquad P_{\psi\psi}(u|v)\, , \,\,\,\, \text{and} \qquad P_{\psi\bar\psi}(u|v) \,. \la{listPs}
\eeq
All of them are constructed out of the physical and mirror S-matrices between flux tube excitations. In turn, as explained in~\cite{Peng,withAdam,long,Fioravanti:2013eia} and sketched in appendix~\ref{summary-S-matrices}, the latter can be found by solving some inhomogeneous versions~\cite{BenDispPaper} of the BES equation~\cite{BES}.
The immediate consequence is that all the pentagon transitions display the same pattern: they can all be written in the form 
\beq
P_{ab}(u|v)=F_{ab}(u,v)\,\exp\Big[2(\kappa_a(u)-i\tilde\kappa_a(u))^{t}\cdot \mathbb{Q} \cdot\mathbb{M}\cdot\kappa_b(v)+2i(\kappa_a(u)+i\tilde\kappa_a(u))^t\cdot \mathbb{Q} \cdot\mathbb{M}\cdot\tilde\kappa_b(v)\Big] \,, \la{FandExponent}
\eeq
with the meaning of each terms in the exponent as given in appendix~\ref{summary-S-matrices}.
As for the rest, the prefactor $F_{ab}$ depends on the coupling, on the kind of excitations, and on their rapidities~$u, v$. However, contrary to the exponent, this contribution can be written down explicitly at any coupling and does not involve the solution to the integral equation (i.e., it does not require dealing with (inverses of) infinite matrices). It is akin to the prefactors in~(\ref{S-Smirror-app}) which can be given in closed form, as in~(\ref{easy-part-real},\ref{easy-part-mirror}) for instance.

The square measure $\mu_a(u)$ for any excitation $a$ can be obtained from the corresponding direct transition $P_{aa}(u|v)$ by applying
\beqa
\mu_{a}(u)=\frac{i}{\underset{v=u}{\operatorname{{\rm residue}}}\, F_{aa}(u,v)} \exp\Big[-2\kappa_a(u)^t \cdot \mathbb{Q} \cdot\mathbb{M}\cdot\kappa_a(u)+2 \tilde\kappa_a(u)^t\cdot \mathbb{Q} \cdot\mathbb{M}\cdot\tilde\kappa_b(u)\Big] \,. \la{muSummary}
\eeqa

In summary, in order to determine the transitions we simply need to specify the prefactors $F_{ab}$ as well as the infinitive vectors $\kappa, \tilde \kappa$ for all the different excitations. This is done in the appendices~\ref{FCPs} and~\ref{summaryBS} for finite coupling, and in~\ref{pert-exp-app} to leading order at weak coupling.

Note that for the fermions we will separately present the results when the rapidity $u$ is in the large sheet or when it is in the small sheet, in which case we use a checked rapidity $\red{\check{u}}$. More precisely we use the notation 
\beqa
P_{\psi\psi}(u|v)&=&F_{\psi\psi}^{LL}(u,v) \,\exp\Big[2(\kappa_\psi^L(u)-i\tilde\kappa_\psi^L(u))^t\cdot \mathbb{Q} \cdot\mathbb{M}\cdot\kappa_\psi^L(v)+\dots \Big] \,,\nn \\
P_{\psi\psi}(u|\Red{\check v})&=&F_{\psi\psi}^{LS}(u,v) \,\exp\Big[2(\kappa_\psi^L(u)-i\tilde\kappa_\psi^L(u))^t\cdot \mathbb{Q} \cdot\mathbb{M}\cdot\kappa_\psi^S(v)+\dots \Big] \,, \la{examples} \\
P_{\psi\psi}(\Red{\check u}|\Red{\check v})&=&F_{\psi\psi}^{SS}(u,v) \,\exp\Big[2(\kappa_\psi^S(u)-i\tilde\kappa_\psi^S(u))^t\cdot \mathbb{Q} \cdot\mathbb{M}\cdot\kappa_\psi^S(v)+\dots \Big]\, , \nn
\eeqa
et cetera. This is of course redundant since the above pentagon transitions are all related to one another by analytic continuation from small to large sheet. However, given that this analytic continuation is a bit involved and does not commute with perturbation theory, it is convenient to summarize all possibilities as if they were independent. We should add that even though the full expressions in (\ref{examples}) are analytically related to each other their building blocks, when taken individually, are not in a such a relationship with respect to one another. For example, $\kappa_\psi^S(u) \neq \kappa_\psi^L(\Red{\check u})$, $F_{\psi\psi}^{LS}(u,v)\neq F_{\psi\psi}^{LL}(u,\Red{\check{v}})$, and so on.
In appendix \ref{continuationAp} we will explain how to perform the relevant analytic continuations in detail thus showing that the checked and un-checked fermionic transitions are indeed related as they should. 

\subsection{Finite coupling proposals}\la{FCPs}
The $F_{ab}$ prefactors are given explicitly as
\begin{eqnarray}
F_{\phi\phi}(u,v)\!\!\!&=&\!\!\! \frac{\Gamma(iu-iv)}{g^2\Gamma(\frac{1}{2}+iu)\Gamma(\frac{1}{2}-iv)}\times \\
&& \exp{\bigg[\int\limits_{0}^{\infty}\frac{dt}{t}(J_{0}(2gt)-1)\frac{J_{0}(2gt)+1-e^{t/2}(e^{-iut}+e^{ivt})}{e^{t}-1}  \bigg]}\,, \notag\\
F_{FF}(u,v)\!\!\!&=&\!\!\! \frac{-\Gamma(iu-iv)}{g^2\Gamma(\frac{3}{2}+iu)\Gamma(\frac{3}{2}-iv)}\sqrt{\left(x^{+}y^{-}-g^2\right)\left(x^{-}y^{+}-g^2\right)\left(x^{+}y^{+}-g^2\right)\left(x^{-}y^{-}-g^2\right)}\times \nn \\
&&\exp{\int\limits_{0}^{\infty}\frac{dt}{t}(J_{0}(2gt)-1)\frac{J_{0}(2gt)+1-e^{-t/2}(e^{-iut}+e^{ivt})}{e^{t}-1} }\, , \\
F_{F\bar{F}}(u,v)\!\!\!&=&\!\!\! \frac{\Gamma(2+iu-iv)\,x^+x^-y^+y^-}{\Gamma(\frac{3}{2}+iu)\Gamma(\frac{3}{2}-iv)\sqrt{\left(x^{+}y^{-}-g^2\right)\left(x^{-}y^{+}-g^2\right)\left(x^{+}y^{+}-g^2\right)\left(x^{-}y^{-}-g^2\right)}}\times \nn \\
&&\exp{\int\limits_{0}^{\infty}\frac{dt}{t}(J_{0}(2gt)-1)\frac{J_{0}(2gt)+1-e^{-t/2}(e^{-iut}+e^{ivt})}{e^{t}-1} }\, , \\
F_{\psi\psi}^{LL}({u},{v})\!\!\!&=&\!\!\! \frac{\Gamma(iu-iv)\sqrt{x y \left(x y-g^2\right)}}{g^2\Gamma(1+iu)\Gamma(1-iv)}\times \la{psiF}\\
&&\exp{\int\limits_{0}^{\infty}\frac{dt}{t}(J_{0}(2gt)-1)\frac{J_{0}(2gt)+1-(e^{-iut}+e^{ivt})}{e^{t}-1} }\, , \notag \\
F^{LL}_{\psi\bar\psi}({u},{v})\!\!\!&=&\!\!\! \frac{\Gamma(1+iu-iv)\sqrt{xy}}{\Gamma(1+iu)\Gamma(1-iv)\sqrt{xy-g^2}}\times\\
&&\exp{\int\limits_{0}^{\infty}\frac{dt}{t}(J_{0}(2gt)-1)\frac{J_{0}(2gt)+1-(e^{-iut}+e^{ivt})}{e^{t}-1} }\,  ,  \notag \\
F^{LS}_{\psi\psi}({u},{ v})\!\!\!&=&\!\!\!\frac{i}{x(v)}/\sqrt{1-\frac{g^2}{x(u)x(v)}}\, , \qquad\qquad  F^{LS}_{\psi\bar\psi}({u},{ v}) =   \sqrt{1-\frac{g^2}{x(u)x(v)}}\, ,   \\ \la{psiF2}
F^{SS}_{\psi\psi}(u,v)\!\!\!&=&\!\!\! \frac{i}{u-v} \sqrt{1-\frac{g^2}{x(u)x(v)}}\, ,   \qquad\qquad F^{SS}_{\psi\bar\psi}({ u},{ v})= 1/\sqrt{1-\frac{g^2}{x(u)x(v)}}\, , \la{psiF3}
\end{eqnarray}
where $x=x(u), y=y(v), x^{\pm}=x(u\pm i/2), y^{\pm}=x(v\pm i/2)$.

For the source terms~\cite{BenDispPaper}, it is enough to give the expressions corresponding to excitations with, say, positive $U(1)$ charge, since $\kappa_{a} = \kappa_{\bar{a}}$ and similarly for $\tilde{\kappa}_a$. We have then
\begin{eqnarray}
\kappa_{\phi}(u)_{j} &=&  -\int\limits_{0}^\infty \frac{dt}{t} \frac{J_j(2gt)(J_0(2gt)-\cos(ut)e^{t/2})}{e^t-1}\, , \\
\tilde\kappa_{\phi}(u)_j&=&\int\limits_{0}^\infty \frac{dt}{t} (-1)^{j+1}\frac{J_j(2gt) \sin(u t) e^{t/2}}{e^t-1}\, , \\
\kappa_{F}(u)_j&=& -\int\limits_{0}^\infty \frac{dt}{t} \frac{J_j(2gt)(J_0(2gt)-\cos(ut)\[e^{t/2}\]^{(-1)^{j}})}{e^t-1}\, , \\
\tilde\kappa_{F}(u)_j&=&\int\limits_{0}^\infty \frac{dt}{t} (-1)^{j+1}\frac{J_j(2gt) \sin(u t) \[e^{t/2}\]^{(-1)^{(j+1)}}}{e^t-1}\, , \\
\kappa_{\psi}^L({ u})_j &=& -\int\limits_{0}^\infty \frac{dt}{t} \frac{J_j(2gt)(\cos(ut)-J_0(2gt))}{e^t-1} - \frac{(-1)^{j/2}\delta^{(+)}_{j}}{2j}\(\frac{g}{x(u)}\)^j\, ,  \la{psikappa}\\
\tilde\kappa^L_{\psi}({ u})_j &=&\int\limits_{0}^\infty \frac{dt}{t} (-1)^{j+1}\frac{J_j(2gt) \sin(u t) }{e^t-1} + \frac{(-1)^{(j-1)/2}\delta^{(-)}_{j} }{2j}\(\frac{g}{x(u)}\)^j\, , \la{psikappatilde}\\
\kappa^S_{\psi}({ u})_j&=& + \frac{(-1)^{j/2}\delta^{(+)}_{j}}{2j}\(\frac{g}{x(u)}\)^j \la{psikappa2}\, ,\\
\tilde\kappa_{\psi}^S({ u})_j&=&- \frac{(-1)^{(j-1)/2}\delta^{(-)}_{j}}{2j}\(\frac{g}{x(u)}\)^j \la{psikappatilde2}  \, ,
\end{eqnarray}
where $\delta^{(\pm)}_{j} = \ft{1}{2}(1\pm (-1)^j)$ and $j = 1, 2,  \ldots$\,.

Finally, the energy and momentum of the excitations are given by 
\begin{eqnarray}
E_{F}({u})=&1+4g\, \[\mathbb{Q}\cdot \mathbb{M}\cdot {\kappa}_F({u})\]_1\, , &\qquad \qquad p_{F}({u})=2u-4g\, \[\mathbb{Q}\cdot \mathbb{M}\cdot \tilde{\kappa}_F({u})\]_1\, ,\nn   \\
E_{\phi}({u})=&1+4g\, \[\mathbb{Q}\cdot \mathbb{M}\cdot {\kappa}_\phi({u})\]_1\, , &\qquad \qquad p_{\phi}({u})=2u-4g\, \[\mathbb{Q}\cdot \mathbb{M}\cdot \tilde{\kappa}_\phi({u})\]_1\, ,\la{dispersion} \\
E_{\psi}({ u})=&1+4g\, \[\mathbb{Q}\cdot \mathbb{M}\cdot {\kappa}^L_\psi({u})\]_1\, , &\qquad \qquad p_{\psi}({ u})=2u-4g\, \[\mathbb{Q}\cdot \mathbb{M}\cdot \tilde{\kappa}^L_\psi({ u})\]_1\, ,\nn \\ 
E_{\psi}(\Red{\check u})=&1+4g\, \[\mathbb{Q}\cdot \mathbb{M}\cdot \kappa^S_\psi(u)\]_1\, , &\qquad \qquad p_{\psi}(\Red{\check u})={\color{white}2u}-4g\, \[\mathbb{Q}\cdot \mathbb{M}\cdot \tilde{\kappa}^S_\psi(u)\]_1\, . \nn
\end{eqnarray}

\subsection{The first bound-state $D_zF_{z-}$} \la{summaryBS}
We report here the final expressions for the measure, energy, and momentum, of the bound-state discussed in the main text. The derivation of the measure will be presented in \cite{toappear} while the energy and momentum were obtained in~\cite{BenDispPaper}. We have
\begin{eqnarray}
&&\mu_{DF}(u) = \frac{\pi u (u^2+1)}{\sinh({\pi u}) (x^{++} x^{--}-g^2)\sqrt{((x^{++})^2-g^2)((x^{--})^2-g^2)}} \times \\
&&\exp\[\int_0^\infty \frac{dt}{t} \frac{(J_0(2gt)-1)(2e^{-t}\cos(u t)-J_0(2gt)-1)}{e^t-1}   \] e^{2 \tilde \kappa_{DF}(u)\cdot \mathbb{Q}\cdot \mathbb{M} \cdot  \tilde \kappa_{DF}(u) - 2  \kappa_{DF}(u)\cdot \mathbb{Q}\cdot \mathbb{M} \cdot  \kappa_{DF}(u)} \,,\notag
\end{eqnarray}
where $x^{++}=x(u+i)$, $x^{--}=x(u-i)$, and the vectors $\kappa_{DF}$ and $\tilde\kappa_{DF}$ are
\beqa
\[\kappa_{DF}(u)\]_j&=& -\int\limits_{0}^\infty \frac{dt}{t} \frac{J_j(2gt)(J_0(2gt)-\cos(ut)\[e^{t}\]^{(-1)^{j}})}{e^t-1}  \,, \\
\[\tilde\kappa_{DF}(u)\]_j&=&\int\limits_{0}^\infty \frac{dt}{t} (-1)^{j+1}\frac{J_j(2gt) \sin(u t) \[e^{t}\]^{(-1)^{(j+1)}}}{e^t-1}  \,.
\eeqa
The energy and momentum of the bound-state are given by 
\beqa
E_{DF}({u})=&2+4g\, \mathbb{Q}\cdot \mathbb{M}\cdot {\kappa}_{DF}({u}) &,\qquad \qquad p_{DF}({u})=2u-4g\, \mathbb{Q}\cdot \mathbb{M}\cdot \tilde{\kappa}_{DF}({u})   \,.
\eeqa

\subsection{Perturbative expansions}\label{pert-exp-app}
Here we present the first non-trivial orders in the weak coupling expansion of the several ingredients used in this paper. 
For the transitions we have
\begin{eqnarray*}
P_{FF}(u|v)&=&-\frac{1}{g^2} \frac{\Gamma(iu-iv)}{\Gamma(-\ft{1}{2}+iu)\Gamma(-\ft{1}{2}-iv)} +\mathcal{O}(g^0)\, ,\\
P_{F\bar{F}}(u|v)&=&\frac{\Gamma(2+iu-iv)}{\Gamma(\ft{3}{2}+iu)\Gamma(\ft{3}{2}-iv)} +\mathcal{O}(g^2)\, ,\\
P_{\phi\phi}(u|v)&=&\frac{1}{g^2} \frac{\Gamma(iu-iv)}{\Gamma(\ft{1}{2}+iu)\Gamma(\ft{1}{2}-iv)} +\mathcal{O}(g^0)\, , \\
P_{\psi\psi}({u}|{v})&=&\frac{1}{g^2} \frac{\Gamma(iu-iv)}{\Gamma(iu)\Gamma(-iv)} +\mathcal{O}(g^0)\, , \\
P_{\psi\bar\psi}({u}|{v})&=&\frac{\Gamma(1+iu-iv)}{\Gamma(1+iu)\Gamma(1-iv)} +\mathcal{O}(g^2)\, ,\\
P_{\psi\psi}({u}|\Red{\check v})&=&-\frac{i}{v}+\mathcal{O}(g^2)\, ,\\
P_{\psi\bar\psi}({u}|\Red{\check v})&=&1+\mathcal{O}(g^2)\, , \\
P_{\psi\psi}(\Red{\check u}|\Red{\check v})&=&\frac{i}{u-v} +\mathcal{O}(g^2)\, , \\
P_{\psi\bar\psi}(\Red{\check u}|\Red{\check v})&=&1+\mathcal{O}(g^2) \, .
\end{eqnarray*}
The measure are given by
\begin{eqnarray}
\mu_{F}({u})&=&-\frac{\pi g^2}{\(u^2+\frac{1}{4}\)\cosh(\pi u)}+\mathcal{O}(g^4)\, ,  \nn\\
\mu_{DF}({u})&=&\frac{\pi u g^2}{\(u^2+1\)\sinh(\pi u)}+\mathcal{O}(g^4)\, ,  \nn\\
\mu_{\phi}({u})&=&\frac{\pi g^2}{ \cosh(\pi u)}+\mathcal{O}(g^4)\, ,\la{measureweak} \\
\mu_{\psi}({u})&=&\frac{\pi g^2}{u \sinh(\pi u)}+\mathcal{O}(g^4)\, , \nn\\
\mu_{\psi}(\Red{\check u})&=&-1+\mathcal{O}(g^2) \, . \nn
\end{eqnarray}
Finally, the dispersion relations read
\beq
\begin{array}{ll} \\
E_{F}({u}) \, \,  \, =1+2g^2 (\psi(\frac{3}{2}-iu)+\psi(\frac{3}{2}+iu)-2\psi(1) ) + \mathcal{O}(g^4)\, ,  &\qquad  p_{F}({u})\, \,  \,\,=2u+\mathcal{O}(g^2)\,, \vspace{.2cm}\\ 
E_{DF}(u)=1+2g^2 (\psi(2-iu)+\psi(2+iu)-2\psi(1) ) + \mathcal{O}(g^4)\, , &\qquad  p_{DF}({u})=2u+\mathcal{O}(g^2)\,,\vspace{.2cm}\\ 
E_{\phi}({u})\, \,  \, =1+2g^2 (\psi(\frac{1}{2}-iu)+\psi(\frac{1}{2}+iu)-2\psi(1) ) + \mathcal{O}(g^4)\, ,  &\qquad  p_{\phi}({u})\, \,  \,\,=2u+\mathcal{O}(g^2)\,,\vspace{.2cm}\\ 
E_{\psi}({u})\, \,  \, =1+2g^2 (\psi(1-iu)+\psi(1+iu)-2\psi(1) ) + \mathcal{O}(g^4)\, , &\qquad  p_{\psi}({u})\, \,  \,\,=2u+\mathcal{O}(g^2)\,,\vspace{.2cm}\\ 
E_{\psi}(\Red{\check u})\, \,  \, = 1+\frac{4 \zeta(3) g^6}{u^2}+ \mathcal{O}(g^8)\, , &\qquad  p_{\psi}(\Red{\check u})\, \, \, =\frac{2g^2}{u}+\mathcal{O}(g^4) \,,
\end{array} \nn
\eeq
where
\beq
\psi(u) = \frac{\partial }{\partial u}  \log\Gamma(u) \nn \,.
\eeq

\subsection{Fermions and their analytic continuation} \la{continuationAp}
The several pentagon transitions, measures, and dispersion relations have a complicated analytic structure as functions of the Bethe rapidity $u$. In particular, they have several cuts. In some cases we need to go through these cuts and investigate what is happening on the following sheets. One such case in the when we integrate over the fermion momentum and use its dispersion relation $p_\psi(u)$ to map the integral over $p$ into an integral over the rapidity $u$. As explained in the main text, the corresponding integral over $u$ is taking place in the two sheeted Riemann surface defined by the Zhukowsky map
\beq
\frac{x(u)}{g}+\frac{g}{x(u)}=\frac{u}{g}\, .
\eeq
rather than on the complex plane. This surface can be viewed as two sheets, that is two copies of the $u$-plane, glued together alone the interval $u^2 < (2g)^2$. In this paper, we always use $x(u)$ to denote the map $u \to x$ with $|x(u)|>g$, that corresponds to the first sheet or \textit{large} sheet (LS). I.e.
\begin{center} \verb" x[u_]=(u+Sqrt[u-2g]Sqrt[u+2g])/2"\,, \end{center}
with $x(u) = u + O(g^2/u)$ at large rapidity. As we perform a monodromy around any one of the two branchpoints at $u=\pm 2g$ we go to the second or \textit{small} sheet (SS) where $|x(u)|<g$. This is equivalent to $x(u) \to g^2/x(u)$. Let us emphasize once more that for us $x(u)$ will always indicates the branch where $|x(u)|>g$ so that in order to maintain continuity we need to flip by hand $x\to g^2/x$ when we go through cuts. This is different from the convention used in~\cite{BenDispPaper} where $x(u)$ was denoting either $\ft{1}{2}(u+\sqrt{u^2-4g^2})$ or $\ft{1}{2}(u-\sqrt{u^2-4g^2})$ depending on whether $|x(u)|$ was bigger or smaller than $g$. A function $f({\red \check u})$ indicates the analytic continuation of $f(u)$ along a path that starts at $u$ in the large sheet, goes through the Zhukowsky cut and is evaluated at $u$ again but now in the small sheet. For example, with our conventions we have
\beq
x({\red \check u}) =g^2/x(u) \,.  \la{x1x}
\eeq

The goal of this section is to illustrate the fermion analytic continuations by analytically continuing the fermion transition $P_{\psi\psi}(u|v)$ into $P_{\psi\psi}({u}|\Red{\check v})$ and then $P_{\psi\psi}(\Red{ \check u}|\Red{\check v})$. In other words, our goal is to show that (\ref{examples}) are indeed analytically continuations of one another. 
To make this task easier it is instructive to first introduce a few notation and some simple identities

\subsubsection*{Some useful vectors and identities}
It is very useful to define two infinite vectors with components
\beqa
\[ v(u) \]_k &\equiv&  -\frac{ \delta_{\text{$k$ even}}}{2k(-1)^{k/2}} \[\(\frac{g}{x(u)}\)^k+\(\frac{g}{x(u)}\)^{-k}\] \,, \la{defv}\\
\[ \tilde v(u) \]_k&\equiv& - \frac{ \delta_{\text{$k$ odd}}}{2k(-1)^{(k-1)/2}} \[\(\frac{g}{x(u)}\)^k+\(\frac{g}{x(u)}\)^{-k}\] \,. \la{defvt}
\eeqa
They have no cuts in $u$ since they are invariant under $x\to g^2/x$. I.e., 
\beq
v(\Red{\check u})=v(u) \, , \qquad \tilde v(\Red{\check u})=\tilde v(u)\,.\la{vNoCuts}
\eeq
Also, we introduce a spectral parameter independent vector $\[j\]_k \equiv -2k J_{k}(2gt)$. 
Then we have two simple but very important identities,
\beq
j \cdot v(u) =   \cos(ut)-J_0(2gt) \la{sumj}  \, , \qquad j \cdot \tilde v(u) =   \sin(ut) \,.
\eeq
The action of the kernel of the vectors defined above will be used below. Using the important identities (\ref{sumj}) and the explicit form of the kernel in (\ref{Mdef}) we immediately obtain 
\beqa
\[ \mathbb{K} \cdot v(u) \]_i &=&- \int \frac{dt}{t} \frac{J_i(2gt)}{e^t-1} \( \cos(ut)-J_0(2gt) \)  \, \la{id1} \\
\[ \mathbb{K} \cdot \tilde v(u) \]_i &=&-(-1)^{i+1} \int \frac{dt}{t} \frac{J_i(2gt)}{e^t-1}  \sin(ut) 
\eeqa
{where $u$ is inside the strip $|{\rm Im}(u)|<i$ such that all integrals are properly convergent.}
\subsubsection*{Continuing $v$ through the cut}
Comparing the vector $\[\kappa_{\psi}(v)\]_j $ given in (\ref{psikappa}) with (\ref{id1}) and (\ref{defv}) we recognize that 
\beqa
\[\kappa^L_{\psi}(v)\]_j &=& \[\(\mathbb{I}+\mathbb{K}\)\cdot v(v)\]_j + \frac{ \delta_{\text{$j$ even}}}{2j(-1)^{j/2}} \(\frac{g}{x(v)}\)^{-j}   \la{secondStep} \,.
\eeqa
Since the vector $v$ has no cuts (\ref{vNoCuts}) and the Zhukoswky variables transform trivially (\ref{x1x})), it is straightforward to analytically continue this expression towards the Zhukoswky cut, 
\beqa
\[\kappa^L_{\psi}(\Red{\check v})\]_j &=& \[\(\mathbb{I}+\mathbb{K}\)\cdot v({v})\]_j + \frac{ \delta_{\text{$j$ even}}}{2j(-1)^{j/2}} \(\frac{g}{x(v)}\)^{+j}   \la{secondStep}\, .
\eeqa
We define
\beq
\[ \kappa_{\psi}^{S}(u)\]_j \equiv + \frac{ \delta_{\text{$j$ even}}}{2j(-1)^{j/2}} \(\frac{g}{x(u)}\)^{+j}  \, , \qquad \[\tilde \kappa_{\psi}^{S}(u)\]_j \equiv - \frac{ \delta_{\text{$j$ odd}}}{2j(-1)^{(j-1)/2}} \(\frac{g}{x(u)}\)^{+j}  \la{kappaS}
\eeq
then the previous expression plus its counterpart for $\tilde \kappa$ can be concisely written as 
\beq
 \kappa^L_{\psi}(\Red{\check{v}}) =  \kappa_{\psi}^{S}(v) + \(\mathbb{I}+\mathbb{K}\)\cdot  v(v)\,, \qquad \tilde \kappa_{\psi}^L(\Red{\check{v}}) = \tilde \kappa_{\psi}^{S}(v) - \(\mathbb{I}+\mathbb{K}\)\cdot \tilde v(v)\, . 
\eeq
Since $\mathbb{M}=(\mathbb{I}+\mathbb{K})^{-1}$ we will obtain considerable simplifications when plugging these expressions in (\ref{FandExponent}). In sum, we established that the analytical continuation of the first line in (\ref{examples}) leads to 
\beqa
P_{\psi\psi}(u|\Red{\check v})&=&F^{LS}_{\psi\psi}(u|v) e^{2 \[\kappa_\psi^L (u)-2 i \tilde\kappa_\psi^L(u) \]\cdot \mathbb{Q} \cdot \mathbb{M}\cdot  \kappa_\psi^S(v)+\[2i \kappa_\psi^L(u) -2 \tilde\kappa_\psi^L(u) \]\cdot \mathbb{Q} \cdot \mathbb{M}\cdot  \tilde\kappa_\psi^S(v)} \la{almost}
\eeqa
where 
\beq
F^{LS}_{\psi\psi}(u|v) \equiv F^{LL}_{\psi\psi}(u|\Red{\check v}) e^{\[2 \kappa_\psi^L (u) -2 i \tilde\kappa_\psi^L(u) \]\cdot \mathbb{Q} \cdot v(v)  -\[ 2i \kappa_\psi^L(u) -2 \tilde\kappa_\psi^L(u) \] \cdot \mathbb{Q} \cdot \tilde v(v)} \la{Fdef}
\eeq
At this point, it is important to note that the vectors $\kappa^S_\psi(u)$ and $\tilde\kappa^S_\psi(u)$ defined above are nothing but the vectors in (\ref{psikappa2}) and (\ref{psikappatilde2}) such that we are already very close to bringing (\ref{almost}) to the desired form, namely to the second line in (\ref{examples}). All we have to do now is to show that the prefactor $F^{LS}_{\psi\psi}(u|v) $ defined in (\ref{Fdef}) agrees with (\ref{psiF2}). This can be done rather straightforwardly as we now explain. We will proceed in a rather pedestrian and painfully explicit fashion. 

To simplify the exponential in (\ref{Fdef}) -- which we denote as \texttt{exp} -- we simply need to preform the infinite sums corresponding to the scalar products in this quantity (recall that $\mathbb{Q}$ is a trivial diagonal matrix). That is quite simple to do. Recall that in $\kappa(u)$ we have two terms:
\beqa
\[\kappa(u)\]_j = -\int\limits_{0}^\infty \frac{dt}{t} \frac{J_j(2gt)}{e^t-1}(\cos(ut)-J_0(2gt)) - \frac{ \delta_{\text{$j$ even}}}{2j(-1)^{j/2}} \(\frac{g}{x(u)}\)^{j} \equiv \[a(u)\]_j+\[b(u)\]_j  \nn
\eeqa
and similar in $\tilde \kappa(v)$. The contribution of the second term in $\kappa$ or $\tilde \kappa$ to \texttt{exp} is particularly simple: 
\beqa
&& \sum_{j=1}^\infty \[ \(2 \[b(u)\]_j  -2 i [\tilde b(u)]_j  \) j(-1)^{j+1}  \[v(v)\]_j  -\( 2i \[b(u)\]_j -2 [\tilde b(u)]_j \)  j(-1)^{j+1}   \[\tilde v(v) \]_j \]\nn \\
&=& \sum_{j=1}^\infty 2j \[ - \[b(u)\]_j      \[v(v)\]_j  + [\tilde b(u)]_j      \[\tilde v(v) \]_j \]\nn \\
&=& \sum_{j=1}^\infty 2j \[ - \frac{\delta_{\text{$j$ even}}}{(2j)^2} \[\(\frac{g}{x(u)}\frac{g}{x(v)}\)^j+\(\frac{g}{x(u)}/\frac{g}{x(v)}\)^{j}\]  - \frac{ \delta_{\text{$j$ odd}}}{(2j)^2}   \[\(\frac{g}{x(u)}\frac{g}{x(v)}\)^j+\(\frac{g}{x(u)}/\frac{g}{x(v)}\)^{j}\] \]\nn \\
&=& -\frac{1}{2} \sum_{j=1}^\infty \frac{1}{j}  \[\(\frac{g}{x(u)}\frac{g}{x(v)}\)^j+\(\frac{g}{x(u)}/\frac{g}{x(v)}\)^{j}\]  = \frac{1}{2} \log\(1-\frac{x(v)}{x(u)}\)\(1-\frac{g^2}{x(u)x(v)}\)
\eeqa
or 
\beqa
\texttt{exp}_\text{\, $b$ part}=\sqrt{\(1-\frac{x(v)}{x(u)}\)\(1-\frac{g^2}{x(u)x(v)}\)}
\eeqa
Then we have $\log \(\texttt{exp}_\text{\, $a$ part}\)=\sum\limits_{j=1}^\infty 2j \[ - \([a(u)]_j -i [\tilde a(u)]_j\)      \[v(v)\]_j  +\( [\tilde a(u)]_j-i  [a(u)]_j\)      \[\tilde v(v) \]_j \]$ given by 
\beqa
\sum_{j=1}^\infty \[ \int\limits_{0}^\infty \frac{dt}{t} \frac{2j J_j(2gt)\[v(v)\]_j }{e^t-1}(e^{-i u t}-J_0(2gt))   + \int\limits_{0}^\infty \frac{dt}{t} \frac{2j  J_j(2gt)   \[\tilde v(v) \]_j}{e^t-1}\sin(ut)     \]\nn 
\eeqa
Now we can use (\ref{sumj}) to preform the sums and end up with 
\beqa
\log \(\texttt{exp}_\text{\, $a$ part}\)=-  \int\limits_{0}^\infty \frac{dt}{t} \frac{ (e^{-i u t}-J_0(2gt))(e^{i vt}-J_0(2gt))   }{e^t-1}
\eeqa
Finally, the exponential $\texttt{exp}=\texttt{exp}_\text{\, $a$ part}\texttt{exp}_\text{\, $b$ part}$ needs to be multiplied by the analytically continued pre-factor,  $F_{\psi\psi}(u,\Red{\check v})$. This one is simply obtained by flipping $y=x(v)\to g^2/x(v)$ in (\ref{psiF}). In total we have therefore
\beqa
F^{LS}_{\psi\psi}(u,v)= \frac{\Gamma(iu-iv)}{\Gamma(1+iu)\Gamma(1-iv)} {\sqrt{\frac{x(u)}{x(v)} \left(\frac{x(u)}{x(v)} -1\right)\(1-\frac{x(v)}{x(u)}\)\(1-\frac{g^2}{x(u)x(v)}\)}}\times \, \,  \\
\exp{\int\limits_{0}^{\infty}\frac{dt}{t}\bigg[\frac{(J_{0}(2gt)-1)(J_{0}(2gt)+1-e^{-iut}-e^{ivt})- (e^{-i u t}-J_0(2gt))(e^{i vt}-J_0(2gt))   }{e^t-1} \bigg]}\, .\notag \la{final}
\eeqa
Finally, it turns out that this integral can be evaluated. The final result is remarkably simple and yields   
\beq
F_{\psi\psi}^{LS}(u,v) = \frac{i}{x(v)}/\sqrt{1-\frac{g^2}{x(u)x(v)}} \la{nice}
\eeq
which, as promised, perfectly agrees with (\ref{psiF2}).

\subsubsection*{Continuation in $u$}
We now continue $u$ in to the second sheet as well. We obtain 
\beqa
P(\Red{\check u}| \Red{\check v})&=&F_{\psi\psi}^{LS}(\Red{\check u},v) e^{2  \kappa^S(v)\cdot \mathbb{Q} \cdot v(u)+2 i \kappa^S(v) \cdot \mathbb{Q} \cdot \tilde v(u)+2i  \tilde\kappa^S(v)\cdot \mathbb{Q} \cdot v(u) +2  \tilde\kappa^S(v) \cdot \mathbb{Q} \cdot \tilde v(u) }  \nn \\
&& e^{2  \kappa^S(v)\cdot \mathbb{Q} \cdot \mathbb{M}\cdot \kappa^S(u)-2 i \kappa^S(v) \cdot \mathbb{Q} \cdot \mathbb{M}\cdot  \tilde\kappa^S(u)+2i  \tilde\kappa^S(v)\cdot \mathbb{Q} \cdot \mathbb{M}\cdot \kappa^S(u) -2  \tilde\kappa^S(v) \cdot \mathbb{Q} \cdot \mathbb{M}\cdot \tilde\kappa^S(u) }  \la{Pfin} 
\eeqa
Which is already of the form of the last line in (\ref{examples}) provided we identify the first line with $F_{\psi\psi}^{SS}({ u},v) $. Indeed, evaluating the first line in (\ref{Pfin}) as above we perfectly reproduce (\ref{psiF3}). 
The measure for small rapidity is now obtained from the residue of the pentagon transition expression at $u=v$ where both rapidities are in the small sheet. That is, 
\beqa
\mu_\psi({\red\check{u}})&=&-  \frac{e^{-2  \kappa^\text{\,small}(u)\cdot \mathbb{Q} \cdot \mathbb{M}\cdot \kappa^\text{\,small} (u)+2  \tilde\kappa^\text{\,small}(u) \cdot \mathbb{Q} \cdot \mathbb{M}\cdot \tilde\kappa^\text{\,small}(u) }}{\sqrt{1-\frac{g^2}{x^2(u)}}}\, .  \la{smallMeasure}
\eeqa

We end this subsection with a few simple observations regarding the analytic properties of the small-fermion transitions and measure, which were alluded to above equation~(\ref{Ismall}). We begin with the transitions. First we note that since $|x(u)|>g$ the square roots in~(\ref{psiF2},\ref{psiF3}) never vanish. Hence, the only singularity in the pre-factors $F$ is the pole at $u=v$ for the direct transition (which ought to be there to yield the corresponding fermion measure). Furthermore, still because of $|x(u)|>g$, the sums in the second line in (\ref{Pfin}) with (\ref{kappaS}) are convergent for any value of $u$ in the second sheet away from the Zhukowsky cut and no singularity arises from this exponential. Similarly, the exponential in (\ref{smallMeasure}) is perfectly regular for any value of $u$ in the full second sheet while the prefactor is always non-vanishing for the same reason as above. In sum, the measure for the small fermion is regular and so is the (inverse of the) fermion transition(s) for any values of the rapidities in the small sheet. At weak coupling, the regularity of the transitions and measure for the small fermions -- in sharp contrast with their large fermion analogues -- can be neatly observed in the corresponding expression in appendix \ref{pert-exp-app}.

\subsubsection*{Starting point for gauge fields}
Similar games could be played for analytically continuing the transitions for other excitations. Of particular interest are the analytic continuation of the gauge fields since we need those to construct the bound-states for example. In this case, the starting point that would replace the fundamental identities (\ref{sumj}) would be 
\beqa
j\cdot  v(u+i/2)  &=&   \cos((u+i/2)t)-J_0(2gt) \nn \\
&=& \cos(ut) e^{t/2} -J_0(2gt) + \frac{1}{2}\(e^{+iut-t/2}-e^{+iut+t/2}\) \la{sumjcompF}
\eeqa
Note that the first term, $ \cos(ut) e^{t/2}$ is dangerous for ${\rm Im}(u)\to 1/2$ because it explodes as $e^{t}$ in this limit but the last term, $ \frac{1}{2}\(e^{+iut-t/2}-e^{+iut+t/2}\)$, is fine. Hence, when we find the combination $ \cos(ut) e^{t/2} -J_0(2gt) $ and need to analytically continue along the upper cut we just add and subtract the last term in (\ref{sumjcompF}), use this equation and then follow more or less what we did above for the fermions. Of course, the shift in $v(u\pm i/2)$ will lead to shifted Zhukowsky variables which are characteristic of the gauge fields (contrary to the fermions where typically unshifted variables appear) and so on...


\begin{thebibliography}{99}


\bibitem{OPEpaper}
  L.~F.~Alday, D.~Gaiotto, J.~Maldacena, A.~Sever and P.~Vieira,
``An Operator Product Expansion for Polygonal null Wilson Loops,''
  JHEP {\bf 1104} (2011) 088
  [arXiv:1006.2788].


\bibitem{heptagonPaper}
A.~Sever and P.~Vieira,
``Multichannel Conformal Blocks for Polygon Wilson Loops,''
JHEP {\bf 1201} (2012) 070
[arXiv:1105.5748].


\bibitem{short}
B.~Basso, A.~Sever and P.~Vieira,
  ``Space-time S-matrix and Flux-tube S-matrix at Finite Coupling,''
  Phys.\ Rev.\ Lett.\  {\bf 111} (2013) 091602
  [arXiv:1303.1396 [hep-th]].


\bibitem{long}
 B.~Basso, A.~Sever and P.~Vieira,
  ``Space-time S-matrix and Flux-tube S-matrix II. Extracting and Matching Data,''
  JHEP {\bf 1401} (2014) 008
  [arXiv:1306.2058 [hep-th]].
 

\bibitem{Dorey}
 P.~Dorey,
 ``Exact S matrices,''
 hep-th/9810026.


\bibitem{Watson}
K.~M.~Watson,
``Some general relations between the photoproduction and scattering of pi mesons,''
Phys.\ Rev.\  {\bf 95} (1954) 228
$\bullet$ F.~A.~Smirnov, ``Form-factors in completely integrable models of quantum field theory,''
Adv.\ Ser.\ Math.\ Phys.\  {\bf 14} (1992) 1. $\bullet$ G.~Mussardo,
 ``Off critical statistical models: Factorized scattering theories and bootstrap program,''
 Phys.\ Rept.\  {\bf 218} (1992) 215.


\bibitem{Lance}
L.~J.~Dixon, J.~M.~Drummond and J.~M.~Henn,
``Bootstrapping the three-loop hexagon,''
JHEP {\bf 1111} (2011) 023
[arXiv:1108.4461]. $\bullet$  L.~J.~Dixon, J.~M.~Drummond, M.~von Hippel and J.~Pennington,
 ``Hexagon functions and the three-loop remainder function,''
 arXiv:1308.2276 [hep-th].


\bibitem{4Loops}
L.~Dixon, J.~ Drummond, C.~Duhr and J.~Pennington, [arXiv:1402.nnnn [hep-th]].

\bibitem{AldayMaldacena}
 L.~F.~Alday and J.~M.~Maldacena,
``Comments on operators with large spin,''
 JHEP {\bf 0711} (2007) 019
[arXiv:0708.0672].


\bibitem{Fioravanti:2008rv}
  D.~Fioravanti, P.~Grinza and M.~Rossi,
  ``Strong coupling for planar N=4 SYM theory: An All-order result,''
  Nucl.\ Phys.\ B {\bf 810} (2009) 563
  [arXiv:0804.2893 [hep-th]].


\bibitem{Freyhult:2007pz}
  L.~Freyhult, A.~Rej and M.~Staudacher,
  ``A Generalized Scaling Function for AdS/CFT,''
  J.\ Stat.\ Mech.\  {\bf 0807} (2008) P07015
  [arXiv:0712.2743 [hep-th]].


\bibitem{GKP}
 S.~S.~Gubser, I.~R.~Klebanov and A.~M.~Polyakov,
 ``A Semiclassical limit of the gauge / string correspondence,''
  Nucl.\ Phys.\ B {\bf 636} (2002) 99
  [hep-th/0204051].


\bibitem{BS}
N.~Beisert and M.~Staudacher, ``Long-range psu(2,2|4) Bethe Ansatze for gauge theory and strings,''
Nucl.\ Phys.\ B {\bf 727} (2005) 1
[hep-th/0504190].


\bibitem{BenDispPaper}
B.~Basso,
``Exciting the GKP string at any coupling,''
Nucl.\ Phys.\ B {\bf 857} (2012) 254
[arXiv:1010.5237].


\bibitem{Straps}
 D.~Gaiotto, J.~Maldacena, A.~Sever and P.~Vieira,
``Pulling the straps of polygons,''
  JHEP {\bf 1112} (2011) 011
  [arXiv:1102.0062].
  

\bibitem{Dorey:2010iy} 
  N.~Dorey and M.~Losi,
  ``Giant Holes,''
  J.\ Phys.\ A {\bf 43}, 285402 (2010)
  [arXiv:1001.4750 [hep-th]].


\bibitem{Belitsky:2006en}
  A.~V.~Belitsky, A.~S.~Gorsky and G.~P.~Korchemsky,
  ``Logarithmic scaling in gauge/string correspondence,''
  Nucl.\ Phys.\ B {\bf 748} (2006) 24
  [hep-th/0601112].
  
\bibitem{BES}
  N.~Beisert, B.~Eden and M.~Staudacher,
  ``Transcendentality and Crossing,''
  J.\ Stat.\ Mech.\  {\bf 0701} (2007) P01021
  [hep-th/0610251].



\bibitem{Wang}
A.~Sever, P.~Vieira and T.~Wang,
``From Polygon Wilson Loops to Spin Chains and Back,''
JHEP {\bf 1212} (2012) 065 [arXiv:1208.0841 [hep-th]].


\bibitem{Belitsky:2011nn}
  A.~V.~Belitsky,
  ``OPE for null Wilson loops and open spin chains,''
  Phys.\ Lett.\ B {\bf 709} (2012) 280
  [arXiv:1110.1063 [hep-th]].


\bibitem{toappear}
B.~Basso, A.~Sever, P.~Vieira,
To appear.

\bibitem{andrei}
 A.~V.~Belitsky, S.~E.~Derkachov and A.~NManashov,
 ``Quantum mechanics of null polygonal Wilson loops,''
 arXiv:1401.7307 [hep-th].

\bibitem{Bootstrapping} 
D.~Gaiotto, J.~Maldacena, A.~Sever and P.~Vieira,
``Bootstrapping Null Polygon Wilson Loops,''
JHEP {\bf 1103}, 092 (2011)
[arXiv:1010.5009 [hep-th]].
  

\bibitem{Papathanasiou:2013uoa}
 G.~Papathanasiou,
 ``Hexagon Wilson Loop OPE and Harmonic Polylogarithms,''
 JHEP {\bf 1311} (2013) 150
 [arXiv:1310.5735 [hep-th]].


\bibitem{BDS}
Z.~Bern, L.~J.~Dixon and V.~A.~Smirnov,
``Iteration of planar amplitudes in maximally supersymmetric Yang-Mills theory at three loops and beyond,''
  Phys.\ Rev.\ D {\bf 72} (2005) 085001
  [hep-th/0505205].


\bibitem{twoLoopsBefore}
  V.~Del Duca, C.~Duhr and V.~A.~Smirnov,
  ``The Two-Loop Hexagon Wilson Loop in N = 4 SYM,''
  JHEP {\bf 1005} (2010) 084
  [arXiv:1003.1702 [hep-th]].
  

\bibitem{twoLoopsAfter}
A.~B.~Goncharov, M.~Spradlin, C.~Vergu and A.~Volovich,
``Classical Polylogarithms for Amplitudes and Wilson Loops,''
Phys.\ Rev.\ Lett.\  {\bf 105}, 151605 (2010)
[arXiv:1006.5703 [hep-th]].


\bibitem{Qbar}
S.~Caron-Huot and S.~He,
``Jumpstarting the All-Loop S-Matrix of Planar N=4 Super Yang-Mills,''
JHEP {\bf 1207} (2012) 174
[arXiv:1112.1060].


\bibitem{AGM}
L.~F.~Alday, D.~Gaiotto and J.~Maldacena,
``Thermodynamic Bubble Ansatz,''
  JHEP {\bf 1109} (2011) 032
  [arXiv:0911.4708].


\bibitem{AMSV}
  L.~F.~Alday, J.~Maldacena, A.~Sever and P.~Vieira,
``Y-system for Scattering Amplitudes,''
  J.\ Phys.\ A {\bf 43} (2010) 485401
  [arXiv:1002.2459].


 \bibitem{arkady} 
 R.~Roiban and A.~A.~Tseytlin,
 ``Spinning superstrings at two loops: Strong-coupling corrections to dimensions of large-twist SYM operators,''
 Phys.\ Rev.\ D {\bf 77}, 066006 (2008)
 [arXiv:0712.2479 [hep-th]].


\bibitem{Zamolodchikov:2013ama}
  A.~Zamolodchikov,
  ``Ising Spectroscopy II: Particles and poles at T>Tc,''
  arXiv:1310.4821 [hep-th].


\bibitem{Zarembo:2011ag}
  K.~Zarembo and S.~Zieme,
  ``Fine Structure of String Spectrum in $AdS_5$ x $S^5$,''
  JETP Lett.\  {\bf 95} (2012) 219
   [Erratum-ibid.\  {\bf 97} (2013) 8,  504]
  [arXiv:1110.6146 [hep-th]].


\bibitem{Basso:2008tx}
  B.~Basso and G.~P.~Korchemsky,
  ``Embedding nonlinear O(6) sigma model into N=4 super-Yang-Mills theory,''
  Nucl.\ Phys.\ B {\bf 807} (2009) 397
  [arXiv:0805.4194 [hep-th]].


\bibitem{Frassek:2013xza}
  R.~Frassek, N.~Kanning, Y.~Ko and M.~Staudacher,
  ``Bethe Ansatz for Yangian Invariants: Towards Super Yang-Mills Scattering Amplitudes,''
  arXiv:1312.1693 [math-ph].
  

\bibitem{spectralstuff}
 L.~Ferro, T.~Lukowski, C.~Meneghelli, J.~Plefka and M.~Staudacher,
  ``Harmonic R-matrices for Scattering Amplitudes and Spectral Regularization,''
  Phys.\ Rev.\ Lett.\  {\bf 110} (2013) 12,  121602
  [arXiv:1212.0850 [hep-th]].
  $\bullet$  L.~Ferro, T.~Lukowski, C.~Meneghelli, J.~Plefka and M.~Staudacher,
  ``Spectral Parameters for Scattering Amplitudes in N=4 Super Yang-Mills Theory,''
  arXiv:1308.3494 [hep-th].
  $\bullet$
  D.~Chicherin, S.~Derkachov and R.~Kirschner,
  ``Yang-Baxter operators and scattering amplitudes in $\mathcal{N} = 4$ super-Yang-Mills theory,''
  arXiv:1309.5748 [hep-th].
  $\bullet$
  D.~Chicherin and R.~Kirschner,
  ``Yangian symmetric correlators,''
  Nucl.\ Phys.\ B {\bf 877} (2013) 484
  [arXiv:1306.0711 [math-ph]].
$\bullet$
N.~Beisert, J.~Broedel and M.~Rosso,
  ``On Yangian-invariant regularisation of deformed on-shell diagrams in N=4 super-Yang-Mills theory,''
  arXiv:1401.7274 [hep-th].


\bibitem{Babujian:2006md} 
  H.~M.~Babujian, A.~Foerster and M.~Karowski,
  ``The Nested SU(N) off-shell Bethe ansatz and exact form-factors,''
  J.\ Phys.\ A {\bf 41}, 275202 (2008)
  [hep-th/0611012].
  $\bullet$
  H.~M.~Babujian, A.~Foerster and M.~Karowski,
  ``Exact form factors of the O(N) $\sigma$-model,''
  JHEP {\bf 1311} (2013) 089
  [arXiv:1308.1459 [hep-th]].


\bibitem{Beisert:2003jj} 
  N.~Beisert,
  ``The complete one loop dilatation operator of N=4 superYang-Mills theory,''
  Nucl.\ Phys.\ B {\bf 676}, 3 (2004)
  [hep-th/0307015].


\bibitem{Golden:2013xva}
  J.~Golden, A.~B.~Goncharov, M.~Spradlin, C.~Vergu and A.~Volovich,
  ``Motivic Amplitudes and Cluster Coordinates,''
  arXiv:1305.1617 [hep-th]. $\bullet$ J.~Golden and M.~Spradlin,
  ``The differential of all two-loop MHV amplitudes in $\mathcal{N}$ = 4 Yang-Mills theory,''
  JHEP {\bf 1309} (2013) 111
  [arXiv:1306.1833 [hep-th]]. $\bullet$ M.~A.~C.~Torres,
 ``Cluster algebras in Scattering Amplitudes with special 2D kinematics,''
 arXiv:1310.6906 [hep-th].
  

\bibitem{GoldenTalk}
  J.~Golden, M.~F.~Paulos, M.~Spradlin and A.~Volovich,
  ``Cluster Polylogarithms for Scattering Amplitudes,''
  arXiv:1401.6446 [hep-th].
 $\bullet$   J.~Golden, M.~F.~Paulos, M.~Spradlin and A.~Volovich,
  ``Cluster Polylogarithms for Scattering Amplitudes,''
  arXiv:1401.6446 [hep-th].
  

\bibitem{SHe}
S.~Caron-Huot and S.~He,
  ``Three-loop octagons and n-gons in maximally supersymmetric Yang-Mills theory,''
  JHEP {\bf 1308} (2013) 101
  [arXiv:1305.2781 [hep-th]].


\bibitem{ArkaniHamed:2010kv}
  N.~Arkani-Hamed, J.~L.~Bourjaily, F.~Cachazo, S.~Caron-Huot and J.~Trnka,
  ``The All-Loop Integrand For Scattering Amplitudes in Planar N=4 SYM,''
  JHEP {\bf 1101} (2011) 041
  [arXiv:1008.2958 [hep-th]].


 \bibitem{ArkaniHamed:2012nw}
  N.~Arkani-Hamed, J.~L.~Bourjaily, F.~Cachazo, A.~B.~Goncharov, A.~Postnikov and J.~Trnka,
  ``Scattering Amplitudes and the Positive Grassmannian,''
  arXiv:1212.5605 [hep-th].


\bibitem{Arkani-Hamed:2013jha}
  N.~Arkani-Hamed and J.~Trnka,
  ``The Amplituhedron,''
  arXiv:1312.2007 [hep-th].


\bibitem{Lipstein:2013xra}
  A.~E.~Lipstein and L.~Mason,
  ``From dlogs to dilogs; the super Yang-Mills MHV amplitude revisited,''
  arXiv:1307.1443 [hep-th].


\bibitem{Frolov:2002av}
  S.~Frolov and A.~A.~Tseytlin,
  JHEP {\bf 0206} (2002) 007
  [hep-th/0204226].


\bibitem{Pmu}
N.~Gromov, V.~Kazakov, S.~Leurent and D.~Volin,
  ``Quantum spectral curve for $AdS_5/CFT_4$,''
  arXiv:1305.1939 [hep-th].


\bibitem{multi-Regge}
J.~Bartels, L.~N.~Lipatov and A.~Sabio Vera,
  ``BFKL Pomeron, Reggeized gluons and Bern-Dixon-Smirnov amplitudes,''
  Phys.\ Rev.\ D {\bf 80} (2009) 045002
  [arXiv:0802.2065 [hep-th]].
  $\bullet$
 J.~Bartels, L.~N.~Lipatov and A.~Sabio Vera,
  ``N=4 supersymmetric Yang Mills scattering amplitudes at high energies: The Regge cut contribution,''
  Eur.\ Phys.\ J.\ C {\bf 65} (2010) 587
  [arXiv:0807.0894 [hep-th]].
  $\bullet$
  J.~Bartels, L.~N.~Lipatov and A.~Prygarin,
  ``Collinear and Regge behavior of 2 -> 4 MHV amplitude in N = 4 super Yang-Mills theory,''
  arXiv:1104.4709 [hep-th].
 $\bullet$
 J.~Bartels, A.~Kormilitzin, L.~N.~Lipatov and A.~Prygarin,
  ``BFKL approach and $2 \to 5$ maximally helicity violating amplitude in ${\cal N}=4$ super-Yang-Mills theory,''
  Phys.\ Rev.\ D {\bf 86} (2012) 065026
  [arXiv:1112.6366 [hep-th]].
  $\bullet$
  L.~J.~Dixon, C.~Duhr and J.~Pennington,
  ``Single-valued harmonic polylogarithms and the multi-Regge limit,''
  JHEP {\bf 1210} (2012) 074
  [arXiv:1207.0186 [hep-th]].
  $\bullet$
  S.~Caron-Huot,
  ``When does the gluon reggeize?,''
  arXiv:1309.6521 [hep-th].
  $\bullet$
  J.~Bartels, A.~Kormilitzin and L.~Lipatov,
  ``Analytic structure of the $n=7$ scattering amplitude in $\mathcal{N}=4$ SYM theory at multi-Regge kinematics: Conformal Regge Pole Contribution,''
  arXiv:1311.2061 [hep-th].


\bibitem{Bartels:2013dja}
J.~Bartels, J.~Kotanski and V.~Schomerus,
  ``Excited Hexagon Wilson Loops for Strongly Coupled N=4 SYM,''
  JHEP {\bf 1101} (2011) 096
  [arXiv:1009.3938 [hep-th]].
$\bullet$
  J.~Bartels, J.~Kotanski, V.~Schomerus and M.~Sprenger,
  ``The Excited Hexagon Reloaded,''
  arXiv:1311.1512 [hep-th].


\bibitem{Bartels:2013dja2}
J.~Bartels, V.~Schomerus and M.~Sprenger,
  ``Multi-Regge Limit of the n-Gluon Bubble Ansatz,''
  JHEP {\bf 1211} (2012) 145
  [arXiv:1207.4204 [hep-th]].


\bibitem{withAdam}
  B.~Basso and A.~Rej,
  ``Bethe Ansaetze for GKP strings,''
  arXiv:1306.1741 [hep-th].


\bibitem{Fioravanti:2013eia}
  D.~Fioravanti, S.~Piscaglia and M.~Rossi,
  ``On the scattering over the GKP vacuum,''
  arXiv:1306.2292 [hep-th].


\bibitem{MoreDispPaper}
 B.~Basso and A.~V.~Belitsky,
``Luescher formula for GKP string,''
Nucl.\ Phys.\ B {\bf 860} (2012) 1
[arXiv:1108.0999].


\bibitem{Berg:1977dp}
  B.~Berg, M.~Karowski, P.~Weisz and V.~Kurak,
  ``Factorized U(n) Symmetric s Matrices in Two-Dimensions,''
  Nucl.\ Phys.\ B {\bf 134} (1978) 125.


\bibitem{Zamolodchikov:1989cf}
  A.~B.~Zamolodchikov,
  ``Thermodynamic Bethe Ansatz in Relativistic Models. Scaling Three State Potts and Lee-yang Models,''
  Nucl.\ Phys.\ B {\bf 342} (1990) 695.


\bibitem{Ambjorn:2005wa}
  J.~Ambjorn, R.~A.~Janik and C.~Kristjansen,
  ``Wrapping interactions and a new source of corrections to the spin-chain/string duality,''
  Nucl.\ Phys.\ B {\bf 736} (2006) 288
  [hep-th/0510171].
$\bullet$ G.~Arutyunov and S.~Frolov,
  ``On String S-matrix, Bound States and TBA,''
  JHEP {\bf 0712} (2007) 024
  [arXiv:0710.1568 [hep-th]].


\bibitem{Giombi:2009gd}
  S.~Giombi, R.~Ricci, R.~Roiban, A.~A.~Tseytlin and C.~Vergu,
  ``Quantum AdS(5) x S5 superstring in the AdS light-cone gauge,''
  JHEP {\bf 1003} (2010) 003
  [arXiv:0912.5105 [hep-th]].
  

\bibitem{superloopskinner}
 L.~J.~Mason, D.~Skinner,
``The Complete Planar S-matrix of N=4 SYM as a Wilson Loop in Twistor Space,''
  JHEP {\bf 1012},  (2010) 018
  [arXiv:1009.2225].


\bibitem{superloopsimon}
 S.~Caron-Huot,
 ``Notes on the scattering amplitude / Wilson loop duality,''
 [arXiv:1010.1167].


\bibitem{Bourjaily:2013mma} 
  J.~L.~Bourjaily, S.~Caron-Huot and J.~Trnka,
 ``Dual-Conformal Regularization of Infrared Loop Divergences and the Chiral Box Expansion,''
  arXiv:1303.4734 [hep-th].


\bibitem{Beisert}
N.~Beisert,
  ``The SU(2|2) dynamic S-matrix,''
  Adv.\ Theor.\ Math.\ Phys.\  {\bf 12} (2008) 945
  [hep-th/0511082].


  

\bibitem{Benna:2006nd}
  M.~K.~Benna, S.~Benvenuti, I.~R.~Klebanov and A.~Scardicchio,
  ``A Test of the AdS/CFT correspondence using high-spin operators,''
  Phys.\ Rev.\ Lett.\  {\bf 98} (2007) 131603
  [hep-th/0611135].
  $\bullet$
  L.~F.~Alday, G.~Arutyunov, M.~K.~Benna, B.~Eden and I.~R.~Klebanov,
  ``On the Strong Coupling Scaling Dimension of High Spin Operators,''
  JHEP {\bf 0704} (2007) 082
  [hep-th/0702028 [HEP-TH]].


\bibitem{Peng}
N.~Dorey and P.~Zhao,
``Scattering of Giant Holes,''
JHEP {\bf 1108} (2011) 134
[arXiv:1105.4596 [hep-th]].


\bibitem{ZamolodchikovS}
A.~B.~Zamolodchikov and A.~B.~Zamolodchikov, ``Relativistic Factorized S Matrix in Two-Dimensions Having O(N) Isotopic Symmetry,''
Nucl.\ Phys.\ B {\bf 133} (1978) 525
[JETP Lett.\  {\bf 26} (1977) 457].

\end{thebibliography}
\end{document}